\documentclass[rnote]{aa} % for the research notes

\usepackage{graphicx}
\usepackage{txfonts}
\usepackage{aalongtable,lscape}

\begin{document}

\title{Stars and brown dwarfs in the $\sigma$~Orionis cluster} 
\subtitle{IV. IDS/INT and OSIRIS/GTC spectroscopy and \textit{Gaia} DR2 astrometry\thanks{Full Tables~\ref{table.observations.int} to~\ref{table.results.spt} are only available at the CDS via anonymous ftp to {\tt cdsarc.u-strasbg.fr} (130.79.128.5) or via {\tt http://cdsweb.u-starsbg.fr/cgi-bin/qcat?J/A+A/***/A***}}}
\titlerunning{GTC+INT spectroscopy and \textit{Gaia} DR2 astrometry in $\sigma$~Orionis}
\author{J. A. Caballero\inst{1}, A. de~Burgos\inst{2,1}, F. J. Alonso-Floriano\inst{3,4}, A. Cabrera-Lavers\inst{5,6,7}, D. Garc\'{\i}a-\'Alvarez\inst{5,6,7}, D. Montes\inst{4}}
\authorrunning{J. A. Caballero et al.}
\institute{
Centro de Astrobiolog\'ia (CSIC-INTA), ESAC, camino bajo del castillo s/n, 
28691 Villanueva de la Ca\~nada, Madrid, Spain
\and
Isaac Newton Group of Telescopes, Apartado de correos 321, 38700 Santa Cruz de La Palma, La Palma, Spain
\and
Leiden Observatory, Leiden University, P.O. Box 9513, 
2300 RA Leiden, 2300 RA Leiden, The Netherlands
\and
Departamento de F\'isica de la Tierra y Astrof\'isica \& IPARCOS-UCM (Instituto de F\'isica de Part\'iculas y del Cosmos de la UCM),
Facultad de Ciencias F\'isicas, Universidad Complutense de Madrid,
28040 Madrid, Spain
\and
Instituto de Astrof\'isica de Canarias, Avenida V\'ia L\'actea, 
38205 La Laguna, Tenerife, Spain
\and
Grantecan S.\,A., Centro de Astrof\'isica de La Palma, Cuesta de San Jos\'e,
38712 Bre\~na Baja, La Palma, Spain 
\and
Departamento de Astrof\'isica, Universidad de La Laguna, 
38205 La Laguna, Tenerife, Spain
}
\date{Received 30 May 2019 / Accepted 12 August 2019}
% \abstract{}{}{}{}{} 
% 5 {} token are mandatory
\abstract 
% context heading (optional), leave it empty if necessary 
{Only a few open clusters are as important for the study of stellar and substellar objects, and their formation and evolution, as the young $\sigma$~Orionis cluster. 
However, a complete spectroscopic characterisation of its whole stellar population is still missing.} 
{We filled most of that gap with a large spectroscopic and astrometric survey of targets towards $\sigma$~Orionis.
Eventually, it will be one of the open clusters with the lowest proportion of interlopers and the largest proportion of confirmed cluster members with known uncontrovertible youth features.} 
{We acquired {317} low-resolution optical spectra with 
%\LEt{Please spell out all acronyms the first time they appear in the paper, followed by the abbreviation in parentheses, both in the abstract and again in the main text. After that, please only use the abbreviation. See A and A language guide Section 5.2.4 www.aanda.org/language-editing}
{the Intermediate Dispersion Spectrograph (IDS)} at the 2.5\,m Isaac Newton Telescope (INT) {and the Optical System for Imaging and low Resolution Integrated Spectroscopy (OSIRIS)} at the 10.4\,m Gran Telescopio Canarias (GTC). 
We measured equivalent widths of Li~{\sc i}, H$\alpha$, and other key lines from these spectra, and determined spectral types.
We complemented this information with \textit{Gaia} DR2 astrometric data and other features of youth (mid-infrared excess, X-ray emission) compiled with Virtual Observatory tools and from the literature.}
{Of the {168} observed targets, we determined for the first time spectral types of {39} stars and  equivalent widths of Li~{\sc i} and H$\alpha$ of {34} and {12} stars, respectively. 
We identified {11} close ($\rho \lesssim$ 3\,arcsec) binaries resolved by \textit{Gaia}, of which three are new,
{14} strong accretors, of which {four} are new and another {four} have H$\alpha$ emission shifted by over 120\,km\,s$^{-1}$,
{two juvenile star candidates in the sparse population of the Ori~OB1b association, and one} spectroscopic binary candidate.
Remarkably, we found {51} non-cluster-members, {35} of which were previously considered as $\sigma$~Orionis members and taken into account in high-impact works on, for example, disc frequency and initial mass function.}
{}
 %maximum 6
\maketitle

\section{Introduction}
\label{section.introduction}

\begin{table*}
  \centering
  \caption[]{Observations of standard stars for the OSIRIS spectral type classification.} 
    \label{table.standards.gtc}
      \begin{tabular}{llcccccc}
\hline
\hline
\noalign{\smallskip}
Name    &GJ &$\alpha$   &$\delta$       & Spectral      &       Ref.$^a$ & Date of               &$t_{\rm exp}$          \\ %
                &       &(J2000)        &(J2000)        & type      &                    & observation   &[s]                            \\ %
\noalign{\smallskip}
\hline
\noalign{\smallskip}
HD~88230            &380        &10 11 22.14    &+49 27 15.3    &K7 V & DK91      &20 Dic 2012       &2 $\times$ 0.5 \\ 
BD+33~1505          &270        &07 19 31.27    &+32 49 48.3    &M0.0 V & AF15      &09 Dic 2012  &2 $\times$ 60  \\ 
BD+02~2098      &328    &08 55 07.62    &+01 32 47.4    &M0.0 V & L\'ep13   &20 Dic 2012  &2 $\times$ 60  \\ 
HD~36395        &205    &05 31 27.40    &--03 40 38.0   &M1.5 V & AF15      &09 Dic 2012 &2 $\times$ 90  \\ 
HD~209290       &846    &22 02 10.26    &+01 24 00.6    &M0.5 V & AF15      &09 Dic 2012    &1 $\times$ 120 \\
LP~379--51$^b$  &3790   &13 31 50.57    &+23 23 20.3    &M2.5 V & Lep13     &16 Mar 2012 &2 $\times$ 30  \\ 
Ross~1022$^b$   &3795   &13 38 37.05    &+25 49 49.7    &M3.0 V & Lep13     &16 Mar 2012 &3 $\times$ 30  \\ 
FN~Vir$^b$              &493.1  &13 00 33.51    &+05 41 08.2    &M4.5 V & Dav15     &16 Mar 2012  &2 $\times$ 60  \\ 
LP~799--7$^b$   &3820   &13 59 10.46    &--19 50 03.5   &M4.5 V & Ria06     &16 Mar 2012 &3 $\times$ 60  \\ 
LP~380--6$^b$   &1179A  &13 48 13.41    &+23 36 48.8    &M5.0 V & New14     &16 Mar 2012 &2 $\times$ 90  \\ 
LP~731--58$^b$  &3622   &10 48 12.58    &--11 20 08.2   &M6.5 V & AF15      &22 Mar 2012 &2 $\times$ 60  \\ 
\noalign{\smallskip}
\hline
      \end{tabular}
      \begin{list}{}{}
      \item[$^{a}$] {\bf References}. 
      DK91: \citet{1991ApJS...77..417K}; 
      Ria06: \citet{2006AJ....132..866R}; 
      Lep13: \citet{2013AJ....145..102L}; 
      New14: \citet{2014AJ....147...20N}; 
      Dav15: \citet{2015AJ....149..106D}; 
      AF15: \citet{2015A&A...577A.128A}.
      \item[$^{b}$] {\bf Notes}. Targets previously observed by \citet{2012A&A...546A..59C}.
      \end{list}
 \end{table*}

 There are a few nearby young clusters, such as the Pleaides and the Hyades, whose stellar (and substellar) populations have been investigated in detail for decades, which translates into hundreds of known Pleiads and Hyads \citep[][to cite just
 a few examples]{1958ApJ...128...31J,1962ApJ...135..736H,1971ApJ...166..593C,1984ApJ...278..679V,1986ApJ...302L..49B,1993AJ....106.1059S,1998A&A...331...81P,1998ApJ...499L.199S,2018A&A...613A..63B,2018ApJ...856...40M}.
As a result, these clusters are cornerstones for the study of the formation and evolution of stars.

The very young $\sigma$~Orionis open cluster ($\tau$ $\sim$3\,Ma, $d \sim${388}\,pc), near the Horsehead nebula in the Ori OB1b association, is one of the most attractive and most visited regions for night-sky observers, both professional and amateur. 
This cluster was discovered by \citet{1967PASP...79..433G}, re-discovered by \citet{1996PhDT........63W}, introduced to a new era by \citet{1999ApJ...521..671B} and \citet{2000Sci...290..103Z}, and finally reviewed by \citet{2008hsf1.book..732W} and \citet{2008A&A...478..667C}.  
At that stage, $\sigma$~Orionis became one of the best-studied young open clusters together with the Pleiades and the Hyades, and at the same level of importance 
%\LEt{"at the same level of knowledge is not a commonly understood formulation".}
as other very young open clusters and star-forming regions, such as the Trapezium, Taurus-Auriga, or $\rho$~Ophiuchi, which are affected by (spatially variable) extinction.
A major event in the history of $\sigma$~Orionis studies was the computation of a continuous mass function from about 20\,$M_\odot$ to only about 0.005\,$M_\odot$ by \citet{2012ApJ...754...30P}.
For that, these latter authors complemented their very deep VISTA survey with the Mayrit catalogue of $\sigma$~Orionis stars and brown dwarfs \citep{2008A&A...478..667C}, which remains the most comprehensive database of cluster members.
This catalogue was the first item of the paper series ``Stars and brown dwarfs in the $\sigma$~Orionis cluster'', which was continued by {\citet{2010A&A...514A..18C}, who conducted a pre-{\em Gaia} proper-motion analysis,} and by \citet{2012A&A...546A..59C}, who reviewed the most relevant publications on the cluster between 2008 and 2012, published a number of low-resolution optical spectra of member stars and brown dwarfs that lacked previous spectroscopy, and updated the Mayrit catalogue.
However, the results presented by \citet{2012A&A...546A..59C} were superseded by the spectroscopic census of \citet{2014ApJ...794...36H}, which constitutes the largest homogeneous spectroscopic data set of the $\sigma$~Orionis cluster to date. 

In the last lustrum, the Mayrit catalogue has been used for searching for pulsations and photometric variability in substellar objects \citep{2014ApJ...796..129C,2017A&A...608A..66E}, 
extending spectroscopic surveys to the lowest masses and the largest separations to the cluster centre \citep{2015AJ....150..100K,2017ApJ...842...65Z,2018ApJS..236...27C}, characterising discs with theoretical models and ALMA and {\em Spitzer} observations \citep{ 2017RMxAA..53..275A,2017AJ....153..240A,2018MNRAS.478.2700W, 2018ApJ...867..116P}, 
identifying large-scale Herbig-Haro jets driven by proto-brown dwarfs \citep{2017ApJ...844...47R,2019MNRAS.486.4114R},
and carefully analysing the massive multiple stellar system that gives the name to the cluster \citep{2015ApJ...799..169S,2016AJ....152..213S},
not counting the studies performed by the first author and his colleagues on multiplicity \citep{2014Obs...134..273C,2016Obs...136..226C,2018AN....339...60C} and parallaxes and proper motions \citep{2017AN....338..629C,2018RNAAS...2b..25C}. 

In spite of all these studies, there are still a number of photometric cluster member candidates in both stellar and substellar regimes that lack membership confirmation.
{Furthermore, beyond the cluster core and at angular separations greater than 20\,arcmin from the cluster centre at the eponymous $\sigma$~Ori Trapezium-like system \citep{2008MNRAS.383..375C}, the contamination by distinct populations of young stars in the Ori~OB1b association continues to increase.
Neighbouring populations have been reported towards the younger Horsehead and Flame nebulae to the east and the older $\epsilon$~Orionis cluster to the north and west, but there are also hints of a sparse population of 5--30\,Ma-old stars in the cluster foreground.
Such distinct populations may have different photometric, kinematic, and spectroscopic properties from those of $\sigma$~Orionis \citep{2006MNRAS.371L...6J, 2007A&A...462L..23S, 2008A&A...488..167S, 2008A&A...485..931C, 2008MNRAS.385.2210M, 2014ApJ...794...36H, 2018AJ....156...84K, 2019MNRAS.486.4114R}, but disentangling them in the cluster halo ($\rho \approx$ 20--30\,arcmin) requires a careful analysis.}
Here we use low-resolution optical spectroscopy, complemented with {\em Gaia} astrometry, for studying the $\sigma$~Orionis membership of a large sample of stars and brown dwarfs.

%========================================================================== 

\section{Data and analysis}
\label{section.data_analysis}

\subsection{Sample}
\label{subsection.sample}

\begin{table*}
  \centering
  \caption[]{Date of observations and instrument configurations.} 
    \label{table.inst_specs}
      \begin{tabular}{lllcccc}
\hline
\hline
\noalign{\smallskip}
Date of                 &Instrument     &Grism  &Slit   width           &Resolution         & $\Delta \lambda$      & $\lambda_{\rm central}$ \\  %
observations    &                       &                               &[arcsec]                       &                               & [\AA]           & [\AA] \\ %
\noalign{\smallskip}
\hline
\noalign{\smallskip}
22--27 Feb 2007                 &IDS            &R1200Y &1.515  &1300           &5750--6850     & 6300 \\  % 
28 Feb 2007                             &IDS            &R150V  &1.515  &460            &3700--7500     & 5500 \\  % 
04--22 Mar 2012                 &OSIRIS         &R1000B &1.23   &660            &3700--7800     & 5750 \\  % 
Nov--Dec 2012, Jan 2013 &OSIRIS         &R1000B &1.23   &660            &3700--7800     & 5750 \\  % 
\noalign{\smallskip}
\hline
      \end{tabular}
 \end{table*}

We obtained {317} low-resolution, long-slit optical spectra of {168} different sources comprising mostly high-to-low-mass stars selected from different surveys.
Our sample includes OB, Herbig Ae/Be, T~Tauri, variable, and binary stars, {as well as brown dwarfs,} in $\sigma$~Orionis, but also young stars in neighbouring star-forming regions, fore- and background stars, and even one active galaxy. 
Most of them have youth features such as X-ray emission, mid-infrared excess, strong broad H$\alpha$ emission, and Li~{\sc i} absorption, and have been identified as genuine $\sigma$~Orionis cluster members using different techniques (e.g. \cite{2008hsf1.book..732W,2008A&A...478..667C,2017AN....338..629C,2014ApJ...794...36H} -- see below), while others are new member candidates or photometric candidates. 
As detailed in Sect.~\ref{subsection.observations}, {142} of those targets were observed with the Isaac Newton Telescope (INT) and {47} with the Gran Telescopio Canarias (GTC).
For comparison purposes, {21} of them were observed with both telescopes. 
With GTC, we also observed {11} nearby standard stars with spectral types between K7\,V and M6.5\,V, which are listed in Table \ref{table.standards.gtc}.
The GTC spectra corresponding to {11} sources towards $\sigma$~Orionis and {six} standard stars were previously published by \citet{2012A&A...546A..59C}.
% Note: Galaxy 2E 1456 is not listed in Table so03 and therefore there are only 10 asterisks 

%__________________________________________________ 

\subsection{Spectroscopy}
\label{subsection.spectroscopy}

%__________________________________________________ 

\subsubsection{Observations and reduction}
\label{subsection.observations}

\paragraph{The IDS at the INT.} 
Observations at intermediate and low resolution were carried out using the Intermediate Dispersion Spectrograph (IDS) at the 2.5\,m INT located at the Observatorio del Roque de los Muchachos in La Palma, Canary Islands, Spain. 
The IDS is a long-slit spectrograph attached at the INT Cassegrain focus equipped with a 235\,mm focal length camera with two different possible 4k$\times$2k CCD detectors. 
For this work we used the EEV10 CCD. 
This latter is a blue-sensitive ($>$50\,\% efficiency from 400 to 700\,nm) CCD that provides a spatial scale of 0.40\,arcsec\,pixel$^{-1}$ and an unvignetted 3.3\,arcmin (500 pixels) slit length.
All observations with IDS were performed in parallactic angle on six consecutive nights from 22 to 28 February 2007.
We used two different gratings: R1200Y, which provides $>$60\% efficiency at 6000\,{\AA} and a spectral resolution of $\sim$1300, on the first five nights, and R150V, which provides a similar efficiency with a wider wavelength range but three times lower resolution, on the last night. 
Our instrument configuration is in Table~\ref{table.inst_specs}. 

The aim of the telescope proposal was to obtain optical spectroscopy of the 100 brightest young stars in $\sigma$~Orionis, as well as of the maximum number possible of bright interloper stars towards the cluster.
We list the names, equatorial coordinates, and main observing parameters of the {142} observed targets in Table~\ref{table.observations.int}.
We provide both the Mayrit identifier (\cite{2008A&A...478..667C}) and the alternative (discovery) name for $\sigma$~Orionis members. 
We took equatorial coordinates from \textit{Gaia} DR2 (\cite{2016A&A...595A...1G,2018A&A...616A...1G}) in all cases except for a few particular ones, indicated in the table. 
We tabulate coordinates of the primary in close binary systems resolved by \textit{Gaia}  (Sect.~\ref{subsection.gaia}). 
Six stars were observed with the two IDS grisms. 
Several stars were also observed a few times for improving the spectral signal-to-noise ratio but avoiding saturation (e.g. $\sigma$~Ori~AB), {for} daily monitoring (e.g. $\sigma$~Ori~E), or {for minimising contamination using} angles different from parallactic angle (e.g. 2MASS J05384652--0235479 = [BHM2009] SigOri-MAD-23, halfway between $\sigma$~Ori~AB and~E).
%\LEt{this sentence is very convoluted and not all of its content is clear. P¨lease consider rewording and possibly breaking the sentence up into more than one sentence.}. 
The weather during the observations was good in general, but with a variable seeing from 0.8 to 2.0\,arcsec.

Data reduction was carried out using standard tasks within the IRAF software environment. 
The reduction of the {193} IDS spectra included bias and flat-field correction, removal of sky background, optimised aperture extraction, wavelength calibration using Cu-Ar and Cu-Ne arc lamps, instrumental response correction (calibrated with flux standard stars observed on the same nights for another programme), and hot-pixel and cosmic-ray removal. 
We checked the logbooks of the IDS observations for mistakes in target names, as well as for relevant remarks regarding signal-to-noise ratios or observed features. 
For stars with multiple observations and low signal-to-noise-ratio spectra (and the same grism), we combined the individual spectra for a higher signal. 
The spectra of two stars, 2MASS J05384652--0235479 and 2MASS J05381494--0219532, were of poor quality because of contamination from nearby sources and were discarded from the following analysis. 
% 2MASS J05381494--0219532 at NE of NAME SO120532
The spectra of 18 normalised, fully reduced IDS spectra from O9.5+\,V to M6 are shown in Fig.~\ref{figure.ids_sample}.

\paragraph{{The} OSIRIS at the GTC.}
Observations were also carried out using the Optical System for Imaging and low Resolution Integrated Spectroscopy (OSIRIS) tunable imager and spectrograph (\cite{2000SPIE.4008..623C,2010ASSP...14...15C}) at the 10.4 m GTC located at the Observatorio Roque de los Muchachos in La Palma, Canary Islands, Spain. 
OSIRIS has a mosaic of two 4k$\times$2k e2v Marconi CCD44-82-BI detectors, which provides an unvignetted field of view of 7.8 $\times$ 7.8 arcmin$^{2}$ with a spatial scale of 0.127\,arcsec\,pixel$^{-1}$.
To increase the signal-to-noise ratio, we selected the standard operation mode of the instrument, which implements a 2$\times$2 binning mode with a readout speed of 100\,kHz.
All observations were carried out using the R1000B grism, which provides a peak efficiency of 65\% at 5455\,{\AA} and covers the optical wavelength up to 7800\,{\AA}. 
Again, our instrument configuration is in Table~\ref{table.inst_specs}. 

The observations with OSIRIS were performed in service mode on different nights in March, November, and December 2012, and January 2013, through ``D-band'' filler programmes\footnote{D-band proposals (or fillers) require very relaxed observing conditions and cover a wide range of coordinates so that they can be carried out at essentially any moment during the semester. Observations in this band are only executed if the observing conditions do not permit observations of proposals in A, B, or C bands ({\tt http://www.gtc.iac.es/observing/}).} (Table~\ref{table.observations.gtc}). 
Observations before 01 April 2012 were presented by \citet{2012A&A...546A..59C}. 
We list the names, equatorial coordinates, and main observing parameters of the {47} observed targets in Table \ref{table.observations.gtc}. 
As for IDS, we provide both the Mayrit identifiers and the alternative (discovery) name for the cluster members. 
We took coordinates from \textit{Gaia} DR2 except for one source, indicated in the table. 
The aim of this filler programme was to obtain, within the GTC nightly operation schedule, high-quality low-resolution spectra of variable M stars with no previous spectroscopic characterisation and spectral type determination that are relatively bright for a 10m-class telescope (with magnitudes of up to $V \sim$ 19\,mag) but under poor weather conditions. 
However, the weather conditions were not very bad in general except for a few nights with dust and poor seeing. 
Most of the targets were observed at an airmass $\le$ 1.6. %90%

Data reduction was carried out using standard tasks within the IRAF software environment as in \citet{2012A&A...546A..59C}. 
All the {124} spectra of {47} sources towards Orion and {11} standard stars were bias-subtracted, corrected from flat-field using lamp flats from the GTC instrument calibration module, and were calibrated in wavelength using Xe, Ne, and Hg-Ar arc lamps. 
The reduction was followed by a sky-background subtraction and a one-dimension spectrum extraction, which were done taking into account the seeing conditions and exposure times. 
Finally, we applied a hot-pixel and cosmic-ray removal. 
The corresponding instrument-response correction was exactly the same as in \citet{2012A&A...546A..59C}. 
As for IDS, we combined the individual spectra of stars with multiple observations and with low signal-to-noise ratio (and the same grism). 
We discarded from the following analysis the source 2E~1456 (UCM0535--0246), a Seyfert~1 galaxy at $z \sim$ 0.1 previously considered to be a reddened low-mass $\sigma$~Orionis member \citep{2012A&A...546A..59C}.
The spectra of five normalised, fully reduced OSIRIS spectra onto the spectral grid of the {11} standard stars are shown in Fig.~\ref{figure.osiris_sample}.

%__________________________________________________ 

\subsubsection{Equivalent widths}
\label{subsection.ew}

After discarding the two stars with poor-quality spectra and the galaxy, we kept {165} stars and brown dwarfs for the analysis.
In the spectra of all of these objects, we measured equivalent widths (EWs) with errors of key spectral line targets using the IRAF $\texttt{splot}$ task and the ``equivalent width'' function (conversely, we measured pseudo-EWs with respect to the pseudo-continuum in the K and M star spectra). 
For both IDS and OSIRIS spectra, we measured EWs of the Balmer lines H$\alpha$ $\lambda$6562.80\,{\AA}, H$\beta$ $\lambda$4861.33\,{\AA}, and the helium triplet He~{\sc i} D$_{3}$ $\lambda$5875.67\,{\AA} in emission, and the lithium line Li~{\sc i} $\lambda$6707.80\,{\AA} in absorption.
For OSIRIS we also measured the calcium doublet Ca H\&K $\lambda\lambda$3933.66,3968.47\,{\AA} and the Balmer line H$\gamma$ $\lambda$4340.47\,{\AA}, which were present in most of the spectra, but also the Balmer H$\delta$, H$\zeta$, H$\eta$, H10, H11, H12 lines, the Mg~{\sc i} b triplet, the Na~{\sc i} D$_{1-2}$ doublet (measured as a {singlet} because of the low resolution) and the He~{\sc i} line at $\lambda$6678.15\,{\AA} when measurable.
Measured EWs of H$\beta$, He~{\sc i} D$_{3}$, H$\alpha$, Li~{\sc i} (IDS and OSIRIS), Ca~K, and Ca~H (OSIRIS only) are shown in Tables~\ref{table.results.ids} and~\ref{table.results.gtc}.

%__________________________________________________ 

\subsubsection{Spectral types}
\label{subsection.spt}

For each target, we determined spectral types using the spectrum with the highest signal-to-noise ratio. 
For the targets observed at the INT with both R1200Y and R150V grisms we used only R1200Y.
%\LEt{Please check that I have retained your intended meaning.} 
Depending on the instrument and stellar effective temperature, we  followed three different strategies for determining the spectral type of all the stars.

\begin{figure}
\centering
\includegraphics[width=0.51\textwidth]{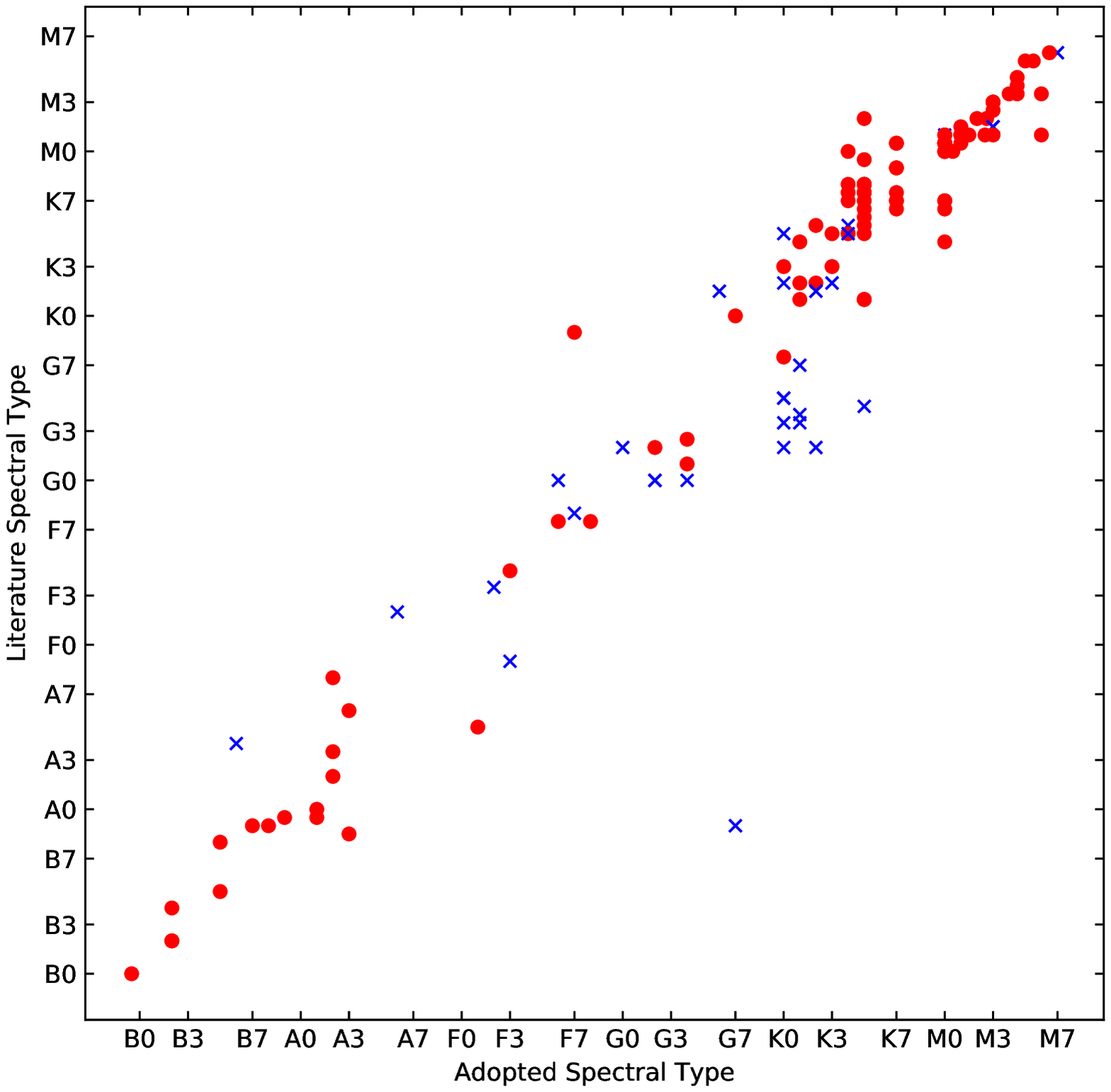}
\caption{Literature and adopted spectral types.
Red filled circles and blue crosses stand for cluster members and non-members, respectively.
{Literature spectral types were retrieved from
\citet{1958ApJ...127..172G},
\citet{1981AJ.....86.1057G},
\citet{1995A&AS..110..367N},
\citet{1996PhDT........63W},
\citet{1999mctd.book.....H},
\citet{2002A&A...384..937Z},
\citet{2008A&A...488..167S},
\citet{2008A&A...491..515C,2012A&A...546A..59C},
\citet{2013A&A...551A.107M}, and
\citet{2014ApJ...794...36H}.}
The G7 star previously classified as A0 is ``IDS~05335--0238~D''; 
the early spectral type may actually correspond to a close-by early A star at a few arcseconds (\cite{2014ApJ...794...36H}).
The F7 star previously classified as K0 is Mayrit~968292;
the late-G/early-K estimation from photometry by {\citet{2008A&A...478..667C}} was incorrect.} % G8--K0\,V  
\label{figure.spt-adop-vs-lit}
\end{figure}

\paragraph{MILES.} 
We derived spectral types of the {140} analysable stars observed with IDS by comparing their spectra with synthetic ones generated with the MILES ``Spectra by Stellar Parameters'' webtool\footnote{\tt http://www.iac.es/proyecto/miles/pages/webtools.php} (\cite{2006MNRAS.371..703S,2010MNRAS.404.1639V,2011A&A...532A..95F}). 
This tool allows the generation of synthetic stellar spectra of stars from 36\,000\,K down to 2750\,K, which approximately correspond to O9 to M6 main sequence spectral types. 
The MILES spectra fully cover the range of spectral types in our sample and the wavelengths of most IDS spectra with a comparable spectral resolution (MILES: 0.90\,{\AA}\,pixel$^{-1}$; R1200Y IDS: 0.48\,{\AA}\,pixel$^{-1}$). 
However, MILES oversamples the spectra taken with the R150V grism, %($\sim$9\% of the sample), 
which has a dispersion of 3.66\,{\AA}\,pixel$^{-1}$. 
For generating our own library of synthetic spectra, we used MILES with solar metallicities and surface gravities varying with effective temperature as described by \citet{1980ARA&A..18..115P}. 
For the relation between temperatures and spectral types we used the relationships of \citet{2013ApJS..208....9P} for K and M spectral types and those of \cite{Allen} for earlier types. 
To perform the spectral type identification, first we classed the {140} spectra into 14 groups, sorted the spectra by approximate effective temperature within each group, and assigned a real (non-synthetic) MILES spectrum to each group.
This grouping helped us to narrow down the interval of effective temperatures and surface gravities of each individual star, {and to determine spectral types with a precision of about one subtype via a $\chi^2$ minimisation as in \citet{2015A&A...577A.128A}}.
The results are listed under ``SpT MILES" in Table~\ref{table.results.ids}.

\paragraph{Standard stars.} 
We derived spectral types of the {46} stars and brown dwarfs observed with OSIRIS towards $\sigma$~Orionis using standard stars from K7 to M6.5 (Table \ref{table.standards.gtc}), with a spectral typing precision of one subtype in this range. 
We assigned spectral types by fitting by eye the problem star to our grid of standard stars, which were observed with exactly the same instrument configuration (Fig.~\ref{figure.ids_sample}).
We took care to properly fit the (pseudo-)continuum and the strongest lines and bands in absorption, for which accurate wavelengths were taken from the ``The Atomic Line List'' webtool\footnote{\tt http://www.pa.uky.edu/$\sim$peter/newpage/}.
%\LEt{Please check that I have retained your intended meaning.} 
Unexpectedly, several stars had spectral types earlier than K7.
A $\chi^2$ fitting as {with MILES did} not improve our {visual} determination.
The results are listed under ``SpT Standards'' in Table \ref{table.results.gtc}.

\paragraph{PyHammer.} 
In addition to the previous strategies, we also used the ``PyHammer" Python Spectral Typing Suite\footnote{\tt http://github.com/BU-hammerTeam/PyHammer} from \citet{2007AJ....134.2398C}.
PyHammer uses its own empirical templates of spectral types and metallicities to estimate the spectral type of a star by measuring prominent line indices and performing a weighted least-squares minimisation. 
It covers spectral types from O5 to L3 and metallicities from --2.0\,dex to +1.0\,dex. 
We applied the PyHammer algorithm to all our spectra from both IDS and OSIRIS. 
The resulting spectral types are listed in Tables \ref{table.results.ids} and \ref{table.results.gtc} under ``SpT PyHammer''. 
According to \citet{2007AJ....134.2398C}, the accuracy provided by The Hammer, the basis on which PyHammer was built, is of around two subtypes. 
Having not investigated the differences between automatic and visual classifications of PyHammer, we assumed that the results are similar to those with The Hammer and OSIRIS/GTC as reported by \citet{2015MNRAS.446.3878M}, who pointed out the improved accuracy of the tool for late M dwarfs.

\paragraph{Adopted spectral type.} 
We report the finally adopted spectral type in Table \ref{table.results.spt}.
In particular, we followed these criteria.
%\LEt{please remove the bullet point format here and give this information in paragraph form and in full sentences.}:
{We used the standard stars strategy for stars} with spectral types K7--M6.5 observed with OSIRIS.
{For stars} earlier than K7 observed with OSIRIS {we used literature information} (Mayrit~783254, 931117, 1042077, and [HHM2007]~648 from \cite{2014ApJ...794...36H}), PyHammer (Mayrit~1223121), or classical visual inspection of spectral features (Haro~5--17 and [HHM2007]~829).
{We used MILES for stars} later than B0 observed with IDS (including those observed with OSIRIS).
{Finally, we used literature information for stars} earlier than B0 observed with IDS ($\sigma$~Ori~AB itself; \cite{2015ApJ...799..169S}).

Of the {165} analysed stars, we present spectral types for the first time for {39} of them, and improve previous determinations in a few cases.
In Fig.~\ref{figure.spt-adop-vs-lit}, we compare our spectral types to those of the literature.
For example, there are half a dozen mid-K  non-cluster-members that were classified as mid-G stars by \citet{2014ApJ...794...36H}.
Figure 1 shows another two extreme outlier examples.
Apart from these differences, the agreement with previous spectral determinations is generally within one or two subtype uncertainties. 

%__________________________________________________ 

\subsection{\textit{Gaia} DR2}
\label{subsection.gaia}

\begin{figure}
\centering
\includegraphics[width=0.49\textwidth]{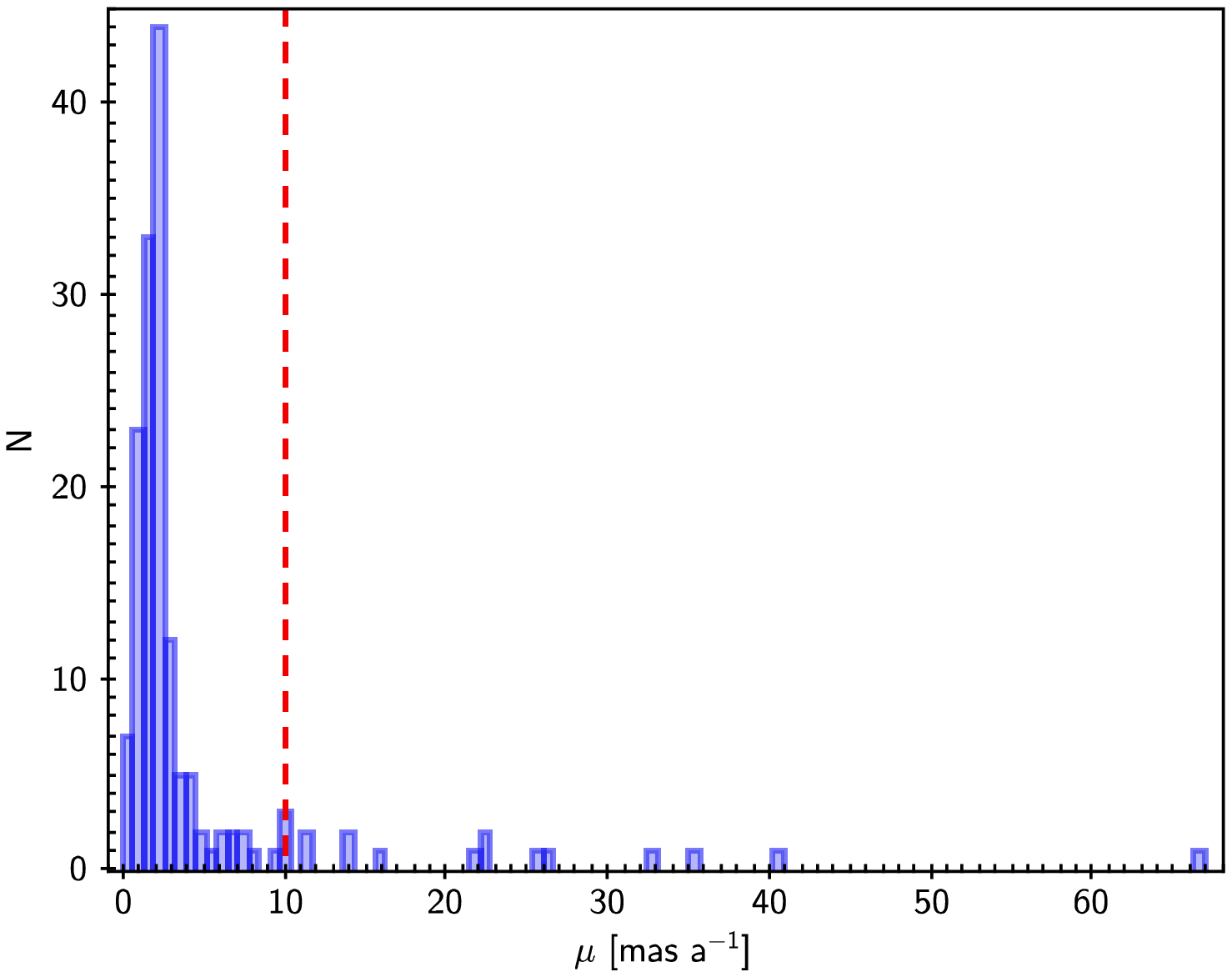}
\includegraphics[width=0.49\textwidth]{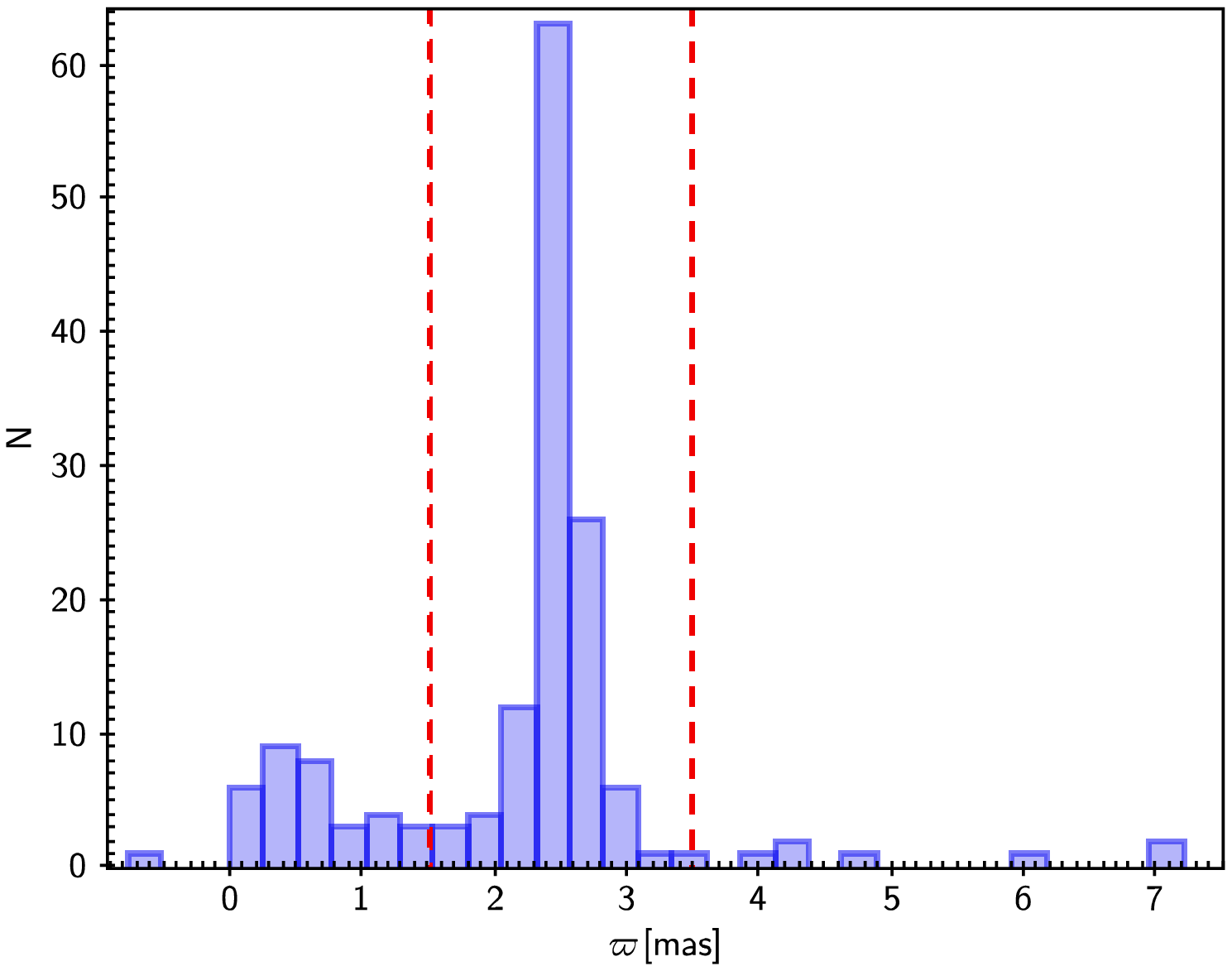}
\caption{Histograms of {\em Gaia} DR2 total proper motions ({\em top}) and parallaxes ({\em bottom}).
Vertical dashed lines indicate the {conservative} $\sigma$~Orionis boundary limits explained in Sect.~\ref{subsection.membership}.
The bin sizes of these histograms have been adjusted according to \cite{Freedman1981}.} % 
\label{figure.hist_mu}
\end{figure}

At the moment of observation, equatorial coordinates were taken from the Two-Micron All-Sky Survey (2MASS; \cite{2006AJ....131.1163S}).
However, we tabulate here the latest {\em Gaia} DR2 coordinates (\cite{2018A&A...616A...1G}).
We used both Topcat\footnote{\tt http://www.star.bris.ac.uk/$\sim$mbt/topcat/} and Aladin\footnote{\tt http://aladin.u-strasbg.fr/AladinDesktop/} Virtual Observatory tools with a small cross-match radius {of 1.5\,arcsec}, and compared our results with {those} of \citet{2017AN....338..629C}. % RNAAS
Apart from the equatorial coordinates (Sect.~\ref{subsection.observations}), we also retrieved parallaxes, proper motions, radial velocities, and broadband magnitudes of the {165} stars and brown dwarfs. %166?
For the {nine} stars without {\em Gaia} DR2 parallaxes and proper motions, we took them from the existing literature (\cite{1995gcts.book.....V,2005yCat.1297....0Z,2007A&A...474..653V,2017A&A...600L...4A}). 
Both proper motions and parallaxes are listed in Tables~\ref{table.observations.int} (IDS) and~\ref{table.observations.gtc} (OSIRIS), and {are} discussed in Sect.~\ref{subsection.membership}. 
The histograms of total proper motions and parallaxes for the complete sample are shown in Fig.~\ref{figure.hist_mu}.
\textit{Gaia} DR2 radial velocities are discussed in Sect.~\ref{subsection.binaries_dr2}.
Magnitudes will be presented in a forthcoming publication.

%__________________________________________________ 

\subsection{Youth features}
\label{subsection.youth}

\begin{table*}
\centering
  \caption[]{Previous cluster member candidates discarded in this work$^{a}$.} 
    \label{table.no_members}
      \begin{tabular}{llll}
\hline
\hline
\noalign{\smallskip}
Name    & Mayrit    & $[$HHM2007$]$ & Reason for discarding             \\  %
\noalign{\smallskip}
\hline
\noalign{\smallskip} 
  HD 294276             &...            &20         & No $\mu$, no $\varpi$ \\ % G4
  2MASS J05372885--0255555  & 1650224 &...          & No $\varpi$, no Li \\ % K0
  HD 294274             &...            &168        & No $\mu$ \\ % G2
  SO210868              & 958292        &...        & No $\varpi$ \\ % G4
  $[$HHM2007$]$ 244     & 882239    &244        & No $\mu$, no Li \\ % K1
  2MASS J05375789--0259536  & 1596206   &...        & No $\varpi$ \\ % G7
  TYC 4771--720--1      &...            &289        & No $\mu$ \\ % G6
  $[$W96$]$ 4771--0950   & 717307       &...        & No $\mu$ \\ %  F2
  $[$HHM2007$]$ 385     & 733222        &385        & No $\mu$, no $\varpi$, no Li \\ % M3
  TYC 4771--873--1        & 1064335     &...        & No $\varpi$ \\ %  F2
  $[$SE2004$]$ 10       & 1564345   &...        & No $\varpi$, no Li \\ % K1
  IRAS 05358--0238       & 377264       &...        & No $\varpi$, no Li \\ % M7
  2MASS J05382265--0257421  & 1343194   &...        & No $\varpi$, no Li \\ % K0
  StHa 50               & 459340        &...        & No $\varpi$ \\ % B6
  $[$W96$]$ pJ053834--0239 & 258215  &...        & No $\varpi$, no Li \\ % M0
  IDS 05335--0238 D      & 240322       &...        & No $\mu$ \\ % G7
  $[$HHM2007$]$ 648     &...            &648        & No $\varpi$ \\ % G5
  $[$SE2004$]$ 30       & 1045356   &...        & No $\varpi$, no Li \\ % K4
  $[$W96$]$ pJ053844--0233 & 123000  &...        & No $\varpi$, no Li \\ % K0
  $[$HHM2007$]$ 829     &...            &829        & No $\mu$, no $\varpi$ \\ % K0
  $[$HHM2007$]$ 846     &...            &846        & No $\varpi$, no Li \\ % M3
  $[$HHM2007$]$ 961     &...            &961        & No $\mu$, no $\varpi$ \\ % G6
  $[$SE2004$]$ 50       & 945030    &...        & No $\varpi$, no Li \\ % K1
  TYC 4771--661--1          &...        &1001       & No $\mu$, no $\varpi$ \\ % G6
  $[$HHM2007$]$ 1009    & 735131        &1009       & No $\varpi$, no Li \\ % K1
  $[$HHM2007$]$ 1092    & 861056    &1092       & No $\varpi$, no Li \\ % K0
  $[$HHM2007$]$ 1129    & 1165138   &1129       & No $\varpi$, no Li \\ % K1
  HD 294299             & 1037054       &1163       & No $\varpi$ \\ % A6
  $[$HHM2007$]$ 1189    & 936072    &1189       & No $\varpi$, no Li \\ % K2
  $[$HHM2007$]$ 1251    & 1107114       &1251       & No $\varpi$, no Li \\ % K4
  $[$HHM2007$]$ 1256    & 1110113   &1256       & No $\varpi$ \\ % G
  $[$HHM2007$]$ 1269    & 1169117       &1269       & No $\varpi$, no Li \\ % K0
  $[$HHM2007$]$ 1347    & 1338116   &1347       & No $\varpi$, no Li \\ % K0
  HD 294301             & 1468100   &...        & No $\mu$ \\ % F2
  HD 294297             & 1659068   &...        & No $\mu$, no $\varpi$ \\ % G2
\noalign{\smallskip}
\hline
      \end{tabular}
      \begin{list}{}{}
      \item[$^{a}$] {\bf Notes}. This table may be completed with Mayrit~1285339 (HD~294268, [HHM2017]~411) and Mayrit~1275190 based on discordant radial velocities (Sect.~\ref{subsection.binaries_dr2}).
      \end{list}
\end{table*}

We complemented our IDS, OSIRIS, and {\em Gaia} DR2 data with an exhaustive literature compilation of features of youth to support our determination of membership to the $\sigma$~Orionis cluster. 
Below we describe each of the features that we looked for, together with EW(Li~{\sc i}) and EW(H$\alpha$) from our spectra. 

\paragraph{Lithium.}
Li~{\sc i} $\lambda$6707.80\,{\AA} is the main indicator of youth in stars later than mid-F spectral type (\cite{1989ARA&A..27..351B,1994A&A...288..860C,2001MNRAS.328...45M,2002A&A...384..937Z}). 
{We highlight the fact that for open clusters as young as $\sigma$~Orionis ($\tau \sim$ 3\,Ma) this is true only for low-mass stars with K and M spectral types, as Sun-like stars (spectral types F and G) do not have a deep convection zone and therefore lithium depletion occurs on longer time scales (\cite{2014prpl.conf..219S}).}
We measured EW(Li~{\sc i}) in absorption in the IDS and OSIRIS spectra of {78} stars (Table \ref{table.results.spt}). 
% 78 different stars, 93 in total (some observed with both IDS/OSIRIS)
Because of the low resolution of our spectra, especially for those obtained with OSIRIS and IDS R150V, we conservatively set a lower limit at EW(Li~{\sc i}) = 0.1\,{\AA} for claiming a true detection.
We complemented our measurements with previous EW(Li~{\sc i}) values published by \citet{2002A&A...384..937Z}, \citet{2005MNRAS.356...89K}, \citet{2008A&A...488..167S}, and \citet{2014ApJ...794...36H}. 
{Observations with higher spectral resolution and signal-to-noise ratio would be needed to disentangle the two populations of G- and K-type stars with very young ages in $\sigma$~Orionis and with juvenile ages of up to 30\,Ma in the cluster foreground.}

\paragraph{Balmer series.}
In very young stars, H$\alpha$ $\lambda$6562.80\,{\AA} can be in emission because of chromospheric emission or accretion from a circumstellar disc (\cite{1979ApJ...234..579C,1986ApJS...61..531S,1989ARA&A..27..351B,2003AJ....126.2997B}).
We complemented our measurements with previous  EW(H$\alpha$) values or remarks published by \citet{2002A&A...384..937Z}, \citet{2005MNRAS.356...89K}, \citet{2008A&A...478..667C}, \citet{2008A&A...488..167S}, {and} \citet{2014ApJ...794...36H}. 

\paragraph{OB stars.}
Massive stars of early O and B spectral types stay in the main sequence for only a few million years. 
There are no evolved (i.e. giant) early-type stars in $\sigma$~Orionis, providing evidence for its extreme youth.
Except for one anomalous example of a background B-type star, all O- and B-type stars in our sample, which were also the brightest targets that required the shortest exposure times, belong to the cluster (\cite{2007A&A...470..903C,2008A&A...478..667C} -- but see \cite{2008AJ....135.1616S}). 

\paragraph{Mid-infrared.} 
The flux excess in the mid-infrared (MIR) is produced by the presence of circum(sub)stellar discs around stars and brown dwarfs, and is often detected at the early stages of star formation (\cite{2001ApJ...558L..51M,2006AJ....131.1574L,2008ApJ...676.1109F}).
We searched for MIR indicators and Class I, II, and evolutionary and transition disc classification in \citet{2007ApJ...662.1067H}, \citet{2008ApJ...688..362L}, and mostly compiled by \citet{2008A&A...478..667C}.

\paragraph{X-rays.} 
Although \citet{1967PASP...79..433G} and \citet{1982A&A...109..213L} had already identified the star clustering, the importance of $\sigma$~Orionis as a milestone star-forming region began with the discovery by \citet{1996PhDT........63W} of an agglomerate of intense X-ray-emitting stars densely concentrated around the eponymous $\sigma$~Ori multiple star system.
In general, the origin of X-ray emission lies on fast rotation and deep convective zones or wind collision (\cite{1981ApJ...248..279P,2001A&A...377..538S,2003SSRv..108..577F,2005ApJS..160..401P}), depending on stellar mass.
Here, we compiled X-ray detections from \citet{2006A&A...446..501F}, \citet{2008ApJ...683..796S}, \citet{2008A&A...491..961L}, \citet{2009AJ....137.5012C}, and \citet{2010A&A...521A..45C}.

\paragraph{Proper motion and parallax.} 
Per se, proper motion ($\mu$) and parallax ($\varpi$) are not youth features. 
However, all cluster members share common characteristics. 
In particular, they are located approximately in the same region of the sky and at the same distance, they move in the same direction with the same traverse velocity, have the same age and metallicity, and are born from the same parental cloud.
Members within the $\mu$ and $\varpi$ boundaries of the cluster are likely to have youth features. 
Vice versa, stars with $\mu$ and $\varpi$ outside the cluster boundaries are likely to be devoid of youth features.

%========================================================================== 

\section{Results and discussion}
\label{section.results_discussion}

\subsection{Cluster membership}
\label{subsection.membership}

\begin{figure}
\centering
\includegraphics[width=0.51\textwidth]{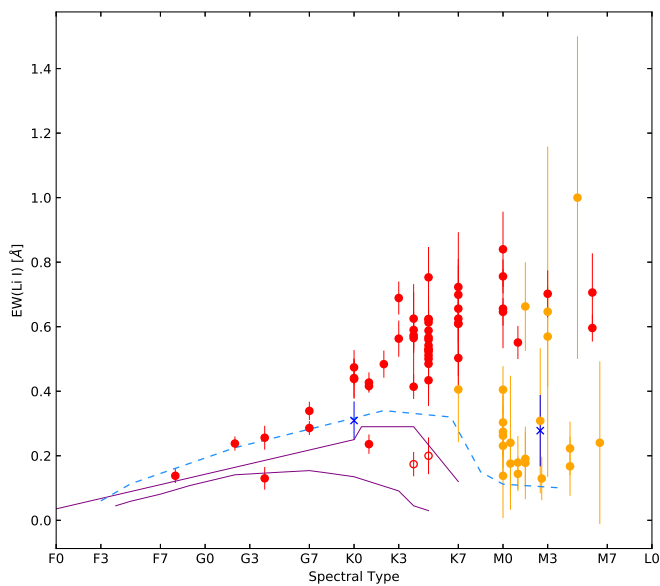}
\caption{Measured Li~{\sc i} EWs as a function of adopted spectral type.
Filled red circles show young targets observed with IDS R1200Y;
open red circles show young targets observed with IDS R150V (controvertible);
filled orange circles show young targets observed with OSIRIS R100B;
blue crosses show non-member stars with Li~{\sc i} (Haro~5-17 and Haro~5-46);
solid magenta lines show upper and lower EW(Li~{\sc i}) envelopes of the Pleiades ($\tau \sim$ 120\,Ma -- \cite{1993AJ....106.1059S,2001A&A...379..976M}); and the
dashed cyan line shows the upper EW(Li~{\sc i}) envelope of IC~2602 ($\tau \sim$ 30\,Ma -- \cite{1997A&A...325..647N}).} 
\label{figure.spt_li}
\end{figure}

We used all available information to determine membership in $\sigma$~Orionis of the {165} spectroscopically observed stars and brown dwarfs.
We followed a step-by-step process, summarised in Table~\ref{table.results.spt}:

\begin{itemize}
    \item First, we discarded four T~Tauri stars discovered by \citet{1953BOTT....1g..11H} in neighbouring star-forming regions: 
    Haro 5-17 in the eastern outskirts of the $\epsilon$~Orionis cluster (\cite{1931AnLun...2....1C,2008A&A...485..931C}), Haro 5-40 and Haro 5-44 near the Horsehead nebula (\cite{2003AJ....125.2108P,2005A&A...437..177H,2006A&A...456..565G}), and Haro 5-46 close to the Flame nebula (\cite{1988A&A...191...44M,1989ApJ...342..883B,2003A&A...404..249B}). 
    The four H$\alpha$-emitter stars are located to the west of the Ori OB1b association, but at angular separations much greater than the extension of the halo of the $\sigma$~Orionis cluster at 0.5\,deg from the cluster centre, which is defined by the $\sigma$~Ori Trapezium-like system \citep{2008MNRAS.383..375C}.
    The remaining {161} stars and brown dwarfs were located at less than 0.5\,deg from the centre. %162?

    \item We next applied the same astrometric criteria with {\em Gaia} DR2 data for cluster membership as in \citet{2018RNAAS...2b..25C}, which represented an improvement with respect to those presented in \citet{2010A&A...514A..18C}.
    In particular, we discarded {51} stars with total proper motions $\mu >$ 10\,mas\,a$^{-1}$ and/or parallaxes less than 1.5\,mas or greater than 3.5\,mas.
    The parallax criterion translated into discarding all stars with distances outside the 290--670\,pc interval ({assuming $d = 1/\varpi$;} the precise $\sigma$~Orionis distance of \cite{2016AJ....152..213S} was 387.5$\pm$1.3\,pc). 
    {Careful inspection of the six {\em Gaia} DR2 sources with parallax uncertainties greater than 20\,\% led to the recovery of Mayrit~1073209, an M3.0 star with youth features but with poor {\em Gaia} DR2 astrometric quality flags ({\tt astrometric\_gof\_al}, {\tt astrometric\_chi2\_al}, {\tt astrometric\_excess\_noise} and, especially, the [re-normalised] unit-weight errors {\tt UWE} and {\tt RUWE}\footnote{\tt https://www.cosmos.esa.int/web/gaia/dr2-known-issues}).}
    
    \item Of the {110} remaining stars we identify OB spectral types in {11} cases, Li~{\sc i} in absorption in {87} ({34} for the first time) cases, H$\alpha$ in emission in {67} ({12} for the first time) cases, MIR excess in {46 cases}, and X-ray emission in {63 cases}.
    A few photometric cluster member candidates continue in the same state. 
\end{itemize}

As expected, none of the astrometrically discarded stars displayed lithium in absorption, MIR excess, or X-ray emission.
Many do not follow the spectro-photometric cluster sequence, either.
However, there are two distant stars with faint and strong H$\alpha$ emission: IRAS~05358--0238 and StHa~50, respectively.
The former is an evolved star {at $d$ = 1.7$^{+0.9}_{-0.5}$\,kpc \citep{2018AJ....156...58B}} with a weak silicate feature and structure in the 10--12\,$\mu$m range \citep{2004A&A...418..663O}, while the later is an unusual, isolated Herbig Ae/Be star in the background {at $d$ = 2.01$^{+0.21}_{-0.18}$\,kpc \citep{2017AN....338..629C,2018AJ....156...58B}}.
In spite of having been identified as a  non-cluster-member over a decade ago {\citep{2008A&A...478..667C}}, IRAS~05358--0238 has since been observed with expensive facilities such as SCUBA and {\em Herschel} (\cite{2013MNRAS.435.1671W,2016ApJ...829...38M}).

Of the {51} foreground and background stars, {16} have previously been classified as such (e.g. \cite{1977ApJS...34..115W,2004AJ....128.2316S,2007A&A...466..917C}).
However, remarkably, {35} %27+8
 stars, which are listed in Table~\ref{table.no_members}, were classified as cluster-member candidates in the extensive surveys of \citet{2007ApJ...662.1067H} and {\citet{2008A&A...478..667C}}.
Although additional cluster member candidates were discarded afterwards (e.g. in previous items of this series of papers or by \cite{2014ApJ...794...36H}), such a large number of new non-members has serious implications for a number of key science cases in the $\sigma$~Orionis cluster.
We qualitatively outline below some of the most relevant implications:
($i$) \citet{2007ApJ...662.1067H} determined frequencies of discs in a sample of 336 $\sigma$~Orionis Herbig Ae/Be and T~Tauri star candidates ranging from 10\,\% to 35\,\%, approximately, depending on stellar mass.
A stellar population smaller by 10\,\% may translate into an increase in the frequency of discs in intermediate-mass T~Tauri stars to almost 40\,\%;
($ii$) \citet{2012ApJ...754...30P} determined the most comprehensive mass function (mass spectrum) of the cluster from about 20\,M$_\odot$ to about 0.005\,M$_\odot$ (see also \cite{2011sca..conf..108C}).
These latter authors built this {mass} spectrum with the help of the Mayrit catalogue (\cite{2008A&A...478..667C}) with the latest updates available that moment.
The lack of {27} stars with estimated masses around 0.5\,M$_\odot$ can make a sharper contrast between the steep Salpeter's region at more than 1\,M$_\odot$ and the almost flat slope of the mass spectrum at the lowest masses (\cite{2001MNRAS.322..231K,2003PASP..115..763C});
($iii$) Most of the discarded stars are located relatively far from the cluster centre\footnote{The Mayrit number indicates the angular separation to and position angle from the cluster centre; e.g. Mayrit~1045067 is at $\rho \approx$ 1045\,arcsec and $\theta \approx$ 67\,deg from $\sigma$~Ori~AB.}.
Therefore, while the cluster core may follow the same power law in the radial distribution presented by \citet{2008MNRAS.383..375C}, the halo at angular separations $\rho \gtrsim$ 20\,arcsec could be even more rarefied.
In other words, $\sigma$~Orionis could be more compact than previously thought, which would ease the separation of its young population from other nearby young populations (Horsehead and Flame nebulae, $\epsilon$~Orionis);
{($iv$)} To date, the only metallicity determination based on a significantly large sample of stars in the cluster was performed by \citet{2008A&A...490.1135G}.
Looking back to their stellar sample with our current knowledge, the average cluster [Fe/H] value may be different from what they derived, albeit still near solar. 

Regarding cluster members and member candidates in particular, while there are a few stars with the four youth features (Li~{\sc i}, H$\alpha$, MIR, X-rays), apart from $\mu$ and $d$ within the suitable intervals, there are also some member candidates that only follow the spectro-photometric sequence of the cluster.
This is the case especially for relatively inactive A- and early-{F-type} stars, which have hot effective temperatures that prevent the formation of Li~{\sc i} and of deep convective zones (and, thus, strong magnetic fields) and, {in addition}, have short disc-dissipation times.
The list of photometric members without spectroscopic confirmation includes Mayrit~11238 {($\sigma$~Ori~C itself; A2\,V)}.
{Besides}, Mayrit~1227243 {(HD~294275; A1\,V)} was classified by \citet{2017AN....338..629C} as a non-cluster-member based on a TGAS (\cite{2017A&A...600L...4A}) parallax that was quite different from the new {\em Gaia} DR2 one, which is compatible with cluster membership, and therefore it stands as a photometric member candidate.

{In addition}, we identified for the first time spectral features of youth (especially, but not only, Li~{\sc i}) in six stars that  were previously photometric cluster-member candidates: 
Mayrit 144349, 1082115, 1042077, 1273081, and 1476077.
We also assigned new Mayrit designations to two stars:
Mayrit~1045067AB {([BMZ2001] S~Ori J053948.1--022914; M3.0)}, for which lithium was not appreciable in our OSIRIS spectrum but for which low-gravity features were detected by \citet{2005MNRAS.356.1583B}, and Mayrit~1042077 {([HHM2007]~1250; K7)}, with a new Li~{\sc i} detection.
Some faint targets with controvertible detections of youth features, such as Mayrit~1298302 and Mayrit~1500066, require further investigation.

Figure~\ref{figure.spt_li} shows the measured EW(Li~{\sc i}) as a function of adopted spectral {type}.
%\LEt{please consider expanding here as the meaning is not explicitely clear.}. 
The earliest $\sigma$~Orionis star for which we measured EW(Li~{\sc i}) is Mayrit~1285339 (HD~294268; F8\,e).
%, of F8\,e spectral type, which corresponds to $T_{\rm eff} \sim$ 6200\,K.
\citet{2018ApJ...867..116P} classified this object as a relatively evolved star in the sparse population of the Orion OB1 association.
Our classification as a true cluster member is in accordance with the fact that the star hosts a transitional disc \citep{2013MNRAS.435.1671W,2013ApJ...770...94B,2017AJ....153..240A,2017A&A...600A..62P}.
{However, since there are Herbig Ae/Be and classical T~Tauri stars (with different kinds of discs) in the sparse population of the Orion~OB1 association \citep[e.g.][]{2005AJ....129..856H,2019AJ....157...85B}, there is still a possibility that Mayrit~1285339 actually belongs to that sparse population and not to the $\sigma$~Orionis cluster. This argument is supported mainly by the radial velocity analysis of \citet{2018ApJ...867..116P}.
Mayrit~1285339 and} some G- and K-type stars observed with IDS R1200Y and EW(Li~{\sc i}) within the envelopes of the Pleiades cluster are discussed in Sect.~\ref{subsection.binaries_dr2}.

%__________________________________________________ 

\subsection{Strong accretors}
\label{subsection.accretors}

\begin{table*}
\centering
  \caption[]{Strongly accreting stars and brown dwarfs with EW(H$\alpha$) $<$ --50\,\AA.}
    \label{table.strong_accretors}
      \begin{tabular}{llllccl}
\hline
\hline
\noalign{\smallskip}
Mayrit  &Alternative    &H$\alpha$ emitter      &Variable       &SpT        &EW(H$\alpha$)       &Remarks        \\  %
                &name               &name                           &name           &adopted     &[\AA]                  &                               \\ %
\noalign{\smallskip}
\hline
\noalign{\smallskip}
Mayrit 1329304 & Haro 5--5 & Haro 5--5 &...&M2.5&--118$_{-15}^{+14}$&EW(Li~{\sc i}) this work; \\
&&&&&&red-shifted H$\alpha$ \\ %
\noalign{\smallskip}
Mayrit 873229AB & Haro 5--7 & ESO--Ha 1646 &NSV 2489&M4.5&--59$_{-6}^{+5}$& Red-shifted H$\alpha$       \\  %
\noalign{\smallskip}
Mayrit 757219 & Haro 5--8 & Kiso A--0976 322 &SVS 1241&M1.0&--91$_{-9}^{+7}$&  \\  %
\noalign{\smallskip}
Mayrit 329261AB & [SWW2004] 207 & ... &...&M4.5&--148$_{-20}^{+16}$&EW(H$\alpha$) this work       \\  %
\noalign{\smallskip}
Mayrit 1207349 & Haro 5--9 & Kiso A--0976 330 &V2731 Ori&M0.0&--54$_{-3}^{+3}$&        \\  %
\noalign{\smallskip}
Mayrit 1316178 &S Ori J053847.2--025756 & ... &...&M6.5&--80$_{-13}^{+10}$&EW(H$\alpha$) this work       \\  %
\noalign{\smallskip}
Mayrit 687156 & [WB2004] 26 & ESO--Ha 1693 &...&M4.5&--50$_{-6}^{+4}$&EW(Li~{\sc i}) this work; \\
&&&&&&red-shifted H$\alpha$ \\ %
\noalign{\smallskip}
Mayrit 871071 & Haro 5--27 & Kiso A--0976 356 &V510 Ori&K2&--123$_{-3}^{+2}$&Source of HH 444       \\  %
\noalign{\smallskip}
Mayrit 1279052 & Haro 5--30 & Haro 5--30 &...&M5.5&--100$_{-25}^{+20}$&        \\  %
\noalign{\smallskip}
Mayrit 1041082& Haro 5--32 & Kiso A--0976 359 &V604 Ori&M3.0&--89$_{-19}^{+12}$&Blue-shifted H$\alpha$       \\  %
\noalign{\smallskip}
Mayrit 1196092 & S Ori J054004.5--023642 & ... &...&M6.5&--125$_{-15}^{+15}$&EW(Li~{\sc i}) this work;   \\  %
&&&&&& {brown dwarf} \\ %
\noalign{\smallskip}
Mayrit 1364078 &V2754 Ori & ... &V2754 Ori&M6.5&--73$_{-19}^{+10}$& \\  %
\noalign{\smallskip}
...& Haro 5--40 & Kiso A--0976 375 &...&M4.0&--147$_{-21}^{+11}$& EW(H$\alpha$) this work; IC 434       \\  %
\noalign{\smallskip}
...& Haro 5--44 & Kiso A--0904 111 &V612 Ori&M3:&--160$_{-35}^{+28}$& EW(H$\alpha$) this work;  \\ %
&&&&&&red-shifted H$\alpha$; IC 434 \\ %
\noalign{\smallskip}
\hline
      \end{tabular}
%      \begin{list}{}{}
%      \item[$^{a}$] When red-shifted or blue-shifted is indicated the spectrum is $\pm$ 2.5\,{\AA} shifted from the H$\alpha$ $\lambda$6562.80\,{\AA}.
%      \end{list}
\end{table*}

\begin{figure}
\centering
\includegraphics[width=0.49\textwidth]{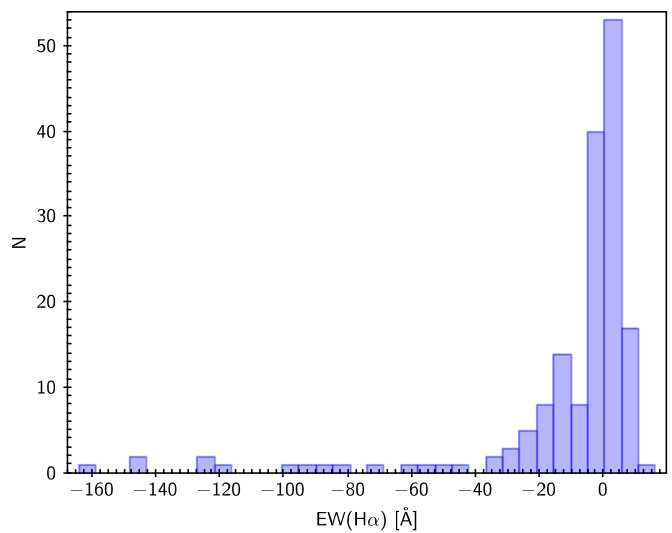}
\caption{Histogram of the number of sources as a function of the measured EW(H$\alpha$). 
The bin size of this histogram has been adjusted according to \cite{Freedman1981}.} % 
\label{figure.hist_ha}
\end{figure}

\begin{figure}
\centering
\includegraphics[width=0.49\textwidth]{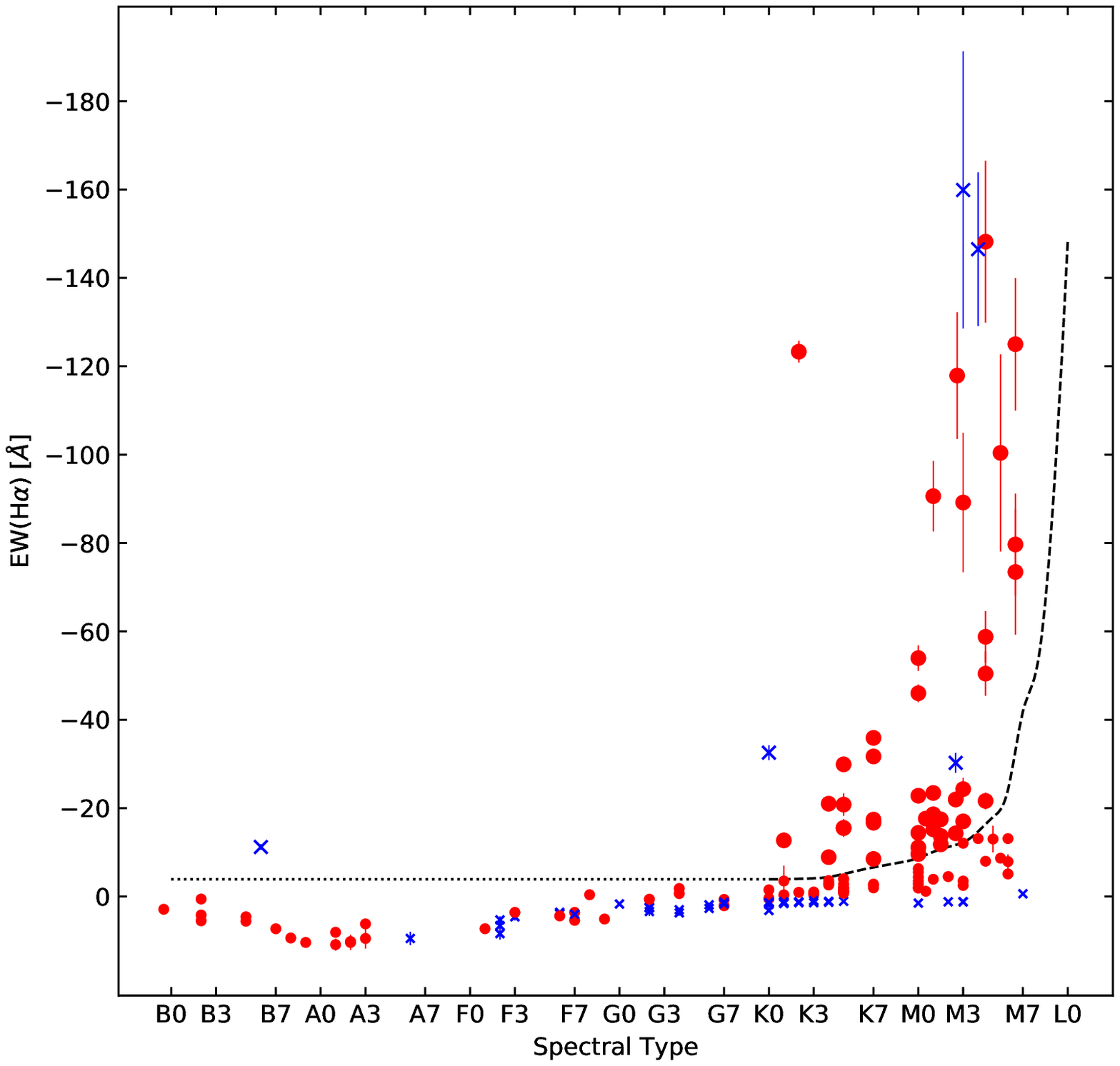}
\label{figure.spt_ha}
\caption{Measured EW(H$\alpha$) of our sample of stars as a function of adopted spectral type.
The black dashed line shows the curve fitting from K0 to L0 of the EWs used in \citet{2003AJ....126.2997B} as an empirical criterion for identifying classical T~Tauri stars. The black dotted line shows the coarse extrapolation of the \citet{2003AJ....126.2997B} criterion to spectral types earlier than K0 (EW(H$\alpha$) = --3.9\,{\AA}). 
Big and small red filled circles indicate cluster member stars and brown dwarfs with EWs above and below the black lines, respectively. 
Big and small blue crosses show  non-cluster-member stars with EWs above and below the black lines, respectively.
The early K-type $\sigma$~Orionis star with strong emission is Mayrit~871071 (Haro~5--27, V510~Ori), the driving source of the jet HH~444 (\cite{1998Natur.396..343R, 2001A&A...371.1118L}).}
\end{figure}

\begin{figure}[]
\includegraphics[width=0.24\textwidth]{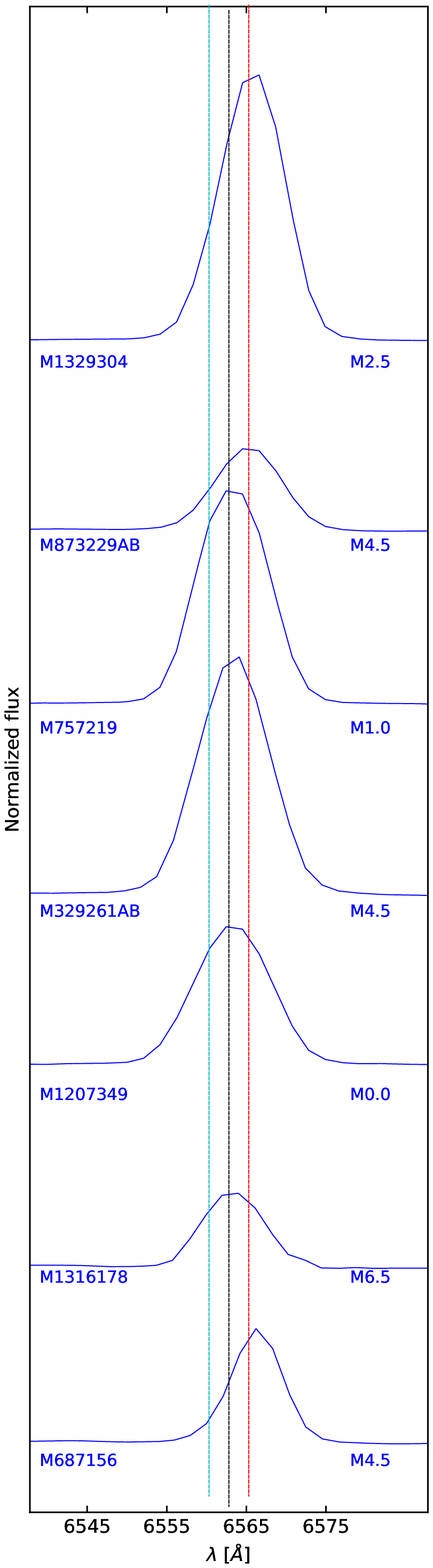}
\includegraphics[width=0.24\textwidth]{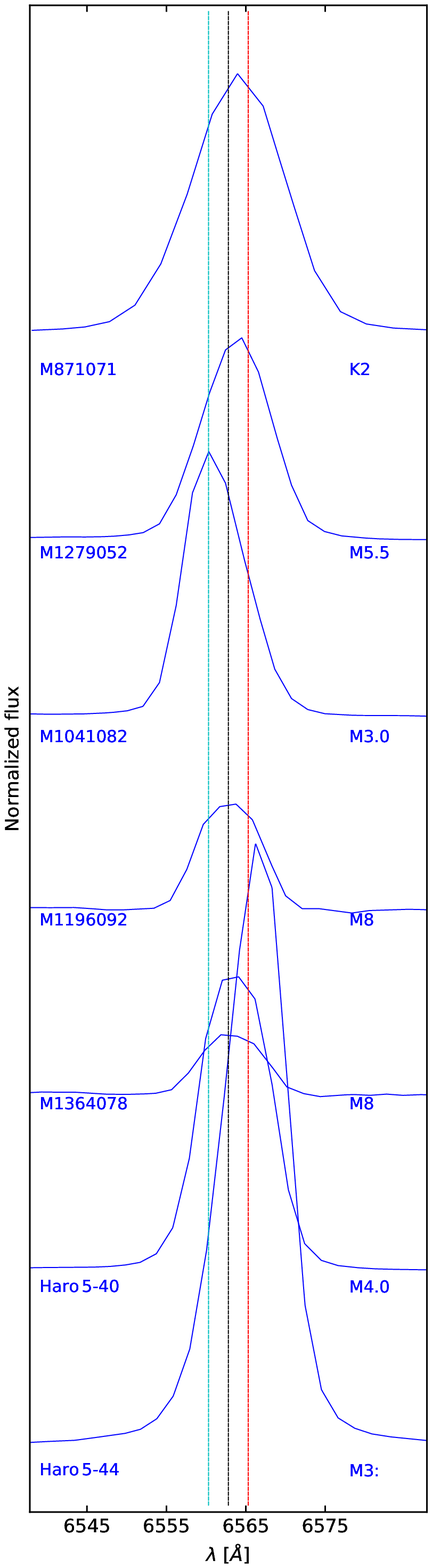}
\caption{Normalised spectra around H$\alpha$ $\lambda$6562.80\,{\AA} of the {14} strong accretors in Table~\ref{table.strong_accretors}.
The vertical lines show the H$\alpha$ line
%\LEt{the significqnce of "in air" is unclear here and above; please consider either expanding or using alternative wording.} 
shifted by --2.5\,{\AA} (cyan), 0.0\,{\AA} (black), and +2.5\,{\AA} (red).} % 
\label{figure.strong_accretors}
\end{figure}

The  H$\alpha$ line is the atomic line for which we have the largest number of measurements, either in emission or in absorption.
Fig.~\ref{figure.hist_ha} shows the distribution of our EW(H$\alpha$) measurements.
The vast majority of these lie in the interval from +17\,{\AA} in absorption (for late B and early A stars) to --40\,{\AA} in emission.
However, there is a tail of very strong emitters, with EW(H$\alpha$) {down} to --160\,{\AA}, which correspond to objects that undergo accretion from a circum(sub)stellar disc.
In particular, there are {40} $\sigma$~Orionis stars and brown dwarfs in our sample that satisfy the \citet{2003AJ....126.2997B} criterion for separating classical T~Tauri objects with accretion from weak-line T~Tauri objects with chromospheric activity.
One  non-cluster-member, the background Herbig Ae/Be star StHa~50, is above the coarse extrapolation of that criterion to spectral types earlier than K.
The four discarded T~Tauri stars discovered by \citet{1953BOTT....1g..11H} are also above the accretional/chromospheric activity boundary {(Fig.~\ref{figure.spt_ha})}.

Of the {40} accreting cluster members, {12} are {young cluster members} with EW(H$\alpha$) $<$ --50\,{\AA}, and are listed in Table~\ref{table.strong_accretors}, together with two Haro objects in IC~434. 
All of these 12\ {also} have EW(H$\beta$) $<$ --12\,{\AA} and significant emission in all observable Balmer lines, except for the M6.5-type brown dwarf Mayrit~1196092 ($J \approx$ 15.3\,mag), whose blue spectrum is too noisy.
Table~\ref{table.strong_accretors} includes: 
the brown dwarf mentioned immediately above;
%\LEt{this is confusing as there is no single brown dwarfe mentioned immioately above but "brown dwarf\uline{s}".};
two objects at the stellar/substellar boundary at $J \approx$ 14.5\,mag (\cite{2007A&A...470..903C}), namely Mayrit~1316178 and Mayrit~1364078, which also have M6.5 adopted spectral types;
seven photometrically variable sources (\cite{1960PZ.....13..166F,2004A&A...424..857C});
three stars with Li~{\sc i} in absorption detected for the first time; and
three stars and one brown dwarf with EW(H$\alpha$) measured for the first time (including the two Haro stars in IC~434).

{After} correcting heliocentric radial velocities using the IRAF $\texttt{bcvcorr}$ task and smoothing the spectra with grade-two splines, we measured the shift of the centroid of the H$\alpha$ line (Fig.~\ref{figure.strong_accretors}) and found five stars with displacements with respect to the air wavelength greater than 2.5\,{\AA}, which translates into gas velocities greater than about 120\,km\,s$^{-1}$.
These five stars are ideal targets for searches for new faint jets in $\sigma$~Orionis as discovered recently by \citet{2019MNRAS.486.4114R}.

Remarkably, nine of the 14 stars in Table~\ref{table.strong_accretors} were discovered in objective-prism photographic plates by a small team of Mexican astronomers back in the middle of the twentieth century \citep{1953BOTT....1g..11H}. %,2004PASP..116.1035W}.

%__________________________________________________ 

\subsection{Binaries}
\label{subsection.binaries_dr2}

\begin{table*}
\centering
  \caption[]{Young binary stars in our sample resolved by {\em Gaia}.} 
    \label{table.binaries_specs}
      \begin{tabular}{lcccccl}
\hline
\hline
\noalign{\smallskip}
Mayrit  &$\rho$ &$\theta$       &$\Delta B_{p}$ &$\Delta G$     &$\Delta R_{p}$ & Reference$^{a}$        \\  %
                &[arcsec]       &[deg]          &[mag]                  &[mag]          &[mag]          &                 \\ %
\noalign{\smallskip}
\hline
\noalign{\smallskip} 
%1415279AB$^{c}$&X              &X              &X              &X              &X              & Cab14           \\ %INT, Not resolved Gaia, Lee94
873229AB                &0.78   &313.6  &...    &2.74   &...    & (This work)\\ %GTC, Resolved Gaia
329261AB                &0.90   &80.5   &...    &3.00   &...    & This work     \\ %GTC, Resolved Gaia
168291AB                &3.25   &55.4   &1.35   &1.82   &2.03   & Cab10b        \\ %INT, Resolved Gaia
%344337AB$^{c}$ &X              &X              &X              &X              &X              & Cab14           \\ % Not resolved Gaia, Sac08
%114305AB$^{c}$ &X              &X              &X              &X              &X              & Cab14           \\ % Not resolved Gaia, Bej01
1248183AB               &0.76   &224.5  &...    &3.33   &...    & This work     \\ %INT, Resolved Gaia
%$\sigma$~Ori$^{b}$     &X      &X              &X              &X              &X              & ????            \\ % ???????
359179AB                &0.99   &226.9  &...    &1.55   &...    & Cab18         \\ %INT, Resolved Gaia
%528005AB               &7.67   &286.1  &4.19   &2.99   &2.57   & Cab14         \\ %INT,GTC, Not resolved Gaia, Cab05
%332168AB$^{c}$ &X              &X              &X              &X              &X              & Cab14           \\ % Not resolved Gaia, Sac08
707162AB                &1.07   &88.7   &...    &1.46   &...    & Cab10a        \\ %INT, Resolved Gaia 
%306125AB$^{c}$ &0.47   &189    &X              &X              &X              & Cab14           \\ % Not resolved Gaia, Cab05
1626148AB               &0.64   &133.7  &--0.07 &0.15   &--0.02 & (This work)\\ %INT, Resolved Gaia
960106AB                &3.25   &233.6  &...    &6.10   &...    & Cab18         \\ %INT, Resolved Gaia
1106058AB               &2.68   &219.5  &3.75   &3.88   &3.15   & Cab08         \\ %INT, Resolved Gaia
1045067AB               &1.35   &197.4  &0.57   &0.041  &0.14   & This work     \\ %GTC, Resolved Gaia
1245057AB               &2.14   &4.3    &0.31   &0.0019 &--0.11 & Cab08         \\ %IDS,GTC Resolved Gaia
\noalign{\smallskip}
\hline
      \end{tabular}
      \begin{list}{}{}
      \item[$^{a}$] {\bf References.}
      Cab08: \cite{2008A&A...478..667C}; 
      Cab10a: \citet{2010ASSP...14...79C}; 
      Cab10b: \citet{2010A&A...514A..18C}; 
      %Cab14: \citet{2014Obs...134..273C}; 
      Cab18: \citet{2018AN....339...60C}.
      %\item[$^{b}$] $\sigma$~Ori ...
      %\item[$^{c}$] Stars not resolved by Gaia. The data in the table was taken from \citet{2014Obs...134..273C}.
      \end{list}
 \end{table*}

\begin{figure}
\centering
\includegraphics[width=0.51\textwidth]{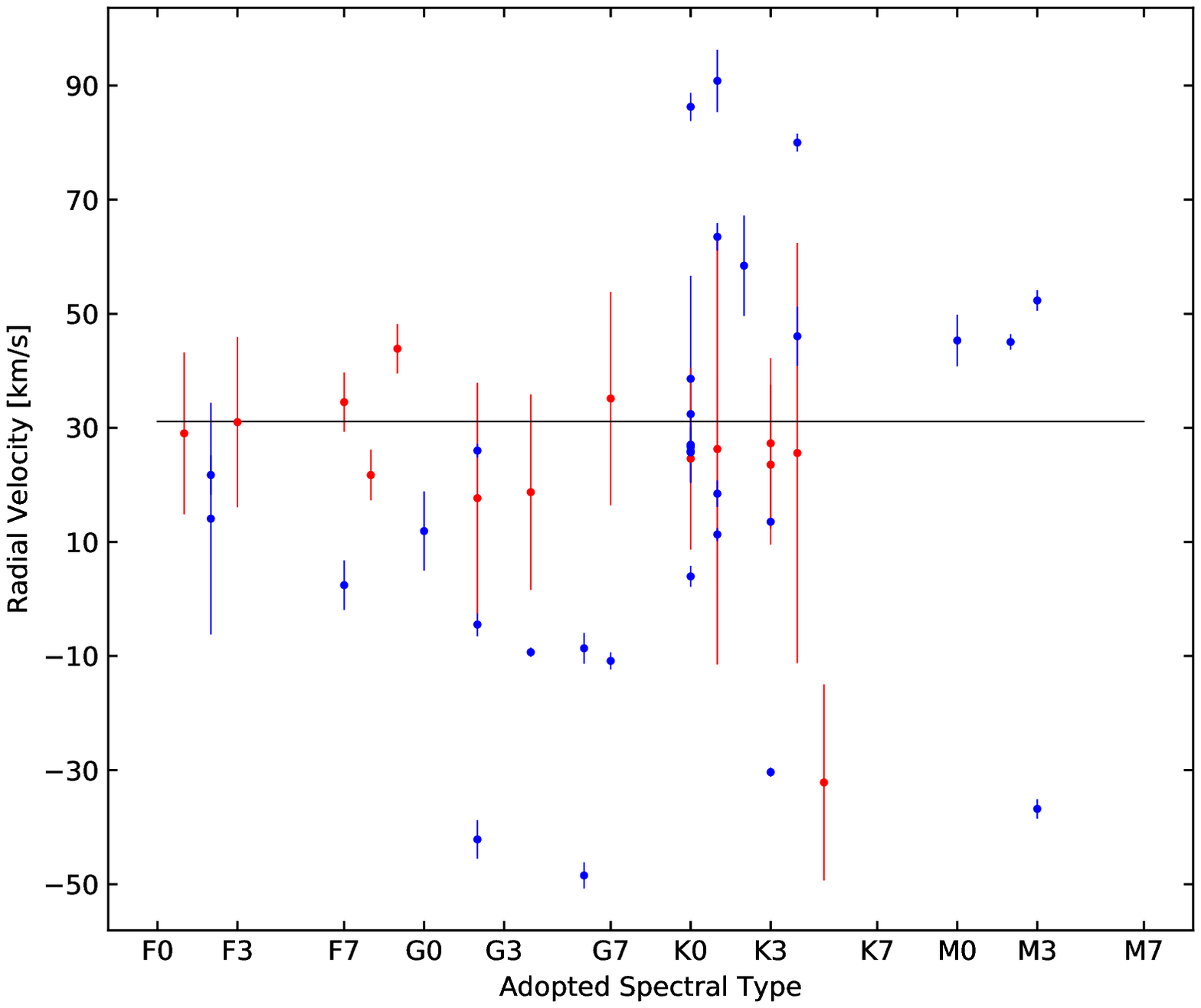}
\caption{\textit{Gaia} DR2 radial velocity as a function of adopted spectral type. 
Red {and blue} circles highlight $\sigma$~Orionis stars {and} non-cluster-member stars, {respectively};
coloured vertical error bars show 3$\sigma$ uncertainties;
and the black horizontal line shows the $\sigma$~Ori systemic velocity $\gamma$ = +31.10$\pm$0.16\,km\,s$^{-1}$.} % 
\label{figure.radial_vel}
\end{figure}

On the one hand, in Fig.~\ref{figure.spt_li} there are two mid-type stars with relatively small EW(Li~{\sc i}) for their effective temperatures.
These are Mayrit~863116 (HD~294300; G4) and Mayrit~1415279AB (OriNTT~429; K1).
The latter is a spectroscopic binary discovered by \citet{1994AJ....108.1445L}.
As discussed in detail by \citet{2010ASSP...14...79C}\footnote{This is a summary in English of Caballero (2006, PhD thesis, Universidad de La Laguna, Spain).}, this double star shows a lower EW(Li~{\sc i}) than its actual value when observed in low spectral resolution (the combined spectrum of the red- and blueshifted components is smoothed by the low resolution).

On the other hand, there are another two classical T~Tauri stars in {Fig.~\ref{figure.spt_li}} with youth features and proper motions consistent with $\sigma$~Orionis membership, but with abnormal \textit{Gaia} DR2 parallaxes and low-quality astrometric flags.
These {are} Mayrit~359179AB (V595~Ori) and Mayrit~873229AB (Haro~5--7), which were classified by \citet{2018AN....339...60C} as a resolved binary and a binary candidate, respectively, with angular separations $\rho <$1\,arcsec {(Mayrit~873229AB is also a double-line spectroscopic binary; \cite{2008MNRAS.385.2210M})}.

{Based on these two facts, close binarity could} explain part of the scatter in the EW(Li~{\sc i}) versus spectral-type diagram in Fig.~\ref{figure.spt_li} (and in colour-magnitude diagrams), as well as inconsistencies between spectroscopic and astrometric data.
%\LEt{Please check that I have retained your intended meaning.}.
Multiplicity in the $\sigma$~Orionis cluster was reviewed in detail by \citet{2014Obs...134..273C} and updated by \citet{2018AN....339...60C}.
Following the procedure of the latter publication, we looked in \textit{Gaia} DR2 for close binaries ($\rho <$ 4\,arcsec) among the investigated stars. 
Our results are summarised in Table~\ref{table.binaries_specs}.

With the help of the Topcat Virtual Observatory tool, we identified and measured angular separation $\rho$, position angle $\theta$, and magnitude difference in the {\em Gaia} broadband filter $\Delta G$ for {11} $\sigma$~Orionis stars (at {\em Gaia} DR2 epoch J2015.5).
We were also able to measure magnitude differences in the blue and red {\em Gaia} filters $\Delta B_p$ and $\Delta R_p$ for 5 of those 11.
Of the {11} binaries, {6} had been discovered by \citet{2008A&A...478..667C,2010ASSP...14...79C,2010A&A...514A..18C} and \citet{2018AN....339...60C}.
Another two, indicated with parenthesis under ``Reference'' in Table~\ref{table.binaries_specs}, were presented as binary candidates by {\citet{2018AN....339...60C}}, but these latter authors were only able to impose upper limits on angular separation.
In summary, we discovered three new close binaries, namely
Mayrit 329261AB, Mayrit 1045067AB, and Mayrit 1248183AB, and measured $\rho$ and $\theta$ for the first time for another two close binaries, namely Mayrit~873229AB and Mayrit~1626148AB. 
Of these five objects, only the equal-mass Mayrit 1045067AB binary has an angular separation {greater} than 0.90\,arcsec.
The eponymous $\sigma$~Ori multiple system, which contains at least five stars with spectral types between B2\,V and O9.5\,V (\cite{2014Obs...134..273C,2015ApJ...799..169S,2018A&A...615A.161M}, and references therein), is not resolved by {\em Gaia}. 

As illustrated by Fig.~\ref{figure.radial_vel}, we also retrieved {\em Gaia} DR2 radial velocities for {14} $\sigma$~Orionis stellar members (and for {32}  non-cluster-members), and found three young stars with radial velocities discordant with that of the cluster by more than 3$\sigma$:
 Mayrit 822170 {([W96]~4771--0119)}, Mayrit 1275190, and Mayrit 1285339 {(HD~294268)}.
The case of the T~Tauri star Mayrit~822170 is striking, as it has lithium in absorption and H$\alpha$ and X-rays in emission but a radial velocity of --32$\pm$6\,km\,s$^{-1}$, vastly different from the +31.10$\pm$0.16\,km\,s$^{-1}$ determined by \citet{2015ApJ...799..169S} {for the open cluster.
We looked for additional radial-velocity determinations on these three stars in the studies by
\citet{2006MNRAS.371L...6J}, 
\citet{2010ASSP...14...79C},
\citet{2008MNRAS.385.2210M},
\citet{2008A&A...488..167S}, 
\citet{2014ApJ...794...36H}, 
and \citet{2018AJ....156...84K}.
Our findings are summarised in Table~\ref{table.RVs}.
The observed discordances are likely due to membership to the sparse foreground population of the Orion OB1 association (Mayrit~1275190 and, especially, Mayrit~1285339; \cite{2018ApJ...867..116P}) or to the presence of an unseen companion (Mayrit 822170)}.
The same may apply to the other young stars with low EW(Li~{\sc i}) for their spectral type or abnormal radial velocities and parallaxes.
{If both Mayrit~1275190 and Mayrit~1285339, at 21.3--21.4\,arcmin to the cluster centre, actually belonged to a distinct population, the list of previous cluster member candidates discarded in this work (Table~\ref{table.no_members}) would increase to 37 and in addition $\sigma$~Orionis would be spatially more compact.

There are six tight binary systems among the 11 investigated {\em Gaia} DR2 sources with the highest {\tt RUWE} values (Sect.~\ref{subsection.membership}):
$\sigma$~Ori\,AB ($\rho \approx$ 0.266\,arcsec -- \cite{1892AN....130..257B,  2008AJ....136..554T, 2011ApJ...742...55S}), 
Mayrit~528005AB ([W96]~4771--899, $\rho \approx$ 0.40\,arcsec -- \cite{2005AN....326.1007C}), 
and four of the five young binary stars in our sample resolved by {\em Gaia} (Table~\ref{table.binaries_specs}) with angular separations between 0.6 and 1.0\,arcsec.
All six have {\tt RUWE} values greater than 3.8, much larger than the maximum value of 1.4 recommended for reliable astrometry by the {\em Gaia} Data Processing and Analysis Consortium (c.f., Sect.~\ref{subsection.membership}).
The fifth resolved binary, with $\rho \approx$ 0.90\,arcsec, $\Delta G \approx$ 3.00\,mag, and {\tt RUWE} $\approx$ 1.23, is Mayrit~329261AB ([SWW2004]~207), for which \citet{2004AJ....128.2316S} reported a very nearby source with similar optical magnitudes ([SWW2004]~103).
The other five stars sorted by {\tt RUWE} values from 21.8 to 5.0 are Mayrit~203283, 1082115, 1073209, 521199, and 157155.
All five of them have at least Li~{\sc i} in absorption and are thought to be single, with one of them, namely Mayrit~521199 (TX~Ori), displaying all the attributes of a classical T~Tauri star.
Alternatively, these five objects could be tight binaries yet unresolved by {\em Gaia}, with angular separations of 0.5\,arcsec or less.}

\begin{table}
\centering
  \caption[]{Three Mayrit stars with discordant radial velocities.} 
    \label{table.RVs}
      \begin{tabular}{lcccccl}
\hline
\hline
\noalign{\smallskip}
Mayrit  &       & $V_r$ [km\,s$^{-1}$]  &       \\  %
\noalign{\smallskip}
                & Her14$^{a}$   & Kou18$^{a}$ & {\em Gaia}$^{a}$ \\ %
\noalign{\smallskip}
\hline
\noalign{\smallskip} 
822170          & +34.0$\pm$0.8 & +33.09$\pm$0.47 & --32.15$\pm$5.72 \\ 
1275190 & ...   & +45.81$\pm$0.75 & +43.88$\pm$1.44 \\
1285339 & +22.2$\pm$1.1 & +22.90$\pm$0.63 & +21.73$\pm$1.48 \\
\noalign{\smallskip}
\hline
      \end{tabular}
      \begin{list}{}{}
      \item[$^{a}$] {\bf References.}
      {Her14: \cite{2014ApJ...794...36H};
      Kou18: \cite{2018AJ....156...84K};
      {\em Gaia}: \cite{2018A&A...616A...1G}.}
      \end{list}
 \end{table}

%========================================================================== 

\section{Summary}
\label{section.summary}

We present a new detailed characterisation of {111} $\sigma$~Orionis cluster members with spectral types from O9.5 to M6.5, {4} T~Tauri stars in neighbouring star-forming regions, and {51} foreground and background stars.
Our target sample includes OB stars in Trapezium-like systems, Herbig Ae/Be and T~Tauri stars, and brown dwarfs with ages of about 3\,Ma, in addition to
%\LEt{Please check that I have retained your intended meaning.}  
peculiar interloper stars of different ages and evolutionary stages.

We used low-resolution optical spectroscopy collected with IDS at the INT and OSIRIS at the GTC to identify new youth features (mostly Li~{\sc i} in absorption and Balmer lines in emission) and to determine spectral types, {\em Gaia} DR2 proper motions, parallaxes, and photometry. We also used Virtual Observatory tools to astrometrically discard cluster members and to find close binaries, and literature information, such as X-ray emission, MIR flux excess, and previous measurements of Li~{\sc i} and H$\alpha$, for previous and relevant information. 
The combination of all these data allowed us to carry out an exhaustive membership classification and thereby form a more detailed picture of the $\sigma$~Orionis stellar and substellar  populations.

Perhaps the most remarkable result is the finding that {35} of those {51}  non-cluster-member stars were previously classified and considered as cluster members in highly cited works on disc frequency and characterisation, initial mass function, spatial distribution, and chemical abundances.
These 35 stars represent about 10\,\% of the whole $\sigma$~Orionis stellar population, and very likely a higher fraction at intermediate masses ($\mathcal{M} \sim$ 0.5--1.0\,$M_\odot$), which has a significant impact on the results shown in previous works.

In addition, we have found:
{14} strong accretors with EW($\alpha$) $<$ --50\,{\AA}, of which 2 are new identifications and {4} have measured EW(H$\alpha$) for the first time;
{5} strong accretors with signficant blue- or redshift of the H$\alpha$ line, which can host undetected jets;
{11} astrometric binaries with angular separations of 0.6--3.2\,arcsec, of which {3} are discovered here and {5} have measured $\rho$ and $\theta$ for the first time; 
{2 juvenile star candidates in the sparse population of the Ori~OB1b association; and one spectroscopic candidate} based on {\em Gaia} DR2 radial velocities.
{Another five cluster members could be tight binaries based on large values of {\em Gaia} DR2 re-normalised unit-weight errors.}

This fourth paper of the  series ``Stars and brown dwarfs in the $\sigma$~Orionis cluster'' is a follow-on of the comprehensive analysis of the very young open cluster with perhaps the best-studied stellar and substellar populations.
It is also a preliminary step for the improvement in the determination of key parameters in star formation: metallicity, multiplicity, slope of the mass spectrum, disc and jet frequency, and spatial concentration across the huge mass interval from 20 to 0.005\,$M_\odot$.

%========================================================================== 

\begin{acknowledgements}

We thank {the anonymous referee for the careful review}, J.~Sanz-Forcada for helpful comments on X-rays, and R. Campillo for starting working on the data during his MSc thesis.
This article is based on observations made in the Observatorios de Canarias del IAC with the Gran Telescopio Canarias and Isaac Newton Telescope of the Isaac Newton Group of Telescopes, both installed at the Spanish Observatorio del Roque de los Muchachos of the Instituto de Astrof\'isica de Canarias, in the island of La Palma, Spain, under programs INT15-07A, GTC55-12A, and GTC30-12B.
This research made use of the SIMBAD, operated at Centre de Donn\'ees astronomiques de Strasbourg, France, and NASA's Astrophysics Data System.
Financial support was provided by the Universidad Complutense de Madrid, the Comunidad Aut\'onoma de Madrid, and the Spanish Ministerio de Ciencia e Innovaci\'on under grant AYA2016-79425-C3-2-P.        % (CARMENES-CAB) 

\end{acknowledgements}

%========================================================================== 

   \bibliographystyle{aa} % style aa.bst
   \bibliography{so.bib} % your references Yourfile.bib

\begin{thebibliography}{138}
\expandafter\ifx\csname natexlab\endcsname\relax\def\natexlab#1{#1}\fi

\bibitem[{{Alonso-Floriano} {et~al.}(2015){Alonso-Floriano}, {Morales},
  {Caballero}, {Montes}, {Klutsch}, {Mundt}, {Cort{\'e}s-Contreras}, {Ribas},
  {Reiners}, {Amado}, {Quirrenbach}, \& {Jeffers}}]{2015A&A...577A.128A}
{Alonso-Floriano}, F.~J., {Morales}, J.~C., {Caballero}, J.~A., {et~al.} 2015,
  \aap, 577, A128

\bibitem[{{Altmann} {et~al.}(2017){Altmann}, {Roeser}, {Demleitner}, {Bastian},
  \& {Schilbach}}]{2017A&A...600L...4A}
{Altmann}, M., {Roeser}, S., {Demleitner}, M., {Bastian}, U., \& {Schilbach},
  E. 2017, \aap, 600, L4

\bibitem[{{\'Alvarez-Meraz} {et~al.}(2017){\'Alvarez-Meraz}, {Nagel}, {Rendon},
  \& {Barrag\'an}}]{2017RMxAA..53..275A}
{\'Alvarez-Meraz}, R., {Nagel}, E., {Rendon}, F., \& {Barrag\'an}, O. 2017,
  \rmxaa, 53, 275

\bibitem[{{Ansdell} {et~al.}(2017){Ansdell}, {Williams}, {Manara}, {Miotello},
  {Facchini}, {van der Marel}, {Testi}, \& {van
  Dishoeck}}]{2017AJ....153..240A}
{Ansdell}, M., {Williams}, J.~P., {Manara}, C.~F., {et~al.} 2017, \aj, 153, 240

\bibitem[{{Bailer-Jones} {et~al.}(2018){Bailer-Jones}, {Rybizki}, {Fouesneau},
  {Mantelet}, \& {Andrae}}]{2018AJ....156...58B}
{Bailer-Jones}, C.~A.~L., {Rybizki}, J., {Fouesneau}, M., {Mantelet}, G., \&
  {Andrae}, R. 2018, \aj, 156, 58

\bibitem[{{Barnes} {et~al.}(1989){Barnes}, {Crutcher}, {Bieging}, {Storey}, \&
  {Willner}}]{1989ApJ...342..883B}
{Barnes}, P.~J., {Crutcher}, R.~M., {Bieging}, J.~H., {Storey}, J.~W.~V., \&
  {Willner}, S.~P. 1989, \apj, 342, 883

\bibitem[{{Barrado y Navascu{\'e}s} \&
  {Mart{\'{\i}}n}(2003)}]{2003AJ....126.2997B}
{Barrado y Navascu{\'e}s}, D. \& {Mart{\'{\i}}n}, E.~L. 2003, \aj, 126, 2997

\bibitem[{{B{\'e}jar} {et~al.}(1999){B{\'e}jar}, {Osorio}, \&
  {Rebolo}}]{1999ApJ...521..671B}
{B{\'e}jar}, V.~J.~S., {Osorio}, M.~R.~Z., \& {Rebolo}, R. 1999, \apj, 521, 671

\bibitem[{{Bertout}(1989)}]{1989ARA&A..27..351B}
{Bertout}, C. 1989, \araa, 27, 351

\bibitem[{{Bik} {et~al.}(2003){Bik}, {Lenorzer}, {Kaper}, {Comer{\'o}n},
  {Waters}, {de Koter}, \& {Hanson}}]{2003A&A...404..249B}
{Bik}, A., {Lenorzer}, A., {Kaper}, L., {et~al.} 2003, \aap, 404, 249

\bibitem[{{Boesgaard} \& {Tripicco}(1986)}]{1986ApJ...302L..49B}
{Boesgaard}, A.~M. \& {Tripicco}, M.~J. 1986, \apj, 302, L49

\bibitem[{{Bouvier} {et~al.}(2018){Bouvier}, {Barrado}, {Moraux}, {Stauffer},
  {Rebull}, {Hillenbrand}, {Bayo}, {Boisse}, {Bouy}, {DiFolco}, {Lillo-Box}, \&
  {Morales Calder{\'o}n}}]{2018A&A...613A..63B}
{Bouvier}, J., {Barrado}, D., {Moraux}, E., {et~al.} 2018, \aap, 613, A63

\bibitem[{{Brice{\~n}o} {et~al.}(2019){Brice{\~n}o}, {Calvet}, {Hern{\'a}ndez},
  {Vivas}, {Mateu}, {Jos{\'e} Downes}, {Loerincs}, {P{\'e}rez-Blanco},
  {Berlind}, \& {Espaillat}}]{2019AJ....157...85B}
{Brice{\~n}o}, C., {Calvet}, N., {Hern{\'a}ndez}, J., {et~al.} 2019, \aj, 157,
  85

\bibitem[{{Brown} {et~al.}(2013){Brown}, {Pontoppidan}, {van Dishoeck},
  {Herczeg}, {Blake}, \& {Smette}}]{2013ApJ...770...94B}
{Brown}, J.~M., {Pontoppidan}, K.~M., {van Dishoeck}, E.~F., {et~al.} 2013,
  \apj, 770, 94

\bibitem[{{Burnham}(1892)}]{1892AN....130..257B}
{Burnham}, S.~W. 1892, Astronomische Nachrichten, 130, 257

\bibitem[{{Burningham} {et~al.}(2005){Burningham}, {Naylor}, {Littlefair}, \&
  {Jeffries}}]{2005MNRAS.356.1583B}
{Burningham}, B., {Naylor}, T., {Littlefair}, S.~P., \& {Jeffries}, R.~D. 2005,
  \mnras, 356, 1583

\bibitem[{{Caballero}(2005)}]{2005AN....326.1007C}
{Caballero}, J.~A. 2005, Astronomische Nachrichten, 326, 1007

\bibitem[{{Caballero}(2007)}]{2007A&A...466..917C}
{Caballero}, J.~A. 2007, \aap, 466, 917

\bibitem[{{Caballero}(2008{\natexlab{a}})}]{2008MNRAS.383..375C}
{Caballero}, J.~A. 2008{\natexlab{a}}, \mnras, 383, 375

\bibitem[{{Caballero}(2008{\natexlab{b}})}]{2008A&A...478..667C}
{Caballero}, J.~A. 2008{\natexlab{b}}, \aap, 478, 667

\bibitem[{{Caballero}(2010{\natexlab{a}})}]{2010ASSP...14...79C}
{Caballero}, J.~A. 2010{\natexlab{a}}, Astrophysics and Space Science
  Proceedings, 14, 79

\bibitem[{{Caballero}(2010{\natexlab{b}})}]{2010A&A...514A..18C}
{Caballero}, J.~A. 2010{\natexlab{b}}, \aap, 514, A18

\bibitem[{{Caballero}(2011)}]{2011sca..conf..108C}
{Caballero}, J.~A. 2011, in Stellar Clusters \& Associations: A RIA Workshop on
  Gaia, 108--112

\bibitem[{{Caballero}(2014)}]{2014Obs...134..273C}
{Caballero}, J.~A. 2014, The Observatory, 134, 273

\bibitem[{{Caballero}(2017)}]{2017AN....338..629C}
{Caballero}, J.~A. 2017, Astronomische Nachrichten, 338, 629

\bibitem[{{Caballero}(2018)}]{2018RNAAS...2b..25C}
{Caballero}, J.~A. 2018, Research Notes of the American Astronomical Society,
  2, 25

\bibitem[{{Caballero} {et~al.}(2010){Caballero}, {Albacete-Colombo}, \&
  {L{\'o}pez- Santiago}}]{2010A&A...521A..45C}
{Caballero}, J.~A., {Albacete-Colombo}, J.~F., \& {L{\'o}pez- Santiago}, J.
  2010, \aap, 521, A45

\bibitem[{{Caballero} {et~al.}(2007){Caballero}, {B{\'e}jar}, {Rebolo},
  {Eisl{\"o}ffel}, {Zapatero Osorio}, {Mundt}, {Barrado Y Navascu{\'e}s},
  {Bihain}, {Bailer- Jones}, {Forveille}, \&
  {Mart{\'\i}n}}]{2007A&A...470..903C}
{Caballero}, J.~A., {B{\'e}jar}, V.~J.~S., {Rebolo}, R., {et~al.} 2007, \aap,
  470, 903

\bibitem[{{Caballero} {et~al.}(2004){Caballero}, {B{\'e}jar}, {Rebolo}, \&
  {Zapatero Osorio}}]{2004A&A...424..857C}
{Caballero}, J.~A., {B{\'e}jar}, V.~J.~S., {Rebolo}, R., \& {Zapatero Osorio},
  M.~R. 2004, \aap, 424, 857

\bibitem[{{Caballero} {et~al.}(2016){Caballero}, {Bouy}, \&
  {Lillo-Box}}]{2016Obs...136..226C}
{Caballero}, J.~A., {Bouy}, H., \& {Lillo-Box}, J. 2016, The Observatory, 136,
  226

\bibitem[{{Caballero} {et~al.}(2012){Caballero}, {Cabrera-Lavers},
  {Garc{\'\i}a-{\'A}lvarez}, \& {Pascual}}]{2012A&A...546A..59C}
{Caballero}, J.~A., {Cabrera-Lavers}, A., {Garc{\'\i}a-{\'A}lvarez}, D., \&
  {Pascual}, S. 2012, \aap, 546, A59

\bibitem[{{Caballero} {et~al.}(2009){Caballero}, {L{\'o}pez-Santiago}, {de
  Castro}, \& {Cornide}}]{2009AJ....137.5012C}
{Caballero}, J.~A., {L{\'o}pez-Santiago}, J., {de Castro}, E., \& {Cornide}, M.
  2009, \aj, 137, 5012

\bibitem[{{Caballero} {et~al.}(2018){Caballero}, {Novalbos}, {Tobal}, \&
  {Miret}}]{2018AN....339...60C}
{Caballero}, J.~A., {Novalbos}, I., {Tobal}, T., \& {Miret}, F.~X. 2018,
  Astronomische Nachrichten, 339, 60

\bibitem[{{Caballero} \& {Solano}(2008)}]{2008A&A...485..931C}
{Caballero}, J.~A. \& {Solano}, E. 2008, \aap, 485, 931

\bibitem[{{Caballero} {et~al.}(2008){Caballero}, {Valdivielso}, {Mart{\'\i}n},
  {Montes}, {Pascual}, \& {P{\'e}rez-Gonz{\'a}lez}}]{2008A&A...491..515C}
{Caballero}, J.~A., {Valdivielso}, L., {Mart{\'\i}n}, E.~L., {et~al.} 2008,
  \aap, 491, 515

\bibitem[{{Carlsson} {et~al.}(1994){Carlsson}, {Rutten}, {Bruls}, \&
  {Shchukina}}]{1994A&A...288..860C}
{Carlsson}, M., {Rutten}, R.~J., {Bruls}, J.~H.~M.~J., \& {Shchukina}, N.~G.
  1994, \aap, 288, 860

\bibitem[{{Cepa}(2010)}]{2010ASSP...14...15C}
{Cepa}, J. 2010, Astrophysics and Space Science Proceedings, 14, 15

\bibitem[{{Cepa} {et~al.}(2000){Cepa}, {Aguiar}, {Escalera},
  {Gonzalez-Serrano}, {Joven-Alvarez}, {Peraza}, {Rasilla}, {Rodriguez- Ramos},
  {Gonzalez}, {Cobos Duenas}, {Sanchez}, {Tejada}, {Bland-Hawthorn},
  {Militello}, \& {Rosa}}]{2000SPIE.4008..623C}
{Cepa}, J., {Aguiar}, M., {Escalera}, V.~G., {et~al.} 2000, in Optical and IR
  Telescope Instrumentation and Detectors, Vol. 4008, 623--631

\bibitem[{{Chabrier}(2003)}]{2003PASP..115..763C}
{Chabrier}, G. 2003, \pasp, 115, 763

\bibitem[{{Chaffee} {et~al.}(1971){Chaffee}, {Carbon}, \&
  {Strom}}]{1971ApJ...166..593C}
{Chaffee}, Frederic~H., J., {Carbon}, D.~F., \& {Strom}, S.~E. 1971, \apj, 166,
  593

\bibitem[{{Cody} \& {Hillenbrand}(2014)}]{2014ApJ...796..129C}
{Cody}, A.~M. \& {Hillenbrand}, L.~A. 2014, \apj, 796, 129

\bibitem[{{Collinder}(1931)}]{1931AnLun...2....1C}
{Collinder}, P. 1931, Annals of the Observatory of Lund, 2, B1

\bibitem[{{Cottle} {et~al.}(2018){Cottle}, {Covey}, {Su{\'a}rez},
  {Rom{\'a}n-Z{\'u}{\~n}iga}, {Schlafly}, {Downes}, {Ybarra}, {Hernandez},
  {Stassun}, {Stringfellow}, {Getman}, {Feigelson}, {Borissova}, {Kim},
  {Roman-Lopes}, {Da Rio}, {De Lee}, {Frinchaboy}, {Kounkel}, {Majewski},
  {Mennickent}, {Nidever}, {Nitschelm}, {Pan}, {Shetrone}, {Zasowski},
  {Chambers}, {Magnier}, \& {Valenti}}]{2018ApJS..236...27C}
{Cottle}, J.~N., {Covey}, K.~R., {Su{\'a}rez}, G., {et~al.} 2018, \apjs, 236,
  27

\bibitem[{{Covey} {et~al.}(2007){Covey}, {Ivezi{\'c}}, {Schlegel},
  {Finkbeiner}, {Padmanabhan}, {Lupton}, {Ag{\"u}eros}, {Bochanski}, {Hawley},
  {West}, {Seth}, {Kimball}, {Gogarten}, {Claire}, {Haggard}, {Kaib},
  {Schneider}, \& {Sesar}}]{2007AJ....134.2398C}
{Covey}, K.~R., {Ivezi{\'c}}, {\v{Z}}., {Schlegel}, D., {et~al.} 2007, \aj,
  134, 2398

\bibitem[{Cox(2001)}]{Allen}
Cox, A.~N. 2001, Allen's Astrophysical Quantities (Springer)

\bibitem[{{Cram} \& {Mullan}(1979)}]{1979ApJ...234..579C}
{Cram}, L.~E. \& {Mullan}, D.~J. 1979, \apj, 234, 579

\bibitem[{{Davison} {et~al.}(2015){Davison}, {White}, {Henry}, {Riedel}, {Jao},
  {Bailey}, {Quinn}, {Cantrell}, {Subasavage}, \&
  {Winters}}]{2015AJ....149..106D}
{Davison}, C.~L., {White}, R.~J., {Henry}, T.~J., {et~al.} 2015, \aj, 149, 106

\bibitem[{{Elliott} {et~al.}(2017){Elliott}, {Scholz}, {Jayawardhana},
  {Eisl{\"o}ffel}, \& {H{\'e}brard}}]{2017A&A...608A..66E}
{Elliott}, P., {Scholz}, A., {Jayawardhana}, R., {Eisl{\"o}ffel}, J., \&
  {H{\'e}brard}, E.~M. 2017, \aap, 608, A66

\bibitem[{{Falc{\'o}n-Barroso} {et~al.}(2011){Falc{\'o}n-Barroso},
  {S{\'a}nchez-Bl{\'a}zquez}, {Vazdekis}, {Ricciardelli}, {Cardiel}, {Cenarro},
  {Gorgas}, \& {Peletier}}]{2011A&A...532A..95F}
{Falc{\'o}n-Barroso}, J., {S{\'a}nchez-Bl{\'a}zquez}, P., {Vazdekis}, A.,
  {et~al.} 2011, \aap, 532, A95

\bibitem[{{Favata} \& {Micela}(2003)}]{2003SSRv..108..577F}
{Favata}, F. \& {Micela}, G. 2003, \ssr, 108, 577

\bibitem[{{Fedorovich}(1960)}]{1960PZ.....13..166F}
{Fedorovich}, V.~P. 1960, Peremennye Zvezdy, 13, 166

\bibitem[{{F{\H{u}}r{\'e}sz} {et~al.}(2008){F{\H{u}}r{\'e}sz}, {Hartmann},
  {Megeath}, {Szentgyorgyi}, \& {Hamden}}]{2008ApJ...676.1109F}
{F{\H{u}}r{\'e}sz}, G., {Hartmann}, L.~W., {Megeath}, S.~T., {Szentgyorgyi},
  A.~H., \& {Hamden}, E.~T. 2008, \apj, 676, 1109

\bibitem[{{Franciosini} {et~al.}(2006){Franciosini}, {Pallavicini}, \&
  {Sanz-Forcada}}]{2006A&A...446..501F}
{Franciosini}, E., {Pallavicini}, R., \& {Sanz-Forcada}, J. 2006, A\&A, 446,
  501

\bibitem[{Freedman \& Diaconis(1981)}]{Freedman1981}
Freedman, D. \& Diaconis, P. 1981, Zeitschrift f{\"u}r
  Wahrscheinlichkeitstheorie und Verwandte Gebiete, 57, 453

\bibitem[{{Gaia Collaboration} {et~al.}(2018){Gaia Collaboration}, {Brown},
  {Vallenari}, {Prusti}, {de Bruijne}, {Babusiaux}, {Bailer-Jones}, {Biermann},
  {Evans}, {Eyer}, {Jansen}, {Jordi}, {Klioner}, {Lammers}, {Lindegren},
  {Luri}, {Mignard}, {Panem}, {Pourbaix}, {Randich}, {Sartoretti}, {Siddiqui},
  {Soubiran}, {van Leeuwen}, {Walton}, {Arenou}, {Bastian}, {Cropper},
  {Drimmel}, {Katz}, {Lattanzi}, {Bakker}, {Cacciari}, {Casta{\~n}eda},
  {Chaoul}, {Cheek}, {De Angeli}, {Fabricius}, {Guerra}, {Holl}, {Masana},
  {Messineo}, {Mowlavi}, {Nienartowicz}, {Panuzzo}, {Portell}, {Riello},
  {Seabroke}, {Tanga}, {Th{\'e}venin}, {Gracia-Abril}, {Comoretto},
  {Garcia-Reinaldos}, {Teyssier}, {Altmann}, {Andrae}, {Audard},
  {Bellas-Velidis}, {Benson}, {Berthier}, {Blomme}, {Burgess}, {Busso},
  {Carry}, {Cellino}, {Clementini}, {Clotet}, {Creevey}, {Davidson}, {De
  Ridder}, {Delchambre}, {Dell'Oro}, {Ducourant},
  {Fern{\'a}ndez-Hern{\'a}ndez}, {Fouesneau}, {Fr{\'e}mat}, {Galluccio},
  {Garc{\'\i}a-Torres}, {Gonz{\'a}lez-N{\'u}{\~n}ez}, {Gonz{\'a}lez-Vidal},
  {Gosset}, {Guy}, {Halbwachs}, {Hambly}, {Harrison}, {Hern{\'a}ndez},
  {Hestroffer}, {Hodgkin}, {Hutton}, {Jasniewicz}, {Jean-Antoine-Piccolo},
  {Jordan}, {Korn}, {Krone-Martins}, {Lanzafame}, {Lebzelter}, {L{\"o}ffler},
  {Manteiga}, {Marrese}, {Mart{\'\i}n-Fleitas}, {Moitinho}, {Mora}, {Muinonen},
  {Osinde}, {Pancino}, {Pauwels}, {Petit}, {Recio-Blanco}, {Richards},
  {Rimoldini}, {Robin}, {Sarro}, {Siopis}, {Smith}, {Sozzetti}, {S{\"u}veges},
  {Torra}, {van Reeven}, {Abbas}, {Abreu Aramburu}, {Accart}, {Aerts},
  {Altavilla}, {{\'A}lvarez}, {Alvarez}, {Alves}, {Anderson}, {Andrei},
  {Anglada Varela}, {Antiche}, {Antoja}, {Arcay}, {Astraatmadja}, {Bach},
  {Baker}, {Balaguer-N{\'u}{\~n}ez}, {Balm}, {Barache}, {Barata}, {Barbato},
  {Barblan}, {Barklem}, {Barrado}, {Barros}, {Barstow}, {Bartholom{\'e}
  Mu{\~n}oz}, {Bassilana}, {Becciani}, {Bellazzini}, {Berihuete}, {Bertone},
  {Bianchi}, {Bienaym{\'e}}, {Blanco-Cuaresma}, {Boch}, {Boeche}, {Bombrun},
  {Borrachero}, {Bossini}, {Bouquillon}, {Bourda}, {Bragaglia}, {Bramante},
  {Breddels}, {Bressan}, {Brouillet}, {Br{\"u}semeister}, {Brugaletta},
  {Bucciarelli}, {Burlacu}, {Busonero}, {Butkevich}, {Buzzi}, {Caffau},
  {Cancelliere}, {Cannizzaro}, {Cantat-Gaudin}, {Carballo}, {Carlucci},
  {Carrasco}, {Casamiquela}, {Castellani}, {Castro-Ginard}, {Charlot},
  {Chemin}, {Chiavassa}, {Cocozza}, {Costigan}, {Cowell}, {Crifo}, {Crosta},
  {Crowley}, {Cuypers}, {Dafonte}, {Damerdji}, {Dapergolas}, {David}, {David},
  {de Laverny}, {De Luise}, {De March}, {de Martino}, {de Souza}, {de Torres},
  {Debosscher}, {del Pozo}, {Delbo}, {Delgado}, {Delgado}, {Di Matteo},
  {Diakite}, {Diener}, {Distefano}, {Dolding}, {Drazinos}, {Dur{\'a}n},
  {Edvardsson}, {Enke}, {Eriksson}, {Esquej}, {Eynard Bontemps}, {Fabre},
  {Fabrizio}, {Faigler}, {Falc{\~a}o}, {Farr{\`a}s Casas}, {Federici},
  {Fedorets}, {Fernique}, {Figueras}, {Filippi}, {Findeisen}, {Fonti},
  {Fraile}, {Fraser}, {Fr{\'e}zouls}, {Gai}, {Galleti}, {Garabato},
  {Garc{\'\i}a-Sedano}, {Garofalo}, {Garralda}, {Gavel}, {Gavras}, {Gerssen},
  {Geyer}, {Giacobbe}, {Gilmore}, {Girona}, {Giuffrida}, {Glass}, {Gomes},
  {Granvik}, {Gueguen}, {Guerrier}, {Guiraud}, {Guti{\'e}rrez-S{\'a}nchez},
  {Haigron}, {Hatzidimitriou}, {Hauser}, {Haywood}, {Heiter}, {Helmi}, {Heu},
  {Hilger}, {Hobbs}, {Hofmann}, {Holland}, {Huckle}, {Hypki}, {Icardi},
  {Jan{\ss}en}, {Jevardat de Fombelle}, {Jonker}, {Juh{\'a}sz}, {Julbe},
  {Karampelas}, {Kewley}, {Klar}, {Kochoska}, {Kohley}, {Kolenberg},
  {Kontizas}, {Kontizas}, {Koposov}, {Kordopatis}, {Kostrzewa-Rutkowska},
  {Koubsky}, {Lambert}, {Lanza}, {Lasne}, {Lavigne}, {Le Fustec}, {Le
  Poncin-Lafitte}, {Lebreton}, {Leccia}, {Leclerc}, {Lecoeur-Taibi},
  {Lenhardt}, {Leroux}, {Liao}, {Licata}, {Lindstr{\o}m}, {Lister}, {Livanou},
  {Lobel}, {L{\'o}pez}, {Managau}, {Mann}, {Mantelet}, {Marchal}, {Marchant},
  {Marconi}, {Marinoni}, {Marschalk{\'o}}, {Marshall}, {Martino}, {Marton},
  {Mary}, {Massari}, {Matijevi{\v{c}}}, {Mazeh}, {McMillan}, {Messina},
  {Michalik}, {Millar}, {Molina}, {Molinaro}, {Moln{\'a}r}, {Montegriffo},
  {Mor}, {Morbidelli}, {Morel}, {Morris}, {Mulone}, {Muraveva}, {Musella},
  {Nelemans}, {Nicastro}, {Noval}, {O'Mullane}, {Ord{\'e}novic},
  {Ord{\'o}{\~n}ez-Blanco}, {Osborne}, {Pagani}, {Pagano}, {Pailler},
  {Palacin}, {Palaversa}, {Panahi}, {Pawlak}, {Piersimoni}, {Pineau}, {Plachy},
  {Plum}, {Poggio}, {Poujoulet}, {Pr{\v{s}}a}, {Pulone}, {Racero}, {Ragaini},
  {Rambaux}, {Ramos-Lerate}, {Regibo}, {Reyl{\'e}}, {Riclet}, {Ripepi}, {Riva},
  {Rivard}, {Rixon}, {Roegiers}, {Roelens}, {Romero-G{\'o}mez}, {Rowell},
  {Royer}, {Ruiz-Dern}, {Sadowski}, {Sagrist{\`a} Sell{\'e}s}, {Sahlmann},
  {Salgado}, {Salguero}, {Sanna}, {Santana-Ros}, {Sarasso}, {Savietto},
  {Schultheis}, {Sciacca}, {Segol}, {Segovia}, {S{\'e}gransan}, {Shih},
  {Siltala}, {Silva}, {Smart}, {Smith}, {Solano}, {Solitro}, {Sordo}, {Soria
  Nieto}, {Souchay}, {Spagna}, {Spoto}, {Stampa}, {Steele},
  {Steidelm{\"u}ller}, {Stephenson}, {Stoev}, {Suess}, {Surdej}, {Szabados},
  {Szegedi-Elek}, {Tapiador}, {Taris}, {Tauran}, {Taylor}, {Teixeira},
  {Terrett}, {Teyssand ier}, {Thuillot}, {Titarenko}, {Torra Clotet}, {Turon},
  {Ulla}, {Utrilla}, {Uzzi}, {Vaillant}, {Valentini}, {Valette}, {van Elteren},
  {Van Hemelryck}, {van Leeuwen}, {Vaschetto}, {Vecchiato}, {Veljanoski},
  {Viala}, {Vicente}, {Vogt}, {von Essen}, {Voss}, {Votruba}, {Voutsinas},
  {Walmsley}, {Weiler}, {Wertz}, {Wevers}, {Wyrzykowski}, {Yoldas},
  {{\v{Z}}erjal}, {Ziaeepour}, {Zorec}, {Zschocke}, {Zucker}, {Zurbach}, \&
  {Zwitter}}]{2018A&A...616A...1G}
{Gaia Collaboration}, {Brown}, A.~G.~A., {Vallenari}, A., {et~al.} 2018, \aap,
  616, A1

\bibitem[{{Gaia Collaboration} {et~al.}(2016){Gaia Collaboration}, {Prusti},
  {de Bruijne}, {Brown}, {Vallenari}, {Babusiaux}, {Bailer-Jones}, {Bastian},
  {Biermann}, {Evans}, {Eyer}, {Jansen}, {Jordi}, {Klioner}, {Lammers},
  {Lindegren}, {Luri}, {Mignard}, {Milligan}, {Panem}, {Poinsignon},
  {Pourbaix}, {Randich}, {Sarri}, {Sartoretti}, {Siddiqui}, {Soubiran},
  {Valette}, {van Leeuwen}, {Walton}, {Aerts}, {Arenou}, {Cropper}, {Drimmel},
  {H{\o}g}, {Katz}, {Lattanzi}, {O'Mullane}, {Grebel}, {Holland}, {Huc},
  {Passot}, {Bramante}, {Cacciari}, {Casta{\~n}eda}, {Chaoul}, {Cheek}, {De
  Angeli}, {Fabricius}, {Guerra}, {Hern{\'a}ndez}, {Jean-Antoine-Piccolo},
  {Masana}, {Messineo}, {Mowlavi}, {Nienartowicz}, {Ord{\'o}{\~n}ez-Blanco},
  {Panuzzo}, {Portell}, {Richards}, {Riello}, {Seabroke}, {Tanga},
  {Th{\'e}venin}, {Torra}, {Els}, {Gracia-Abril}, {Comoretto},
  {Garcia-Reinaldos}, {Lock}, {Mercier}, {Altmann}, {Andrae}, {Astraatmadja},
  {Bellas-Velidis}, {Benson}, {Berthier}, {Blomme}, {Busso}, {Carry},
  {Cellino}, {Clementini}, {Cowell}, {Creevey}, {Cuypers}, {Davidson}, {De
  Ridder}, {de Torres}, {Delchambre}, {Dell'Oro}, {Ducourant}, {Fr{\'e}mat},
  {Garc{\'\i}a-Torres}, {Gosset}, {Halbwachs}, {Hambly}, {Harrison}, {Hauser},
  {Hestroffer}, {Hodgkin}, {Huckle}, {Hutton}, {Jasniewicz}, {Jordan},
  {Kontizas}, {Korn}, {Lanzafame}, {Manteiga}, {Moitinho}, {Muinonen},
  {Osinde}, {Pancino}, {Pauwels}, {Petit}, {Recio-Blanco}, {Robin}, {Sarro},
  {Siopis}, {Smith}, {Smith}, {Sozzetti}, {Thuillot}, {van Reeven}, {Viala},
  {Abbas}, {Abreu Aramburu}, {Accart}, {Aguado}, {Allan}, {Allasia},
  {Altavilla}, {{\'A}lvarez}, {Alves}, {Anderson}, {Andrei}, {Anglada Varela},
  {Antiche}, {Antoja}, {Ant{\'o}n}, {Arcay}, {Atzei}, {Ayache}, {Bach},
  {Baker}, {Balaguer-N{\'u}{\~n}ez}, {Barache}, {Barata}, {Barbier}, {Barblan},
  {Baroni}, {Barrado y Navascu{\'e}s}, {Barros}, {Barstow}, {Becciani},
  {Bellazzini}, {Bellei}, {Bello Garc{\'\i}a}, {Belokurov}, {Bendjoya},
  {Berihuete}, {Bianchi}, {Bienaym{\'e}}, {Billebaud}, {Blagorodnova},
  {Blanco-Cuaresma}, {Boch}, {Bombrun}, {Borrachero}, {Bouquillon}, {Bourda},
  {Bouy}, {Bragaglia}, {Breddels}, {Brouillet}, {Br{\"u}semeister},
  {Bucciarelli}, {Budnik}, {Burgess}, {Burgon}, {Burlacu}, {Busonero}, {Buzzi},
  {Caffau}, {Cambras}, {Campbell}, {Cancelliere}, {Cantat-Gaudin}, {Carlucci},
  {Carrasco}, {Castellani}, {Charlot}, {Charnas}, {Charvet}, {Chassat},
  {Chiavassa}, {Clotet}, {Cocozza}, {Collins}, {Collins}, {Costigan}, {Crifo},
  {Cross}, {Crosta}, {Crowley}, {Dafonte}, {Damerdji}, {Dapergolas}, {David},
  {David}, {De Cat}, {de Felice}, {de Laverny}, {De Luise}, {De March}, {de
  Martino}, {de Souza}, {Debosscher}, {del Pozo}, {Delbo}, {Delgado},
  {Delgado}, {di Marco}, {Di Matteo}, {Diakite}, {Distefano}, {Dolding}, {Dos
  Anjos}, {Drazinos}, {Dur{\'a}n}, {Dzigan}, {Ecale}, {Edvardsson}, {Enke},
  {Erdmann}, {Escolar}, {Espina}, {Evans}, {Eynard Bontemps}, {Fabre},
  {Fabrizio}, {Faigler}, {Falc{\~a}o}, {Farr{\`a}s Casas}, {Faye}, {Federici},
  {Fedorets}, {Fern{\'a}ndez-Hern{\'a}ndez}, {Fernique}, {Fienga}, {Figueras},
  {Filippi}, {Findeisen}, {Fonti}, {Fouesneau}, {Fraile}, {Fraser}, {Fuchs},
  {Furnell}, {Gai}, {Galleti}, {Galluccio}, {Garabato}, {Garc{\'\i}a-Sedano},
  {Gar{\'e}}, {Garofalo}, {Garralda}, {Gavras}, {Gerssen}, {Geyer}, {Gilmore},
  {Girona}, {Giuffrida}, {Gomes}, {Gonz{\'a}lez-Marcos},
  {Gonz{\'a}lez-N{\'u}{\~n}ez}, {Gonz{\'a}lez-Vidal}, {Granvik}, {Guerrier},
  {Guillout}, {Guiraud}, {G{\'u}rpide}, {Guti{\'e}rrez-S{\'a}nchez}, {Guy},
  {Haigron}, {Hatzidimitriou}, {Haywood}, {Heiter}, {Helmi}, {Hobbs},
  {Hofmann}, {Holl}, {Holland }, {Hunt}, {Hypki}, {Icardi}, {Irwin}, {Jevardat
  de Fombelle}, {Jofr{\'e}}, {Jonker}, {Jorissen}, {Julbe}, {Karampelas},
  {Kochoska}, {Kohley}, {Kolenberg}, {Kontizas}, {Koposov}, {Kordopatis},
  {Koubsky}, {Kowalczyk}, {Krone-Martins}, {Kudryashova}, {Kull}, {Bachchan},
  {Lacoste-Seris}, {Lanza}, {Lavigne}, {Le Poncin-Lafitte}, {Lebreton},
  {Lebzelter}, {Leccia}, {Leclerc}, {Lecoeur-Taibi}, {Lemaitre}, {Lenhardt},
  {Leroux}, {Liao}, {Licata}, {Lindstr{\o}m}, {Lister}, {Livanou}, {Lobel},
  {L{\"o}ffler}, {L{\'o}pez}, {Lopez-Lozano}, {Lorenz}, {Loureiro},
  {MacDonald}, {Magalh{\~a}es Fernandes}, {Managau}, {Mann}, {Mantelet},
  {Marchal}, {Marchant}, {Marconi}, {Marie}, {Marinoni}, {Marrese},
  {Marschalk{\'o}}, {Marshall}, {Mart{\'\i}n-Fleitas}, {Martino}, {Mary},
  {Matijevi{\v{c}}}, {Mazeh}, {McMillan}, {Messina}, {Mestre}, {Michalik},
  {Millar}, {Miranda}, {Molina}, {Molinaro}, {Molinaro}, {Moln{\'a}r},
  {Moniez}, {Montegriffo}, {Monteiro}, {Mor}, {Mora}, {Morbidelli}, {Morel},
  {Morgenthaler}, {Morley}, {Morris}, {Mulone}, {Muraveva}, {Musella},
  {Narbonne}, {Nelemans}, {Nicastro}, {Noval}, {Ord{\'e}novic},
  {Ordieres-Mer{\'e}}, {Osborne}, {Pagani}, {Pagano}, {Pailler}, {Palacin},
  {Palaversa}, {Parsons}, {Paulsen}, {Pecoraro}, {Pedrosa}, {Pentik{\"a}inen},
  {Pereira}, {Pichon}, {Piersimoni}, {Pineau}, {Plachy}, {Plum}, {Poujoulet},
  {Pr{\v{s}}a}, {Pulone}, {Ragaini}, {Rago}, {Rambaux}, {Ramos-Lerate},
  {Ranalli}, {Rauw}, {Read}, {Regibo}, {Renk}, {Reyl{\'e}}, {Ribeiro},
  {Rimoldini}, {Ripepi}, {Riva}, {Rixon}, {Roelens}, {Romero-G{\'o}mez},
  {Rowell}, {Royer}, {Rudolph}, {Ruiz-Dern}, {Sadowski}, {Sagrist{\`a}
  Sell{\'e}s}, {Sahlmann}, {Salgado}, {Salguero}, {Sarasso}, {Savietto},
  {Schnorhk}, {Schultheis}, {Sciacca}, {Segol}, {Segovia}, {Segransan},
  {Serpell}, {Shih}, {Smareglia}, {Smart}, {Smith}, {Solano}, {Solitro},
  {Sordo}, {Soria Nieto}, {Souchay}, {Spagna}, {Spoto}, {Stampa}, {Steele},
  {Steidelm{\"u}ller}, {Stephenson}, {Stoev}, {Suess}, {S{\"u}veges}, {Surdej},
  {Szabados}, {Szegedi-Elek}, {Tapiador}, {Taris}, {Tauran}, {Taylor},
  {Teixeira}, {Terrett}, {Tingley}, {Trager}, {Turon}, {Ulla}, {Utrilla},
  {Valentini}, {van Elteren}, {Van Hemelryck}, {van Leeuwen}, {Varadi},
  {Vecchiato}, {Veljanoski}, {Via}, {Vicente}, {Vogt}, {Voss}, {Votruba},
  {Voutsinas}, {Walmsley}, {Weiler}, {Weingrill}, {Werner}, {Wevers},
  {Whitehead}, {Wyrzykowski}, {Yoldas}, {{\v{Z}}erjal}, {Zucker}, {Zurbach},
  {Zwitter}, {Alecu}, {Allen}, {Allende Prieto}, {Amorim},
  {Anglada-Escud{\'e}}, {Arsenijevic}, {Azaz}, {Balm}, {Beck}, {Bernstein},
  {Bigot}, {Bijaoui}, {Blasco}, {Bonfigli}, {Bono}, {Boudreault}, {Bressan},
  {Brown}, {Brunet}, {Bunclark}, {Buonanno}, {Butkevich}, {Carret}, {Carrion},
  {Chemin}, {Ch{\'e}reau}, {Corcione}, {Darmigny}, {de Boer}, {de Teodoro}, {de
  Zeeuw}, {Delle Luche}, {Domingues}, {Dubath}, {Fodor}, {Fr{\'e}zouls},
  {Fries}, {Fustes}, {Fyfe}, {Gallardo}, {Gallegos}, {Gardiol}, {Gebran},
  {Gomboc}, {G{\'o}mez}, {Grux}, {Gueguen}, {Heyrovsky}, {Hoar}, {Iannicola},
  {Isasi Parache}, {Janotto}, {Joliet}, {Jonckheere}, {Keil}, {Kim},
  {Klagyivik}, {Klar}, {Knude}, {Kochukhov}, {Kolka}, {Kos}, {Kutka}, {Lainey},
  {LeBouquin}, {Liu}, {Loreggia}, {Makarov}, {Marseille}, {Martayan},
  {Martinez-Rubi}, {Massart}, {Meynadier}, {Mignot}, {Munari}, {Nguyen},
  {Nordlander}, {Ocvirk}, {O'Flaherty}, {Olias Sanz}, {Ortiz}, {Osorio},
  {Oszkiewicz}, {Ouzounis}, {Palmer}, {Park}, {Pasquato}, {Peltzer}, {Peralta},
  {P{\'e}turaud}, {Pieniluoma}, {Pigozzi}, {Poels}, {Prat}, {Prod'homme},
  {Raison}, {Rebordao}, {Risquez}, {Rocca-Volmerange}, {Rosen}, {Ruiz-Fuertes},
  {Russo}, {Sembay}, {Serraller Vizcaino}, {Short}, {Siebert}, {Silva},
  {Sinachopoulos}, {Slezak}, {Soffel}, {Sosnowska}, {Strai{\v{z}}ys}, {ter
  Linden}, {Terrell}, {Theil}, {Tiede}, {Troisi}, {Tsalmantza}, {Tur},
  {Vaccari}, {Vachier}, {Valles}, {Van Hamme}, {Veltz}, {Virtanen}, {Wallut},
  {Wichmann}, {Wilkinson}, {Ziaeepour}, \& {Zschocke}}]{2016A&A...595A...1G}
{Gaia Collaboration}, {Prusti}, T., {de Bruijne}, J.~H.~J., {et~al.} 2016,
  \aap, 595, A1

\bibitem[{{Garrison}(1967)}]{1967PASP...79..433G}
{Garrison}, R.~F. 1967, \pasp, 79, 433

\bibitem[{{Goicoechea} {et~al.}(2006){Goicoechea}, {Pety}, {Gerin}, {Teyssier},
  {Roueff}, {Hily-Blant}, \& {Baek}}]{2006A&A...456..565G}
{Goicoechea}, J.~R., {Pety}, J., {Gerin}, M., {et~al.} 2006, \aap, 456, 565

\bibitem[{{Gonz{\'a}lez Hern{\'a}ndez} {et~al.}(2008){Gonz{\'a}lez
  Hern{\'a}ndez}, {Caballero}, {Rebolo}, {B{\'e}jar}, {Barrado Y
  Navascu{\'e}s}, {Mart{\'{\i}}n}, \& {Zapatero Osorio}}]{2008A&A...490.1135G}
{Gonz{\'a}lez Hern{\'a}ndez}, J.~I., {Caballero}, J.~A., {Rebolo}, R., {et~al.}
  2008, \aap, 490, 1135

\bibitem[{{Greenstein} \& {Keenan}(1958)}]{1958ApJ...127..172G}
{Greenstein}, J.~L. \& {Keenan}, P.~C. 1958, \apj, 127, 172

\bibitem[{{Guetter}(1981)}]{1981AJ.....86.1057G}
{Guetter}, H.~H. 1981, \aj, 86, 1057

\bibitem[{{Habart} {et~al.}(2005){Habart}, {Abergel}, {Walmsley}, {Teyssier},
  \& {Pety}}]{2005A&A...437..177H}
{Habart}, E., {Abergel}, A., {Walmsley}, C.~M., {Teyssier}, D., \& {Pety}, J.
  2005, \aap, 437, 177

\bibitem[{{Haro} \& {Moreno}(1953)}]{1953BOTT....1g..11H}
{Haro}, G. \& {Moreno}, A. 1953, Boletin de los Observatorios Tonantzintla y
  Tacubaya, 1, 11

\bibitem[{{Herbig}(1962)}]{1962ApJ...135..736H}
{Herbig}, G.~H. 1962, \apj, 135, 736

\bibitem[{{Hern{\'a}ndez} {et~al.}(2005){Hern{\'a}ndez}, {Calvet}, {Hartmann},
  {Brice{\~n}o}, {Sicilia-Aguilar}, \& {Berlind}}]{2005AJ....129..856H}
{Hern{\'a}ndez}, J., {Calvet}, N., {Hartmann}, L., {et~al.} 2005, \aj, 129, 856

\bibitem[{{Hern{\'a}ndez} {et~al.}(2014){Hern{\'a}ndez}, {Calvet}, {Perez},
  {Brice{\~n}o}, {Olguin}, {Contreras}, {Hartmann}, {Allen}, {Espaillat}, \&
  {Hernan}}]{2014ApJ...794...36H}
{Hern{\'a}ndez}, J., {Calvet}, N., {Perez}, A., {et~al.} 2014, \apj, 794, 36

\bibitem[{{Hern{\'a}ndez} {et~al.}(2007){Hern{\'a}ndez}, {Hartmann}, {Megeath},
  {Gutermuth}, {Muzerolle}, {Calvet}, {Vivas}, {Brice{\~n}o}, {Allen},
  {Stauffer}, {Young}, \& {Fazio}}]{2007ApJ...662.1067H}
{Hern{\'a}ndez}, J., {Hartmann}, L., {Megeath}, T., {et~al.} 2007, \apj, 662,
  1067

\bibitem[{{Houk} \& {Swift}(1999)}]{1999mctd.book.....H}
{Houk}, N. \& {Swift}, C. 1999, {Michigan catalogue of two-dimensional spectral
  types for the HD Stars; vol. 5} (Department of Astronomy, University of
  Michigan)

\bibitem[{{Jeffries} {et~al.}(2006){Jeffries}, {Maxted}, {Oliveira}, \&
  {Naylor}}]{2006MNRAS.371L...6J}
{Jeffries}, R.~D., {Maxted}, P.~F.~L., {Oliveira}, J.~M., \& {Naylor}, T. 2006,
  \mnras, 371, L6

\bibitem[{{Johnson} \& {Mitchell}(1958)}]{1958ApJ...128...31J}
{Johnson}, H.~L. \& {Mitchell}, R.~I. 1958, \apj, 128, 31

\bibitem[{{Kenyon} {et~al.}(2005){Kenyon}, {Jeffries}, {Naylor}, {Oliveira}, \&
  {Maxted}}]{2005MNRAS.356...89K}
{Kenyon}, M.~J., {Jeffries}, R.~D., {Naylor}, T., {Oliveira}, J.~M., \&
  {Maxted}, P.~F.~L. 2005, \mnras, 356, 89

\bibitem[{{Kirkpatrick} {et~al.}(1991){Kirkpatrick}, {Henry}, \&
  {McCarthy}}]{1991ApJS...77..417K}
{Kirkpatrick}, J.~D., {Henry}, T.~J., \& {McCarthy}, Jr., D.~W. 1991, \apjs,
  77, 417

\bibitem[{{Koenig} {et~al.}(2015){Koenig}, {Hillenbrand}, {Padgett}, \&
  {DeFelippis}}]{2015AJ....150..100K}
{Koenig}, X., {Hillenbrand}, L.~A., {Padgett}, D.~L., \& {DeFelippis}, D. 2015,
  \aj, 150, 100

\bibitem[{{Kounkel} {et~al.}(2018){Kounkel}, {Covey}, {Su{\'a}rez},
  {Rom{\'a}n-Z{\'u}{\~n}iga}, {Hernandez}, {Stassun}, {Jaehnig}, {Feigelson},
  {Pe{\~n}a Ram{\'\i}rez}, \& {Roman-Lopes}}]{2018AJ....156...84K}
{Kounkel}, M., {Covey}, K., {Su{\'a}rez}, G., {et~al.} 2018, \aj, 156, 84

\bibitem[{{Kroupa}(2001)}]{2001MNRAS.322..231K}
{Kroupa}, P. 2001, \mnras, 322, 231

\bibitem[{{Lada} {et~al.}(2006){Lada}, {Muench}, {Luhman}, {Allen}, {Hartmann},
  {Megeath}, {Myers}, {Fazio}, {Wood}, {Muzerolle}, {Rieke}, {Siegler}, \&
  {Young}}]{2006AJ....131.1574L}
{Lada}, C.~J., {Muench}, A.~A., {Luhman}, K.~L., {et~al.} 2006, \aj, 131, 1574

\bibitem[{{Lee} {et~al.}(1994){Lee}, {Martin}, \&
  {Mathieu}}]{1994AJ....108.1445L}
{Lee}, C.~W., {Martin}, E.~L., \& {Mathieu}, R.~D. 1994, \aj, 108, 1445

\bibitem[{{L{\'e}pine} {et~al.}(2013){L{\'e}pine}, {Hilton}, {Mann}, {Wilde},
  {Rojas-Ayala}, {Cruz}, \& {Gaidos}}]{2013AJ....145..102L}
{L{\'e}pine}, S., {Hilton}, E.~J., {Mann}, A.~W., {et~al.} 2013, \aj, 145, 102

\bibitem[{{L{\'o}pez-Mart{\'\i}n} {et~al.}(2001){L{\'o}pez-Mart{\'\i}n},
  {Raga}, {L{\'o}pez}, \& {Meaburn}}]{2001A&A...371.1118L}
{L{\'o}pez-Mart{\'\i}n}, L., {Raga}, A.~C., {L{\'o}pez}, J.~A., \& {Meaburn},
  J. 2001, \aap, 371, 1118

\bibitem[{{L{\'o}pez-Santiago} \& {Caballero}(2008)}]{2008A&A...491..961L}
{L{\'o}pez-Santiago}, J. \& {Caballero}, J.~A. 2008, \aap, 491, 961

\bibitem[{{Luhman} {et~al.}(2008){Luhman}, {Hern{\'a}ndez}, {Downes},
  {Hartmann}, \& {Brice{\~n}o}}]{2008ApJ...688..362L}
{Luhman}, K.~L., {Hern{\'a}ndez}, J., {Downes}, J.~J., {Hartmann}, L., \&
  {Brice{\~n}o}, C. 2008, \apj, 688, 362

\bibitem[{Lyng{\aa}(1982)}]{1982A&A...109..213L}
Lyng{\aa}, G. 1982, \aap, 109, 213

\bibitem[{{Ma{\'\i}z Apell{\'a}niz} {et~al.}(2018){Ma{\'\i}z Apell{\'a}niz},
  {Barb{\'a}}, {Sim{\'o}n-D{\'\i}az}, {Sota}, {Trigueros P{\'a}ez},
  {Caballero}, \& {Alfaro}}]{2018A&A...615A.161M}
{Ma{\'\i}z Apell{\'a}niz}, J., {Barb{\'a}}, R.~H., {Sim{\'o}n-D{\'\i}az}, S.,
  {et~al.} 2018, \aap, 615, A161

\bibitem[{{Manara} {et~al.}(2013){Manara}, {Testi}, {Rigliaco}, {Alcal{\'a}},
  {Natta}, {Stelzer}, {Biazzo}, {Covino}, {Covino}, {Cupani}, {D\'Elia}, \&
  {Randich}}]{2013A&A...551A.107M}
{Manara}, C.~F., {Testi}, L., {Rigliaco}, E., {et~al.} 2013, \aap, 551, A107

\bibitem[{{Mart{\'\i}n} {et~al.}(2018){Mart{\'\i}n}, {Lodieu}, {Pavlenko}, \&
  {B{\'e}jar}}]{2018ApJ...856...40M}
{Mart{\'\i}n}, E.~L., {Lodieu}, N., {Pavlenko}, Y., \& {B{\'e}jar}, V. J.~S.
  2018, \apj, 856, 40

\bibitem[{{Mauc{\'o}} {et~al.}(2016){Mauc{\'o}}, {Hern{\'a}ndez}, {Calvet},
  {Ballesteros-Paredes}, {Brice{\~n}o}, {McClure}, {D'Alessio}, {Anderson}, \&
  {Ali}}]{2016ApJ...829...38M}
{Mauc{\'o}}, K., {Hern{\'a}ndez}, J., {Calvet}, N., {et~al.} 2016, \apj, 829,
  38

\bibitem[{{Maxted} {et~al.}(2008){Maxted}, {Jeffries}, {Oliveira}, {Naylor}, \&
  {Jackson}}]{2008MNRAS.385.2210M}
{Maxted}, P.~F.~L., {Jeffries}, R.~D., {Oliveira}, J.~M., {Naylor}, T., \&
  {Jackson}, R.~J. 2008, \mnras, 385, 2210

\bibitem[{{Metodieva} {et~al.}(2015){Metodieva}, {Antonova}, {Golev},
  {Dimitrov}, {Garc{\'{\i}}a-{\'A}lvarez}, \& {Doyle}}]{2015MNRAS.446.3878M}
{Metodieva}, Y., {Antonova}, A., {Golev}, V., {et~al.} 2015, \mnras, 446, 3878

\bibitem[{{Mezger} {et~al.}(1988){Mezger}, {Chini}, {Kreysa}, {Wink}, \&
  {Salter}}]{1988A&A...191...44M}
{Mezger}, P.~G., {Chini}, R., {Kreysa}, E., {Wink}, J.~E., \& {Salter}, C.~J.
  1988, \aap, 191, 44

\bibitem[{{Montes} {et~al.}(2001{\natexlab{a}}){Montes}, {L{\'o}pez-Santiago},
  {Fern{\'a}ndez-Figueroa}, \& {G{\'a}lvez}}]{2001A&A...379..976M}
{Montes}, D., {L{\'o}pez-Santiago}, J., {Fern{\'a}ndez-Figueroa}, M.~J., \&
  {G{\'a}lvez}, M.~C. 2001{\natexlab{a}}, \aap, 379, 976

\bibitem[{{Montes} {et~al.}(2001{\natexlab{b}}){Montes}, {L{\'o}pez-Santiago},
  {G{\'a}lvez}, {Fern{\'a}ndez-Figueroa}, {De Castro}, \&
  {Cornide}}]{2001MNRAS.328...45M}
{Montes}, D., {L{\'o}pez-Santiago}, J., {G{\'a}lvez}, M.~C., {et~al.}
  2001{\natexlab{b}}, \mnras, 328, 45

\bibitem[{{Muench} {et~al.}(2001){Muench}, {Alves}, {Lada}, \&
  {Lada}}]{2001ApJ...558L..51M}
{Muench}, A.~A., {Alves}, J., {Lada}, C.~J., \& {Lada}, E.~A. 2001, \apj, 558,
  L51

\bibitem[{{Nesterov} {et~al.}(1995){Nesterov}, {Kuzmin}, {Ashimbaeva},
  {Volchkov}, {R{\"o}ser}, \& {Bastian}}]{1995A&AS..110..367N}
{Nesterov}, V.~V., {Kuzmin}, A.~V., {Ashimbaeva}, N.~T., {et~al.} 1995, \aaps,
  110

\bibitem[{{Neuh\"auser} {et~al.}(1997){Neuh\"auser}, {Torres}, {Sterzik}, \&
  {Randich}}]{1997A&A...325..647N}
{Neuh\"auser}, R., {Torres}, G., {Sterzik}, M.~F., \& {Randich}, S. 1997, \aap,
  325, 647

\bibitem[{{Newton} {et~al.}(2014){Newton}, {Charbonneau}, {Irwin},
  {Berta-Thompson}, {Rojas-Ayala}, {Covey}, \& {Lloyd}}]{2014AJ....147...20N}
{Newton}, E.~R., {Charbonneau}, D., {Irwin}, J., {et~al.} 2014, \aj, 147, 20

\bibitem[{{Oliveira} \& {van Loon}(2004)}]{2004A&A...418..663O}
{Oliveira}, J.~M. \& {van Loon}, J.~T. 2004, \aap, 418, 663

\bibitem[{{Pallavicini} {et~al.}(1981){Pallavicini}, {Golub}, {Rosner},
  {Vaiana}, {Ayres}, \& {Linsky}}]{1981ApJ...248..279P}
{Pallavicini}, R., {Golub}, L., {Rosner}, R., {et~al.} 1981, \apj, 248, 279

\bibitem[{{Pe{\~n}a Ram{\'\i}rez} {et~al.}(2012){Pe{\~n}a Ram{\'\i}rez},
  {B{\'e}jar}, {Zapatero Osorio}, {Petr-Gotzens}, \&
  {Mart{\'\i}n}}]{2012ApJ...754...30P}
{Pe{\~n}a Ram{\'\i}rez}, K., {B{\'e}jar}, V.~J.~S., {Zapatero Osorio}, M.~R.,
  {Petr-Gotzens}, M.~G., \& {Mart{\'\i}n}, E.~L. 2012, \apj, 754, 30

\bibitem[{{Pecaut} \& {Mamajek}(2013)}]{2013ApJS..208....9P}
{Pecaut}, M.~J. \& {Mamajek}, E.~E. 2013, \apjs, 208, 9

\bibitem[{{P{\'e}rez-Blanco} {et~al.}(2018){P{\'e}rez-Blanco}, {Mauc{\'o}},
  {Hern{\'a}ndez}, {Calvet}, {Espaillat}, {McClure}, {Brice{\~n}o}, {Robinson},
  {Feldman}, {Villarreal}, \& {D'Alessio}}]{2018ApJ...867..116P}
{P{\'e}rez-Blanco}, A., {Mauc{\'o}}, K., {Hern{\'a}ndez}, J., {et~al.} 2018,
  \apj, 867, 116

\bibitem[{{P{\'e}ricaud} {et~al.}(2017){P{\'e}ricaud}, {Di Folco}, {Dutrey},
  {Guilloteau}, \& {Pi{\'e}tu}}]{2017A&A...600A..62P}
{P{\'e}ricaud}, J., {Di Folco}, E., {Dutrey}, A., {Guilloteau}, S., \&
  {Pi{\'e}tu}, V. 2017, \aap, 600, A62

\bibitem[{{Perryman} {et~al.}(1998){Perryman}, {Brown}, {Lebreton}, {Gomez},
  {Turon}, {Cayrel de Strobel}, {Mermilliod}, {Robichon}, {Kovalevsky}, \&
  {Crifo}}]{1998A&A...331...81P}
{Perryman}, M.~A.~C., {Brown}, A.~G.~A., {Lebreton}, Y., {et~al.} 1998, \aap,
  331, 81

\bibitem[{{Popper}(1980)}]{1980ARA&A..18..115P}
{Popper}, D.~M. 1980, \araa, 18, 115

\bibitem[{{Pound} {et~al.}(2003){Pound}, {Reipurth}, \&
  {Bally}}]{2003AJ....125.2108P}
{Pound}, M.~W., {Reipurth}, B., \& {Bally}, J. 2003, \aj, 125, 2108

\bibitem[{{Preibisch} {et~al.}(2005){Preibisch}, {Kim}, {Favata}, {Feigelson},
  {Flaccomio}, {Getman}, {Micela}, {Sciortino}, {Stassun}, {Stelzer}, \&
  {Zinnecker}}]{2005ApJS..160..401P}
{Preibisch}, T., {Kim}, Y.-C., {Favata}, F., {et~al.} 2005, \apjs, 160, 401

\bibitem[{{Reipurth} {et~al.}(1998){Reipurth}, {Bally}, {Fesen}, \&
  {Devine}}]{1998Natur.396..343R}
{Reipurth}, B., {Bally}, J., {Fesen}, R.~A., \& {Devine}, D. 1998, \nat, 396,
  343

\bibitem[{{Riaz} {et~al.}(2017){Riaz}, {Brice{\~n}o}, {Whelan}, \&
  {Heathcote}}]{2017ApJ...844...47R}
{Riaz}, B., {Brice{\~n}o}, C., {Whelan}, E.~T., \& {Heathcote}, S. 2017, \apj,
  844, 47

\bibitem[{{Riaz} {et~al.}(2006){Riaz}, {Gizis}, \&
  {Harvin}}]{2006AJ....132..866R}
{Riaz}, B., {Gizis}, J.~E., \& {Harvin}, J. 2006, \aj, 132, 866

\bibitem[{{Riaz} {et~al.}(2019){Riaz}, {Machida}, \&
  {Stamatellos}}]{2019MNRAS.486.4114R}
{Riaz}, B., {Machida}, M.~N., \& {Stamatellos}, D. 2019, \mnras, 486, 4114

\bibitem[{{Sacco} {et~al.}(2008){Sacco}, {Franciosini}, {Randich}, \&
  {Pallavicini}}]{2008A&A...488..167S}
{Sacco}, G.~G., {Franciosini}, E., {Randich}, S., \& {Pallavicini}, R. 2008,
  \aap, 488, 167

\bibitem[{{Sacco} {et~al.}(2007){Sacco}, {Randich}, {Franciosini},
  {Pallavicini}, \& {Palla}}]{2007A&A...462L..23S}
{Sacco}, G.~G., {Randich}, S., {Franciosini}, E., {Pallavicini}, R., \&
  {Palla}, F. 2007, \aap, 462, L23

\bibitem[{{S{\'a}nchez-Bl{\'a}zquez} {et~al.}(2006){S{\'a}nchez-Bl{\'a}zquez},
  {Peletier}, {Jim{\'e}nez-Vicente}, {Cardiel}, {Cenarro},
  {Falc{\'o}n-Barroso}, {Gorgas}, {Selam}, \& {Vazdekis}}]{2006MNRAS.371..703S}
{S{\'a}nchez-Bl{\'a}zquez}, P., {Peletier}, R.~F., {Jim{\'e}nez-Vicente}, J.,
  {et~al.} 2006, \mnras, 371, 703

\bibitem[{{Schaefer} {et~al.}(2016){Schaefer}, {Hummel}, {Gies}, {Zavala},
  {Monnier}, {Walter}, {Turner}, {Baron}, {ten Brummelaar}, {Che},
  {Farrington}, {Kraus}, {Sturmann}, \& {Sturmann}}]{2016AJ....152..213S}
{Schaefer}, G.~H., {Hummel}, C.~A., {Gies}, D.~R., {et~al.} 2016, \aj, 152, 213

\bibitem[{{Sherry} {et~al.}(2004){Sherry}, {Walter}, \&
  {Wolk}}]{2004AJ....128.2316S}
{Sherry}, W.~H., {Walter}, F.~M., \& {Wolk}, S.~J. 2004, \aj, 128, 2316

\bibitem[{{Sherry} {et~al.}(2008){Sherry}, {Walter}, {Wolk}, \&
  {Adams}}]{2008AJ....135.1616S}
{Sherry}, W.~H., {Walter}, F.~M., {Wolk}, S.~J., \& {Adams}, N.~R. 2008, \aj,
  135, 1616

\bibitem[{{Sim{\'o}n-D{\'\i}az} {et~al.}(2011){Sim{\'o}n-D{\'\i}az},
  {Caballero}, \& {Lorenzo}}]{2011ApJ...742...55S}
{Sim{\'o}n-D{\'\i}az}, S., {Caballero}, J.~A., \& {Lorenzo}, J. 2011, \apj,
  742, 55

\bibitem[{{Sim{\'o}n-D{\'\i}az} {et~al.}(2015){Sim{\'o}n-D{\'\i}az},
  {Caballero}, {Lorenzo}, {Ma{\'\i}z Apell{\'a}niz}, {Schneider}, {Negueruela},
  {Barb{\'a}}, {Dorda}, {Marco}, {Montes}, {Pellerin}, {Sanchez- Bermudez},
  {S{\'o}dor}, \& {Sota}}]{2015ApJ...799..169S}
{Sim{\'o}n-D{\'\i}az}, S., {Caballero}, J.~A., {Lorenzo}, J., {et~al.} 2015,
  \apj, 799, 169

\bibitem[{{Skinner} {et~al.}(2008){Skinner}, {Sokal}, {Cohen}, {Gagn{\'e}},
  {Owocki}, \& {Townsend}}]{2008ApJ...683..796S}
{Skinner}, S.~L., {Sokal}, K.~R., {Cohen}, D.~H., {et~al.} 2008, ApJ, 683, 796

\bibitem[{{Skrutskie} {et~al.}(2006){Skrutskie}, {Cutri}, {Stiening},
  {Weinberg}, {Schneider}, {Carpenter}, {Beichman}, {Capps}, {Chester},
  {Elias}, {Huchra}, {Liebert}, {Lonsdale}, {Monet}, {Price}, {Seitzer},
  {Jarrett}, {Kirkpatrick}, {Gizis}, {Howard}, {Evans}, {Fowler}, {Fullmer},
  {Hurt}, {Light}, {Kopan}, {Marsh}, {McCallon}, {Tam}, {Van Dyk}, \&
  {Wheelock}}]{2006AJ....131.1163S}
{Skrutskie}, M.~F., {Cutri}, R.~M., {Stiening}, R., {et~al.} 2006, \aj, 131,
  1163

\bibitem[{{Soderblom} {et~al.}(2014){Soderblom}, {Hillenbrand}, {Jeffries},
  {Mamajek}, \& {Naylor}}]{2014prpl.conf..219S}
{Soderblom}, D.~R., {Hillenbrand}, L.~A., {Jeffries}, R.~D., {Mamajek}, E.~E.,
  \& {Naylor}, T. 2014, in Protostars and Planets VI, ed. H.~{Beuther}, R.~S.
  {Klessen}, C.~P. {Dullemond}, \& T.~{Henning}, 219

\bibitem[{{Soderblom} {et~al.}(1993){Soderblom}, {Jones}, {Balachandran},
  {Stauffer}, {Duncan}, {Fedele}, \& {Hudon}}]{1993AJ....106.1059S}
{Soderblom}, D.~R., {Jones}, B.~F., {Balachandran}, S., {et~al.} 1993, \aj,
  106, 1059

\bibitem[{{Stauffer} \& {Hartmann}(1986)}]{1986ApJS...61..531S}
{Stauffer}, J.~R. \& {Hartmann}, L.~W. 1986, \apjs, 61, 531

\bibitem[{{Stauffer} {et~al.}(1998){Stauffer}, {Schultz}, \&
  {Kirkpatrick}}]{1998ApJ...499L.199S}
{Stauffer}, J.~R., {Schultz}, G., \& {Kirkpatrick}, J.~D. 1998, \apj, 499, L199

\bibitem[{{Stelzer} \& {Neuh{\"a}user}(2001)}]{2001A&A...377..538S}
{Stelzer}, B. \& {Neuh{\"a}user}, R. 2001, \aap, 377, 538

\bibitem[{{Turner} {et~al.}(2008){Turner}, {ten Brummelaar}, {Roberts},
  {Mason}, {Hartkopf}, \& {Gies}}]{2008AJ....136..554T}
{Turner}, N.~H., {ten Brummelaar}, T.~A., {Roberts}, L.~C., {et~al.} 2008, \aj,
  136, 554

\bibitem[{{van Altena} {et~al.}(1995){van Altena}, {Lee}, \&
  {Hoffleit}}]{1995gcts.book.....V}
{van Altena}, W.~F., {Lee}, J.~T., \& {Hoffleit}, E.~D. 1995, {The general
  catalogue of trigonometric [stellar] parallaxes}

\bibitem[{{van Leeuwen}(2007)}]{2007A&A...474..653V}
{van Leeuwen}, F. 2007, \aap, 474, 653

\bibitem[{{Vandenberg} \& {Bridges}(1984)}]{1984ApJ...278..679V}
{Vandenberg}, D.~A. \& {Bridges}, T.~J. 1984, \apj, 278, 679

\bibitem[{{Vazdekis} {et~al.}(2010){Vazdekis}, {S{\'a}nchez-Bl{\'a}zquez},
  {Falc{\'o}n-Barroso}, {Cenarro}, {Beasley}, {Cardiel}, {Gorgas}, \&
  {Peletier}}]{2010MNRAS.404.1639V}
{Vazdekis}, A., {S{\'a}nchez-Bl{\'a}zquez}, P., {Falc{\'o}n-Barroso}, J.,
  {et~al.} 2010, \mnras, 404, 1639

\bibitem[{{Walter} {et~al.}(2008){Walter}, {Sherry}, {Wolk}, \&
  {Adams}}]{2008hsf1.book..732W}
{Walter}, F.~M., {Sherry}, W.~H., {Wolk}, S.~J., \& {Adams}, N.~R. 2008, {The
  {\ensuremath{\sigma}} Orionis Cluster}, ed. B.~{Reipurth}, Vol.~4, 732

\bibitem[{{Warren} \& {Hesser}(1977)}]{1977ApJS...34..115W}
{Warren}, W.~H., J. \& {Hesser}, J.~E. 1977, \apjs, 34, 115

\bibitem[{{Williams} {et~al.}(2013){Williams}, {Cieza}, {Andrews}, {Coulson},
  {Barger}, {Casey}, {Chen}, {Cowie}, {Koss}, {Lee}, \&
  {Sanders}}]{2013MNRAS.435.1671W}
{Williams}, J.~P., {Cieza}, L.~A., {Andrews}, S.~M., {et~al.} 2013, \mnras,
  435, 1671

\bibitem[{{Winter} {et~al.}(2018){Winter}, {Clarke}, {Rosotti}, {Ih},
  {Facchini}, \& {Haworth}}]{2018MNRAS.478.2700W}
{Winter}, A.~J., {Clarke}, C.~J., {Rosotti}, G., {et~al.} 2018, \mnras, 478,
  2700

\bibitem[{{Wolk}(1996)}]{1996PhDT........63W}
{Wolk}, S.~J. 1996, PhD thesis, Harvard--Smithsonian Center for Astrophysics
  Cambridge, Massachusetts, USA

\bibitem[{{Zacharias} {et~al.}(2005){Zacharias}, {Monet}, {Levine}, {Urban},
  {Gaume}, \& {Wycoff}}]{2005yCat.1297....0Z}
{Zacharias}, N., {Monet}, D.~G., {Levine}, S.~E., {et~al.} 2005, VizieR Online
  Data Catalog, I/297

\bibitem[{{Zapatero Osorio} {et~al.}(2000){Zapatero Osorio}, {B{\'e}jar},
  {Mart{\'\i}n}, {Rebolo}, {Barrado y Navascu{\'e}s}, {Bailer-Jones}, \&
  {Mundt}}]{2000Sci...290..103Z}
{Zapatero Osorio}, M.~R., {B{\'e}jar}, V.~J.~S., {Mart{\'\i}n}, E.~L., {et~al.}
  2000, Science, 290, 103

\bibitem[{{Zapatero Osorio} {et~al.}(2002){Zapatero Osorio}, {B{\'e}jar},
  {Pavlenko}, {Rebolo}, {Allende Prieto}, {Mart{\'\i}n}, \& {Garc{\'\i}a
  L{\'o}pez}}]{2002A&A...384..937Z}
{Zapatero Osorio}, M.~R., {B{\'e}jar}, V.~J.~S., {Pavlenko}, Y., {et~al.} 2002,
  \aap, 384, 937

\bibitem[{{Zapatero Osorio} {et~al.}(2017){Zapatero Osorio}, {B{\'e}jar}, \&
  {Pe{\~n}a Ram{\'\i}rez}}]{2017ApJ...842...65Z}
{Zapatero Osorio}, M.~R., {B{\'e}jar}, V.~J.~S., \& {Pe{\~n}a Ram{\'\i}rez}, K.
  2017, \apj, 842, 65

\end{thebibliography}

%========================================================================== 

\appendix

\section{Long tables and large-format figures}

\begin{figure*} 
\centering
\includegraphics[width=0.47\textwidth]{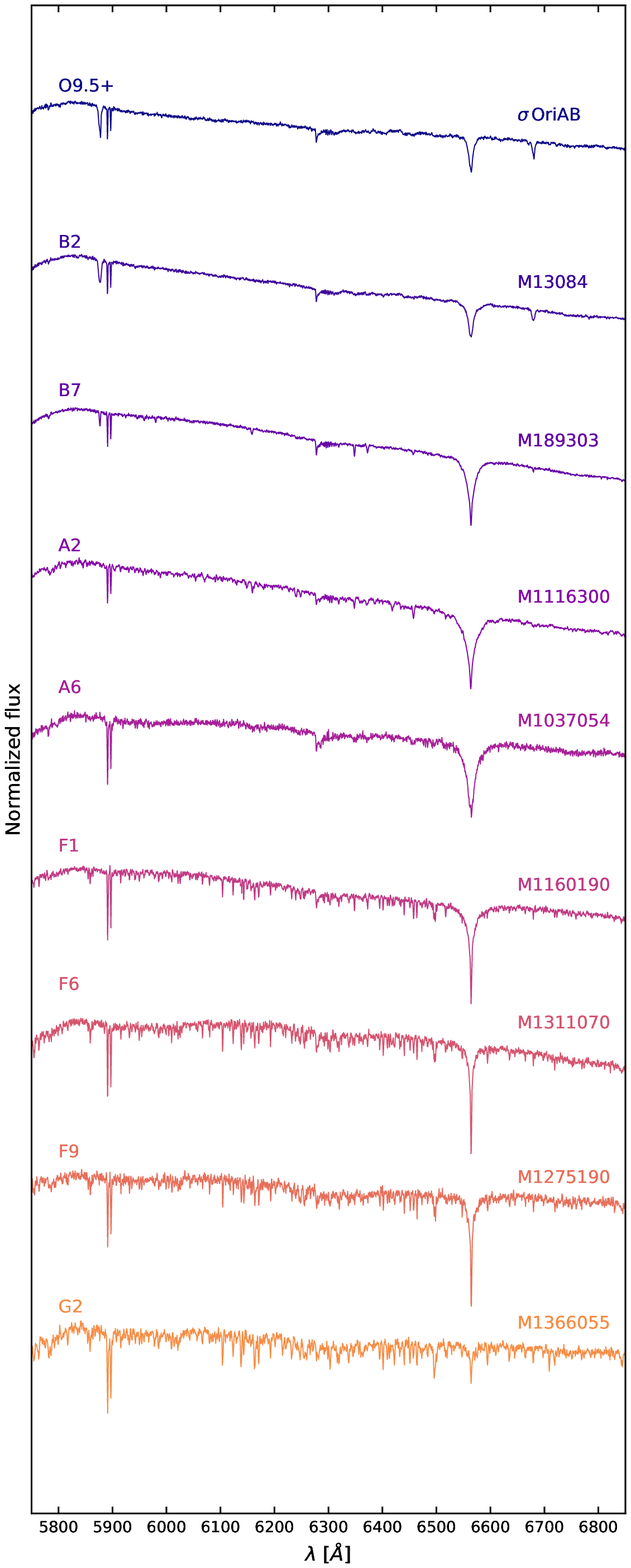}
\includegraphics[width=0.47\textwidth]{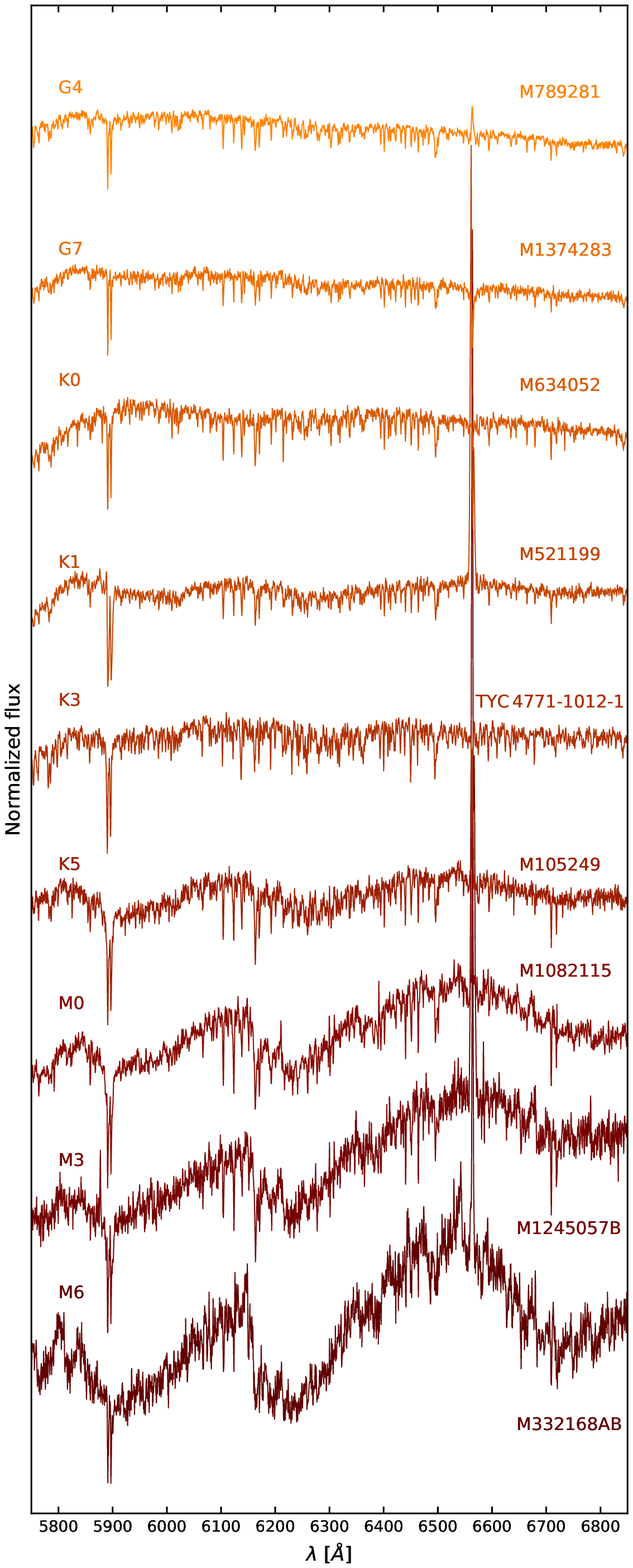}
\caption{R1200Y IDS/INT spectra of 18 stars (most from $\sigma$~Orionis) with spectral types from O9.5+\,V to M6 in the wavelength range used for spectral-type identification.}
\label{figure.ids_sample}
\end{figure*}

\begin{figure}
\centering
\includegraphics[width=0.49\textwidth]{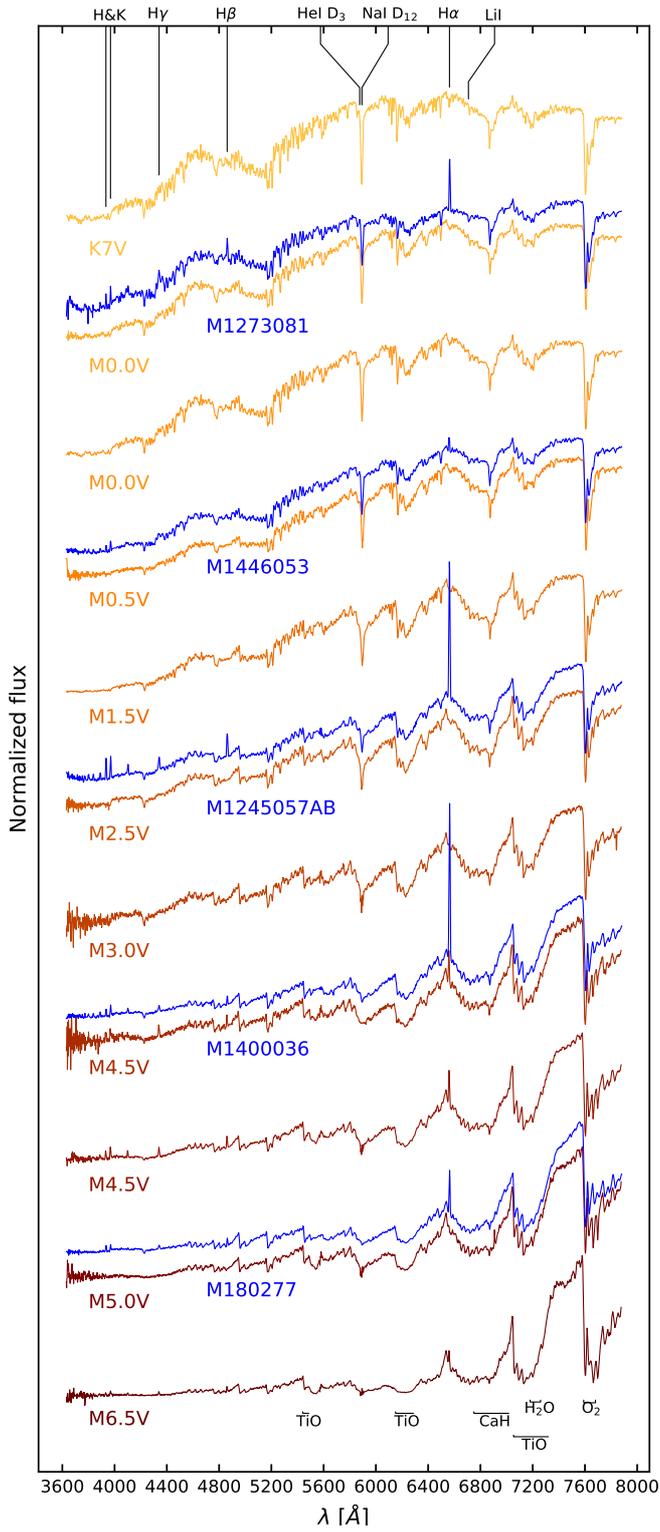}
\caption{OSIRIS/GTC normalised spectra of five of the $\sigma$~Orionis stars of different spectral types (in blue) on a grid of spectra of standard stars used for visual spectral-type identification of the OSIRIS spectra. The main spectral lines and most relevant absorption bands are also marked.} 
\label{figure.osiris_sample}
\end{figure}

{\bf Notes.} Full Tables~\ref{table.observations.int} to~\ref{table.results.spt} are available at the CDS.

%... ... ... ... ... ... ... ... ... ... ... ... ... ... ... ... ... ... ... ... ... ... ... ... ... ... ... ... ... ... ... ... ... ... ... ... ... ... ... ... ... ... ... ... ... ... ... ... ... ... ... ... 
\scriptsize{
\begin{longtable}{llccccccc}
\label{table.observations.int}\\ %
\caption[]{Observations of targets towards $\sigma$~Orionis with IDS/INT.}\\ %
   \hline
   \hline
   \noalign{\smallskip}
Mayrit	&Alternative	&$\alpha$	&$\delta$	&$\mu_{\alpha}\cos{\delta}$	&$\mu_{\delta}$	&$\varpi$	&Date of 		&$t_{\rm exp}$  \\ %
		&name		&(J2000)	&(J2000)	&[mas a$^{-1}$]			&[mas a$^{-1}$]	&[mas]	&observation	&[s]			 \\ %
\noalign{\smallskip}
    \hline
    \noalign{\smallskip}		
 \endfirsthead
\caption[]{Observations of targets towards $\sigma$~Orionis with IDS/INT (cont.).}\\ % 
  \hline
  \hline
  \noalign{\smallskip}		
Mayrit	&Alternative	&$\alpha$	&$\delta$	&$\mu_{\alpha}\cos{\delta}$	&$\mu_{\delta}$	&$\varpi$	&Date of 		&$t_{\rm exp}$  \\ %
		&name		&(J2000)	&(J2000)	&[mas a$^{-1}$]	&[mas a$^{-1}$]		&[mas]	&observation	&[s]			 \\ %
  \noalign{\smallskip}
  \hline
  \noalign{\smallskip}
  \endhead
  \noalign{\smallskip}
  \hline
  \endfoot

... &TYC 4770-1018-1 &05 36 57.15& --02 25 39.9&0.910 $\pm$ 0.086 &--0.437 $\pm$ 0.080 &0.554 $\pm$ 0.045 &23 Feb 2007 &180 \\
1456284 &TYC 4770-1261-1 &05 37 10.49& --02 30 07.1&0.748 $\pm$ 0.083 &--9.226 $\pm$ 0.078 &3.491 $\pm$ 0.040 &23 Feb 2007 &240 \\
 & &&& & & &27 Feb 2007 &450 \\
1415279AB &OriNTT 429 &05 37 11.62& --02 32 08.8&0.639 $\pm$ 0.064 &--0.773 $\pm$ 0.065 &2.725 $\pm$ 0.036 &24 Feb 2007 &300 \\
 & &&& & & &27 Feb 2007 &360 \\
1374283 &SO211394 &05 37 15.37& --02 30 53.4&--0.341 $\pm$ 0.092 &--1.029 $\pm$ 0.091 &2.351 $\pm$ 0.048 &23 Feb 2007 &300 \\
... &HD 294270 &05 37 18.82& --02 31 36.4&16.324 $\pm$ 0.096 &--28.327 $\pm$ 0.094 &2.808 $\pm$ 0.042 &24 Feb 2007 &120 \\
... &HD 294276 &05 37 20.68& --02 49 33.1&6.823 $\pm$ 0.058 &--65.963 $\pm$ 0.062 &7.160 $\pm$ 0.039 &22 Feb 2007 &120 \\
(1650224) &2M J05372885-0255555 &05 37 28.85& --02 55 55.6&0.207 $\pm$ 0.055 &1.337 $\pm$ 0.058 &0.485 $\pm$ 0.035 &23 Feb 2007 &180 \\
... &TYC 4770-1468-1 &05 37 29.88& --02 43 45.7&--2.705 $\pm$ 0.065 &--4.911 $\pm$ 0.063 &0.709 $\pm$ 0.038 &23 Feb 2007 &180 \\
1227243 &HD 294275 &05 37 31.87& --02 45 18.5&--2.188 $\pm$ 0.084 &--3.126 $\pm$ 0.085 &2.849 $\pm$ 0.052 &24 Feb 2007 &120 \\
 & &&& & & &24 Feb 2007 &90 \\
1116300 &HD 37333 &05 37 40.47& --02 26 36.8&--2.763 $\pm$ 0.090 &--3.259 $\pm$ 0.080 &2.847 $\pm$ 0.051 &24 Feb 2007 &30 \\
 & &&& & & &27 Feb 2007 &120 \\
... &TYC 4771-621-1 &05 37 41.79& --02 29 08.2&4.884 $\pm$ 0.070 &--15.200 $\pm$ 0.064 &2.370 $\pm$ 0.042 &24 Feb 2007 &120 \\
968292 &TYC 4771-962-1 &05 37 44.92& --02 29 57.3&--1.260 $\pm$ 0.082 &2.745 $\pm$ 0.071 &2.730 $\pm$ 0.050 &22 Feb 2007 &180 \\
 & &&& & & &27 Feb 2007 &240 \\
... &HD 294274 &05 37 45.36& --02 44 12.5&--10.279 $\pm$ 0.075 &--18.992 $\pm$ 0.075 &3.152 $\pm$ 0.045 &24 Feb 2007 &120 \\
(958292) &SO210868 &05 37 45.57& --02 29 58.5&--0.575 $\pm$ 0.065 &--3.517 $\pm$ 0.058 &0.874 $\pm$ 0.035 &23 Feb 2007 &180 \\
 & &&& & & &27 Feb 2007 &450 \\
797272 &[SWW2004] 125 &05 37 51.61& --02 35 25.7&1.552 $\pm$ 0.097 &--1.186 $\pm$ 0.083 &2.695 $\pm$ 0.049 &28 Feb 2007 &750 \\
789281 &2E 1454 &05 37 53.04& --02 33 34.4&1.767 $\pm$ 0.080 &--1.321 $\pm$ 0.071 &2.712 $\pm$ 0.038 &24 Feb 2007 &240 \\
 & &&& & & &24 Feb 2007 &300 \\
783254 &2E 1455 &05 37 54.40& --02 39 29.8&1.64 $\pm$ 0.12 &--0.42 $\pm$ 0.10 &2.590 $\pm$ 0.064 &22 Feb 2007 &120 \\
(882239) &[HHM2007] 244 &05 37 54.45& --02 43 37.8&0.305 $\pm$ 0.068 &--0.042 $\pm$ 0.069 &0.477 $\pm$ 0.039 &22 Feb 2007 &180 \\
... &HD 294277 &05 37 57.34& --02 53 17.7&5.476 $\pm$ 0.077 &--5.144 $\pm$ 0.075 &1.790 $\pm$ 0.044 &22 Feb 2007 &120 \\
(1596206) &2M J05375789-0259536 &05 37 57.90& --02 59 53.7&1.228 $\pm$ 0.038 &0.802 $\pm$ 0.040 &0.187 $\pm$ 0.026 &23 Feb 2007 &300 \\
 & &&& & & &27 Feb 2007 &600 \\
... &TYC 4771-720-1 &05 37 59.04& --02 41 00.5&--18.766 $\pm$ 0.077 &--18.469 $\pm$ 0.066 &2.944 $\pm$ 0.040 &24 Feb 2007 &120 \\
(717307) &[W96] 4771-0950 &05 38 06.50& --02 28 49.5&5.953 $\pm$ 0.068 &--12.623 $\pm$ 0.060 &1.892 $\pm$ 0.036 &24 Feb 2007 &240 \\
662301 &Kiso A-0904 67 &05 38 06.75& --02 30 22.7&1.035 $\pm$ 0.074 &--0.383 $\pm$ 0.074 &2.354 $\pm$ 0.042 &26 Feb 2007 &450 \\
 & &&& & & &28 Feb 2007 &600 \\
615296 &2E 1459 &05 38 07.84& --02 31 31.4&1.297 $\pm$ 0.065 &0.694 $\pm$ 0.067 &2.485 $\pm$ 0.038 &23 Feb 2007 &240 \\
(733222) &[HHM2007] 385 &05 38 11.75& --02 45 01.2&11.723 $\pm$ 0.060 &--19.198 $\pm$ 0.055 &0.360 $\pm$ 0.036 &22 Feb 2007 &240 \\
1285339 &HD 294268 &05 38 14.12& --02 15 59.8&2.008 $\pm$ 0.069 &--1.631 $\pm$ 0.066 &2.672 $\pm$ 0.051 &24 Feb 2007 &180 \\
(1064335) &TYC 4771-873-1 &05 38 14.43& --02 19 58.8&--2.375 $\pm$ 0.070 &--5.703 $\pm$ 0.059 &1.399 $\pm$ 0.042 &24 Feb 2007 &360 \\
... &2M J05381494-0219532$^{a}$ &05 38 14.95& --02 19 53.2&--2.48 $\pm$ 0.053 &--5.586 $\pm$ 0.044 &1.450 $\pm$ 0.032 &27 Feb 2007 &600 \\
(1564345) &[SE2004] 10 &05 38 17.29& --02 10 52.0&2.597 $\pm$ 0.064 &--5.652 $\pm$ 0.059 &0.611 $\pm$ 0.043 &24 Feb 2007 &300 \\
(377264) &IRAS 05358-0238 &05 38 19.76& --02 36 39.1&1.18 $\pm$ 0.47 &1.92 $\pm$ 0.40 &0.41 $\pm$ 0.31 &24 Feb 2007 &450 \\
 & &&& & & &27 Feb 2007 &1500 \\
 & &&& & & &28 Feb 2007 &60 \\
 & &&& & & &28 Feb 2007 &900 \\
(1343194) &2M J05382265-0257421 &05 38 22.65& --02 57 42.2&0.765 $\pm$ 0.036 &--0.525 $\pm$ 0.034 &0.155 $\pm$ 0.023 &23 Feb 2007 &300 \\
1449349 &V2730 Ori &05 38 26.57& --02 12 17.5&1.91 $\pm$ 0.12 &--1.260 $\pm$ 0.092 &2.855 $\pm$ 0.070 &28 Feb 2007 &750 \\
609206 &Haro 5-10 &05 38 27.26& --02 45 09.7&1.129 $\pm$ 0.064 &--0.675 $\pm$ 0.058 &2.516 $\pm$ 0.039 &28 Feb 2007 &600 \\
521210 &HD 294273 &05 38 27.53& --02 43 32.6&--2.057 $\pm$ 0.077 &0.851 $\pm$ 0.069 &2.367 $\pm$ 0.040 &22 Feb 2007 &180 \\
... &HD 294280 &05 38 28.49&--03 03 33.8&--3.712 $\pm$ 0.078 &2.237 $\pm$ 0.072 &1.254 $\pm$ 0.049 &22 Feb 2007 &120 \\
1207349 &Haro 5-9 &05 38 29.16& --02 16 15.8&0.345 $\pm$ 0.047 &0.769 $\pm$ 0.042 &2.412 $\pm$ 0.036 &26 Feb 2007 &450 \\
1275190 &2M J05383031-0256565 &05 38 30.31& --02 56 56.5&6.860 $\pm$ 0.057 &2.128 $\pm$ 0.056 &2.563 $\pm$ 0.037 &23 Feb 2007 &180 \\
1160190 &HD 294279 &05 38 31.38& --02 55 03.2&--2.424 $\pm$ 0.075 &0.552 $\pm$ 0.072 &2.228 $\pm$ 0.045 &24 Feb 2007 &240 \\
203283$^{e}$ &[W96] rJ053831-0235 &05 38 31.59& --02 35 15.0&--6.785 $\pm$ 2.791&--0.300 $\pm$ 2.835&...&25 Feb 2007 &600 \\
180277 &[W96] rJ053832-0235b &05 38 32.85& --02 35 39.3&0.546 $\pm$ 0.048 &--0.32 $\pm$ 0.047 &2.405 $\pm$ 0.028 &25 Feb 2007 &600 \\
521199 &TX Ori &05 38 33.69& --02 44 14.1&--1.29 $\pm$ 0.46 &--2.70 $\pm$ 0.39 &1.75 $\pm$ 0.26 &23 Feb 2007 &240 \\
165257 &[W96] rJ053833-0236 &05 38 34.06& --02 36 37.6&--0.33 $\pm$ 0.37 &1.11 $\pm$ 0.33 &2.60 $\pm$ 0.22 &28 Feb 2007 &600 \\
189303 &HD 294272 B &05 38 34.23& --02 34 16.0&0.81 $\pm$ 0.13 &--1.26 $\pm$ 0.11 &2.335 $\pm$ 0.071 &22 Feb 2007 &120 \\
168291AB &[HHM2007] 592 &05 38 34.32& --02 35 00.2&1.438 $\pm$ 0.044 &--0.157 $\pm$ 0.040 &2.405 $\pm$ 0.024 &25 Feb 2007 &450 \\
(459340) &StHa 50 &05 38 34.44& --02 28 47.6&0.679 $\pm$ 0.074 &--0.641 $\pm$ 0.072 &0.463 $\pm$ 0.047 &22 Feb 2007 &180 \\
 & &&& & & &27 Feb 2007 &450 \\
(258215) &[W96] pJ053834-0239 &05 38 34.79& --02 39 30.1&1.610 $\pm$ 0.062 &--0.234 $\pm$ 0.058 &0.084 $\pm$ 0.038 &22 Feb 2007 &240 \\
182305 &HD 294272 A &05 38 34.80& --02 34 15.8&1.52 $\pm$ 0.11 &--1.191 $\pm$ 0.098 &2.261 $\pm$ 0.062 &22 Feb 2007 &120 \\
(240322) &IDS 05335-0238 D &05 38 34.85& --02 32 52.2&--5.23 $\pm$ 0.23 &--9.76 $\pm$ 0.21 &1.90 $\pm$ 0.13 &22 Feb 2007 &180 \\
285331 &[W96] rJ053835-0231 &05 38 35.47& --02 31 51.7&1.459 $\pm$ 0.047 &--0.556 $\pm$ 0.043 &2.453 $\pm$ 0.028 &25 Feb 2007 &450 \\
344337AB &2E 1468 &05 38 35.87& --02 30 43.3&1.849 $\pm$ 0.057 &--1.824 $\pm$ 0.049 &2.713 $\pm$ 0.040 &25 Feb 2007 &450 \\
489196 &TY Ori &05 38 35.88& --02 43 51.2&1.447 $\pm$ 0.079 &--0.576 $\pm$ 0.069 &2.431 $\pm$ 0.041 &23 Feb 2007 &240 \\
208324 &HD 294271 &05 38 36.55& --02 33 12.8&1.44 $\pm$ 0.11 &--0.697 $\pm$ 0.096 &2.508 $\pm$ 0.060 &24 Feb 2007 &30 \\
105249 &[W96] rJ053838-0236 &05 38 38.23& --02 36 38.5&1.691 $\pm$ 0.041 &--1.134 $\pm$ 0.043 &2.456 $\pm$ 0.024 &24 Feb 2007 &360 \\
114305AB &[W96] 4771-1147 &05 38 38.49& --02 34 55.1&0.850 $\pm$ 0.063 &--0.220 $\pm$ 0.061 &2.420 $\pm$ 0.037 &24 Feb 2007 &240 \\
... &HD 294278 &05 38 38.76& --02 49 01.4&9.771 $\pm$ 0.066 &--2.581 $\pm$ 0.060 &2.225 $\pm$ 0.041 &23 Feb 2007 &180 \\
(1045356) &[SE2004] 30 &05 38 39.72& --02 18 37.7&1.818 $\pm$ 0.065 &--3.681 $\pm$ 0.064 &0.457 $\pm$ 0.042 &22 Feb 2007 &180 \\
1248183AB &[SWW2004] 145 &05 38 39.82& --02 56 46.2&1.00 $\pm$ 0.20 &--1.01 $\pm$ 0.17 &3.03 $\pm$ 0.11 &25 Feb 2007 &450 \\
348349 &Haro 5-13 &05 38 40.27& --02 30 18.6&0.897 $\pm$ 0.056 &--0.554 $\pm$ 0.051 &2.627 $\pm$ 0.035 &25 Feb 2007 &720 \\
97212 &[W96] rJ053841-0237 &05 38 41.29& --02 37 22.6&1.662 $\pm$ 0.047 &--0.478 $\pm$ 0.044 &2.429 $\pm$ 0.027 &26 Feb 2007 &300 \\
83207 &[W96] P053842-0237 &05 38 42.28& --02 37 14.8&1.198 $\pm$ 0.081 &--0.449 $\pm$ 0.077 &2.462 $\pm$ 0.046 &28 Feb 2007 &600 \\
156353 &[W96] rJ053843-0233 &05 38 43.56& --02 33 25.4&0.39 $\pm$ 0.13 &--1.04 $\pm$ 0.12 &2.149 $\pm$ 0.084 &26 Feb 2007 &450 \\
11238 &$\sigma$ Ori C &05 38 44.12& --02 36 06.3&0.07 $\pm$ 0.12 &--1.10 $\pm$ 0.12 &2.367 $\pm$ 0.088 &22 Feb 2007 &60 \\
 & &&& & & &24 Feb 2007 &30 \\
 & &&& & & &24 Feb 2007 &60 \\
260182 &[W96] 4771-1051 &05 38 44.23& --02 40 19.8&1.517 $\pm$ 0.039 &--0.989 $\pm$ 0.038 &2.495 $\pm$ 0.024 &25 Feb 2007 &450 \\
207358 &[W96] 4771-1055 &05 38 44.24& --02 32 33.7&1.470 $\pm$ 0.078 &--0.700 $\pm$ 0.068 &2.616 $\pm$ 0.046 &24 Feb 2007 &360 \\
$\sigma$ Ori AB$^{d}$ &$\sigma$ Ori Aa, Ab, B, IRS1AB &05 38 44.77& --02 36 00.3&4.6 $\pm$ 1.0&--0.4 $\pm$ 1.0&...&22 Feb 2007 &1 \\
 & &&& & & &22 Feb 2007 &4 \\
 & &&& & & &22 Feb 2007 &2 \\
 & &&& & & &24 Feb 2007 &3 \\
 & &&& & & &24 Feb 2007 &1 \\
 & &&& & & &24 Feb 2007 &1 \\
 & &&& & & &24 Feb 2007 &2 \\
 & &&& & & &24 Feb 2007 &3 \\
(123000) &[W96] pJ053844-0233 &05 38 44.81& --02 33 57.7&5.990 $\pm$ 0.066 &--2.808 $\pm$ 0.065 &0.760 $\pm$ 0.044 &22 Feb 2007 &180 \\
359179AB &Haro 5-14 &05 38 45.39& --02 41 59.4&2.33 $\pm$ 0.38 &0.48 $\pm$ 0.33 &4.709 $\pm$ 0.242 &28 Feb 2007 &600 \\
13084 &$\sigma$ Ori D &05 38 45.63& --02 35 58.9&1.42 $\pm$ 0.13 &--0.36 $\pm$ 0.11 &2.294 $\pm$ 0.079 &22 Feb 2007 &20 \\
 & &&& & & &23 Feb 2007 &5 \\
... &2M J05384652-0235479$^{a}$ &05 38 46.52& --02 35 48.1&22.51 $\pm$ 0.32 &--47.30 $\pm$ 0.20 &0.957 $\pm$ 0.210 &22 Feb 2007 &300 \\
 & &&& & & &22 Feb 2007 &300 \\
 & &&& & & &26 Feb 2007 &600 \\
 & &&& & & &26 Feb 2007 &600 \\
 & &&& & & &26 Feb 2007 &600 \\
 & &&& & & &26 Feb 2007 &600 \\
 & &&& & & &26 Feb 2007 &1500 \\
 & &&& & & &28 Feb 2007 &600 \\
 & &&& & & &28 Feb 2007 &600 \\
42062 &$\sigma$ Ori E &05 38 47.21& --02 35 40.5&1.28 $\pm$ 0.15 &--0.62 $\pm$ 0.14 &2.280 $\pm$ 0.096 &22 Feb 2007 &10 \\
 & &&& & & &22 Feb 2007 &30 \\ 
 & &&& & & &24 Feb 2007 &15 \\
 & &&& & & &24 Feb 2007 &10 \\
 & &&& & & &24 Feb 2007 &5 \\
53049 &$\sigma$ Ori Eb &05 38 47.46& --02 35 25.3&2.221 $\pm$ 0.076 &0.084 $\pm$ 0.071 &2.477 $\pm$ 0.051 &22 Feb 2007 &600 \\
 & &&& & & &28 Feb 2007 &60 \\
528005AB$^{e}$ &[W96] 4771-899 &05 38 48.04& --02 27 14.2&11.410 $\pm$ 2.267&--5.078 $\pm$ 2.267&...&24 Feb 2007 &450 \\
157155 &[W96] rJ053849-0238 &05 38 49.17& --02 38 22.3&2.91 $\pm$ 0.24 &--1.23 $\pm$ 0.22 &2.31 $\pm$ 0.14 &25 Feb 2007 &450 \\
332168AB &[SWW2004] 205 &05 38 49.22& --02 41 25.2&1.49 $\pm$ 0.12 &--0.76 $\pm$ 0.10 &2.413 $\pm$ 0.071 &26 Feb 2007 &300 \\
 & &&& & & &27 Feb 2007 &600 \\
653170 &RU Ori &05 38 52.01& --02 46 43.7&1.630 $\pm$ 0.063 &--1.011 $\pm$ 0.052 &2.471 $\pm$ 0.030 &25 Feb 2007 &600 \\
203039 &[W96] 4771-1049 &05 38 53.38& --02 33 23.1&--0.408 $\pm$ 0.071 &--0.236 $\pm$ 0.068 &2.597 $\pm$ 0.047 &23 Feb 2007 &240 \\
822170 &[W96] 4771-119 &05 38 54.11& --02 49 29.8&1.09 $\pm$ 0.10 &--0.552 $\pm$ 0.092 &2.371 $\pm$ 0.047 &24 Feb 2007 &300 \\
707162AB &[W96] rJ053859-0247 &05 38 59.10& --02 47 13.4&0.911 $\pm$ 0.099 &--0.078 $\pm$ 0.080 &2.189 $\pm$ 0.051 &25 Feb 2007 &450 \\
591158 &[W96] 4771-0026 &05 38 59.56& --02 45 08.2&--2.19 $\pm$ 0.10 &1.399 $\pm$ 0.087 &2.416 $\pm$ 0.046 &25 Feb 2007 &300 \\
 & &&& & & &27 Feb 2007 &600 \\
1082013 &Haro 5-16 &05 39 01.37& --02 18 27.5&1.752 $\pm$ 0.055 &--0.115 $\pm$ 0.053 &2.379 $\pm$ 0.036 &26 Feb 2007 &450 \\
... &[HHM2007] 846 &05 39 01.44& --02 53 43.2&0.033 $\pm$ 0.081 &--1.116 $\pm$ 0.075 &0.025 $\pm$ 0.047 &22 Feb 2007 &120 \\
306125AB &HD 37525AB &05 39 01.49& --02 38 56.4&1.18 $\pm$ 0.21 &--1.02 $\pm$ 0.17 &2.29 $\pm$ 0.13 &24 Feb 2007 &45 \\
 & &&& & & &24 Feb 2007 &30 \\
374056 &[W96] 4771-1075 &05 39 05.41& --02 32 30.4&1.691 $\pm$ 0.047 &--0.478 $\pm$ 0.045 &2.459 $\pm$ 0.031 &25 Feb 2007 &600 \\
397060 &Haro 5-19 &05 39 07.60& --02 32 39.2&1.924 $\pm$ 0.051 &--1.508 $\pm$ 0.048 &2.494 $\pm$ 0.034 &25 Feb 2007 &450 \\
1011159 &[SWW2004] 61 &05 39 08.53& --02 51 46.6&3.15 $\pm$ 0.26 &--1.63 $\pm$ 0.21 &2.33 $\pm$ 0.13 &28 Feb 2007 &600 \\
1288163 &HD 37545 &05 39 09.21& --02 56 34.7&1.036 $\pm$ 0.091 &--0.487 $\pm$ 0.086 &2.163 $\pm$ 0.063 &24 Feb 2007 &90 \\
497054 &Haro 5-20 &05 39 11.52& --02 31 06.6&1.873 $\pm$ 0.059 &--0.339 $\pm$ 0.057 &2.435 $\pm$ 0.040 &28 Feb 2007 &600 \\
403090 &[W96] 4771-1038 &05 39 11.63& --02 36 02.9&1.949 $\pm$ 0.056 &--0.906 $\pm$ 0.052 &2.338 $\pm$ 0.037 &26 Feb 2007 &360 \\
... &TYC 4771-1012-1 &05 39 13.96& --02 10 49.2&4.083 $\pm$ 0.069 &--8.914 $\pm$ 0.064 &0.961 $\pm$ 0.038 &24 Feb 2007 &120 \\
524060 &HD 37564 &05 39 15.06& --02 31 37.6&1.076 $\pm$ 0.094 &--0.363 $\pm$ 0.090 &2.452 $\pm$ 0.063 &24 Feb 2007 &90 \\
 & &&& & & &24 Feb 2007 &45 \\
... &[HHM2007] 961 &05 39 15.64& --02 29 56.9&--0.388 $\pm$ 0.082 &--40.631 $\pm$ 0.080 &7.132 $\pm$ 0.047 &24 Feb 2007 &120 \\
(945030) &[SE2004] 50 &05 39 16.59& --02 22 24.2&2.444 $\pm$ 0.053 &0.652 $\pm$ 0.051 &0.230 $\pm$ 0.035 &23 Feb 2007 &240 \\
 & &&& & & &27 Feb 2007 &450 \\
634052 &[W96] 4771-0598 &05 39 18.07& --02 29 28.5&2.072 $\pm$ 0.069 &--1.369 $\pm$ 0.069 &2.678 $\pm$ 0.045 &23 Feb 2007 &240 \\
596059 &Haro 5-21 &05 39 18.84& --02 30 53.2&1.75 $\pm$ 0.12 &--0.543 $\pm$ 0.075 &2.486 $\pm$ 0.057 &25 Feb 2007 &450 \\
... &TYC 4771-661-1 &05 39 20.46& --02 27 51.1&18.912 $\pm$ 0.061 &--17.227 $\pm$ 0.061 &4.035 $\pm$ 0.043 &24 Feb 2007 &180 \\
(735131) &[HHM2007] 1009 &05 39 21.75& --02 44 03.9&0.859 $\pm$ 0.052 &--2.706 $\pm$ 0.045 &0.502 $\pm$ 0.030 &23 Feb 2007 &300 \\
 & &&& & & &27 Feb 2007 &450 \\
622103 &BG Ori &05 39 25.20& --02 38 22.1&1.945 $\pm$ 0.061 &--0.315 $\pm$ 0.053 &2.607 $\pm$ 0.050 &25 Feb 2007 &450 \\
1403026 &Haro 5-22 &05 39 26.40& --02 15 03.5&0.908 $\pm$ 0.040 &0.475 $\pm$ 0.039 &2.491 $\pm$ 0.026 &28 Feb 2007 &600 \\
(861056) &[HHM2007] 1092 &05 39 32.34& --02 27 57.2&--0.109 $\pm$ 0.042 &0.512 $\pm$ 0.041 &0.667 $\pm$ 0.030 &23 Feb 2007 &300 \\
750107 &[W96] r053932-0239 &05 39 32.57& --02 39 44.0&1.961 $\pm$ 0.070 &--0.667 $\pm$ 0.062 &2.502 $\pm$ 0.056 &24 Feb 2007 &450 \\
863116 &HD 294300 &05 39 36.54& --02 42 17.2&2.78 $\pm$ 0.17 &--3.19 $\pm$ 0.13 &2.79 $\pm$ 0.12 &24 Feb 2007 &90 \\
(1165138) &[HHM2007] 1129 &05 39 36.86& --02 50 24.8&--0.491 $\pm$ 0.068 &0.833 $\pm$ 0.064 &0.612 $\pm$ 0.042 &22 Feb 2007 &180 \\
957055 &[SWW2004] 163 &05 39 37.30& --02 26 56.8&--1.226 $\pm$ 0.044 &--1.463 $\pm$ 0.042 &2.436 $\pm$ 0.028 &26 Feb 2007 &600 \\
... &Haro 5-28 &05 39 38.32& --02 33 05.4&--9.070 $\pm$ 0.041 &10.527 $\pm$ 0.040 &1.519 $\pm$ 0.030 &25 Feb 2007 &600 \\
1400036 &Haro 5-25 &05 39 39.38& --02 17 04.5&2.285 $\pm$ 0.058 &0.28 $\pm$ 0.056 &2.548 $\pm$ 0.035 &25 Feb 2007 &600 \\
871071 &Haro 5-27 &05 39 39.83& --02 31 21.9&1.833 $\pm$ 0.047 &--0.059 $\pm$ 0.043 &2.533 $\pm$ 0.035 &28 Feb 2007 &600 \\
931117 &RW Ori &05 39 39.99& --02 43 09.8&2.19 $\pm$ 0.13 &--0.72 $\pm$ 0.12 &2.610 $\pm$ 0.134 &23 Feb 2007 &240 \\
1233042 &RV Ori &05 39 40.17& --02 20 48.0&2.405 $\pm$ 0.040 &--0.167 $\pm$ 0.041 &2.513 $\pm$ 0.025 &26 Feb 2007 &300 \\
(1037054) &HD 294299 &05 39 40.57& --02 25 46.9&0.212 $\pm$ 0.098 &--0.22 $\pm$ 0.10 &1.301 $\pm$ 0.067 &24 Feb 2007 &300 \\
 & &&& & & &27 Feb 2007 &300 \\
1626148AB${^f}$ &Haro 5-33 &05 39 42.77& --02 58 53.8&2.0 $\pm$ 5.6 &2.6 $\pm$ 5.6&...&23 Feb 2007 &300 \\
... &TYC 4771-934-1 &05 39 43.06& --02 28 45.7&0.268 $\pm$ 0.067 &--8.250 $\pm$ 0.071 &1.118 $\pm$ 0.042 &24 Feb 2007 &120 \\
(936072) &[HHM2007] 1189 &05 39 44.11& --02 31 09.3&3.098 $\pm$ 0.051 &--0.129 $\pm$ 0.043 &0.404 $\pm$ 0.029 &23 Feb 2007 &300 \\
 & &&& & & &27 Feb 2007 &600 \\
960106AB &V1147 Ori &05 39 46.20& --02 40 32.1&--3.68 $\pm$ 0.15 &2.53 $\pm$ 0.12 &2.409 $\pm$ 0.084 &24 Feb 2007 &60 \\
 & &&& & & &24 Feb 2007 &45 \\
1106058AB &2E 1486 &05 39 47.42& --02 26 16.3&1.988 $\pm$ 0.093 &--0.654 $\pm$ 0.092 &2.220 $\pm$ 0.071 &23 Feb 2007 &240 \\
 & &&& & & &27 Feb 2007 &360 \\
969077 &2E 1487 &05 39 47.84& --02 32 24.9&2.017 $\pm$ 0.072 &--0.066 $\pm$ 0.072 &2.788 $\pm$ 0.050 &24 Feb 2007 &360 \\
1082115$^{e}$ &[HHM2007] 1235 &05 39 50.39& --02 43 30.8&--5.485 $\pm$ 2.089&0.468 $\pm$ 2.089&...&27 Feb 2007 &900 \\
 & &&& & & &28 Feb 2007 &600 \\
(1107114) &[HHM2007] 1251 &05 39 52.53& --02 43 22.4&3.515 $\pm$ 0.070 &--0.508 $\pm$ 0.070 &0.788 $\pm$ 0.052 &23 Feb 2007 &240 \\
(1110113) &[HHM2007] 1256 &05 39 53.13& --02 43 08.4&2.727 $\pm$ 0.057 &--6.228 $\pm$ 0.062 &1.038 $\pm$ 0.042 &23 Feb 2007 &300 \\
1245057AB$^{c,e}$ &Haro 5-31 &05 39 54.27& --02 24 40.1&2.28 $\pm$ 0.10 &0.048 $\pm$ 0.094 &2.415 $\pm$ 0.093 &25 Feb 2007 &450 \\
 & &&& & & &27 Feb 2007 &600 \\
 & &&& & & &27 Feb 2007 &600 \\
%1245057A &Haro 5-31A &05 39 54.27& --02 24 40.1&2.28 $\pm$ 0.10 &0.048 $\pm$ 0.094 &2.415 $\pm$ 0.093 &27 Feb 2007 &600 \\
%1245057B &Haro 5-31B &05 39 54.28& --02 24 38.0&2.47 $\pm$ 0.12 &--0.58 $\pm$ 0.11 &2.517 $\pm$ 0.097 &27 Feb 2007 &600 \\
(1169117) &[HHM2007] 1269 &05 39 54.37& --02 44 47.7&2.37 $\pm$ 0.10 &--1.946 $\pm$ 0.077 &0.643 $\pm$ 0.063 &22 Feb 2007 &180 \\
1223121 &Haro 5-34 &05 39 54.66& --02 46 34.1&2.123 $\pm$ 0.051 &--0.725 $\pm$ 0.048 &2.459 $\pm$ 0.030 &24 Feb 2007 &450 \\
1366055 &HD 294298 &05 39 59.31& --02 22 54.4&--0.726 $\pm$ 0.053 &--1.202 $\pm$ 0.055 &2.401 $\pm$ 0.035 &24 Feb 2007 &120 \\
... &UCAC2 30800287&05 40 02.17& --02 53 42.4&0.036 $\pm$ 0.079 &1.159 $\pm$ 0.079 &0.060 $\pm$ 0.050 &22 Feb 2007 &120 \\
1250070 &[HHM2007] 1337 &05 40 03.38& --02 29 01.5&2.463 $\pm$ 0.051 &--0.268 $\pm$ 0.047 &2.518 $\pm$ 0.033 &26 Feb 2007 &450 \\
1652046 &Haro 5-35 &05 40 03.66& --02 16 46.2&2.547 $\pm$ 0.059 &0.044 $\pm$ 0.060 &2.454 $\pm$ 0.044 &26 Feb 2007 &300 \\
1541051 &[NYS99] C-05 &05 40 05.11& --02 19 59.2&2.160 $\pm$ 0.072 &0.025 $\pm$ 0.074 &2.458 $\pm$ 0.053 &26 Feb 2007 &300 \\
(1338116) &[HHM2007] 1347 &05 40 05.30& --02 45 38.1&--0.623 $\pm$ 0.076 &0.183 $\pm$ 0.072 &0.694 $\pm$ 0.049 &22 Feb 2007 &180 \\
1311070 &[HHM2007] 1352 &05 40 06.97& --02 28 30.1&--0.654 $\pm$ 0.068 &--1.904 $\pm$ 0.072 &2.814 $\pm$ 0.043 &22 Feb 2007 &180 \\
1273081 &[HHM2007] 1359 &05 40 08.67& --02 32 43.3&2.09 $\pm$ 0.17 &--0.24 $\pm$ 0.15 &2.593 $\pm$ 0.099 &27 Feb 2007 &900 \\
 & &&& & & &28 Feb 2007 &600 \\
1269083 &Haro 5-39 &05 40 08.89& --02 33 33.8&2.217 $\pm$ 0.062 &--0.229 $\pm$ 0.053 &2.482 $\pm$ 0.035 &26 Feb 2007 &300 \\
... &HD 294307 &05 40 12.46& --02 52 57.7&0.079 $\pm$ 0.092 &--22.179 $\pm$ 0.084 &4.170 $\pm$ 0.060 &24 Feb 2007 &120 \\
 & &&& & & &27 Feb 2007 &300 \\
1396071 &[NYS99] C-09 &05 40 12.74& --02 28 19.9&--1.533 $\pm$ 0.047 &--0.454 $\pm$ 0.049 &2.489 $\pm$ 0.034 &23 Feb 2007 &300 \\
1564058 &Haro 5-38 &05 40 12.87& --02 22 02.2&--2.544 $\pm$ 0.051 &--4.255 $\pm$ 0.052 &2.404 $\pm$ 0.034 &23 Feb 2007 &240 \\
1359077 &HD 37686 &05 40 13.09& --02 30 53.2&1.87 $\pm$ 0.10 &--1.51 $\pm$ 0.10 &2.406 $\pm$ 0.061 &24 Feb 2007 &90 \\
1500066$^{b,e}$ &2M J05401607-0225446 &05 40 16.07&--02 25 44.6&--6.205 $\pm$ 2.131&--3.027 $\pm$ 2.122&...&28 Feb 2007 &750 \\
1748052 &Haro 5-37 &05 40 17.02& --02 18 11.3&2.629 $\pm$ 0.064 &--0.196 $\pm$ 0.066 &2.493 $\pm$ 0.045 &28 Feb 2007 &600 \\
1548068 &HD 37699 &05 40 20.19& --02 26 08.2&1.872 $\pm$ 0.096 &--0.974 $\pm$ 0.095 &2.404 $\pm$ 0.065 &23 Feb 2007 &45 \\
1476077 &2M J05402076-0230299 &05 40 20.77& --02 30 30.0&1.345 $\pm$ 0.050 &--0.166 $\pm$ 0.049 &2.435 $\pm$ 0.028 &28 Feb 2007 &600 \\
(1468100) &HD 294301 &05 40 21.12& --02 40 25.6&3.461 $\pm$ 0.070 &--9.564 $\pm$ 0.082 &2.540 $\pm$ 0.040 &22 Feb 2007 &180 \\
1471085 &Kiso A-0904 105 &05 40 22.56& --02 33 47.0&2.767 $\pm$ 0.063 &--0.481 $\pm$ 0.061 &2.368 $\pm$ 0.035 &24 Feb 2007 &360 \\
(1659068) &HD 294297 &05 40 27.54& --02 25 43.2&24.995 $\pm$ 0.065 &--25.087 $\pm$ 0.066 &6.107 $\pm$ 0.046 &24 Feb 2007 &90 \\
1679078 &[NYS99] C-16 &05 40 34.50& --02 30 22.4&--1.328 $\pm$ 0.065 &--0.560 $\pm$ 0.073 &2.422 $\pm$ 0.039 &23 Feb 2007 &240 \\
\end{longtable}
}
\begin{list}{}{}
\small{
\item[$^{a}$] 2MASS J05381494--0219532 and 2MASS J05384652--0235479 had poor quality spectra and, therefore, were discarded from the analysis.
\item[$^{b}$] Mayrit 1500066 {was not tabulated by} \textit{Gaia} DR2 and its equatorial coordinates are from 2MASS.
\item[$^{c}$] We took three spectra of Mayrit 1245057AB: one of component A, one of component B, and of the two of them inside the slit in parallactic angle.
\item[$^{d}$] Proper motions from \citet{2005yCat.1297....0Z} (UCAC4).
\item[$^{e}$] Proper motions from \citet{2017A&A...600L...4A} (HSOY).
\item[$^{f}$] {We took two spectra of Mayrit 1626148AB: one of component A and one of component B.}
}
\end{list}

%... ... ... ... ... ... ... ... ... ... ... ... ... ... ... ... ... ... ... ... ... ... ... ... ... ... ... ... ... ... ... ... ... ... ... ... ... ... ... ... ... ... ... ... ... ... ... ... ... ... ... ... 
\scriptsize{
\begin{longtable}{llccccccc}
\label{table.observations.gtc}\\ %
\caption[]{Observations of targets towards $\sigma$~Orionis with OSIRIS/GTC.}\\ %
   \hline
   \hline
   \noalign{\smallskip}
Mayrit	&Alternative	&$\alpha$	&$\delta$	&$\mu_{\alpha}\cos{\delta}$	&$\mu_{\delta}$	&$\varpi$	&Date of 		&$t_{\rm exp}$  \\ %
		&name		&(J2000)	&(J2000)	&[mas a$^{-1}$]			&[mas a$^{-1}$]	&[mas]	&observation	&[s]			 \\ %
\noalign{\smallskip}
    \hline
    \noalign{\smallskip}		
 \endfirsthead
\caption[]{Observations of targets towards $\sigma$~Orionis with OSIRIS/GTC (cont.).}\\ % 
  \hline
  \hline
  \noalign{\smallskip}
Mayrit	&Alternative	&$\alpha$	&$\delta$	&$\mu_{\alpha}\cos{\delta}$	&$\mu_{\delta}$	&$\varpi$	&Date of 		&$t_{\rm exp}$  \\ %
		&name		&(J2000)	&(J2000)	&[mas a$^{-1}$]			&[mas a$^{-1}$]	&[mas]	&observation	&[s]			 \\ %
  \noalign{\smallskip}
  \hline
  \noalign{\smallskip}
  \endhead
  \noalign{\smallskip}
  \hline
  \endfoot
1329304&Haro 5-5&05 37 30.95&--02 23 42.8&1.765 $\pm$ 0.082&--1.16 $\pm$ 0.079&2.731 $\pm$ 0.041&13 Nov 2012&2 $\times$ 200\\ %
1298302\textbf{$^{d,e}$}&[SWW2004] 137&05 37 31.54&--02 24 27.0&--3.393 $\pm$ 2.085&--1.398 $\pm$ 2.115&...&04 Mar 2012&1 $\times$ 600\\ %
1129222&[WB2004] 10&05 37 53.98&--02 49 54.5&1.77 $\pm$ 0.25&--1.41 $\pm$ 0.21&2.53 $\pm$ 0.13&27 Dic 2012&2 $\times$ 250\\ %
783254&2E 1455&05 37 54.40&--02 39 29.8&1.64 $\pm$ 0.12&--0.42 $\pm$ 0.10&2.590 $\pm$ 0.064&04 Mar 2012&3 $\times$ 30\\ %
...&2E 1456&05 37 56.31&--02 45 13.2&0.8 $\pm$ 1.0&1.1 $\pm$ 1.1&--0.67 $\pm$ 0.59&17 Mar 2012&1 $\times$ 1200\\ %
873229AB&Haro 5-7&05 38 01.08&--02 45 38.0&1.73 $\pm$ 0.57&--0.94 $\pm$ 0.52&4.23 $\pm$ 0.40&13 Nov 2012&2 $\times$ 200\\ %
662301&Kiso A-0904 67&05 38 06.75&--02 30 22.7&1.035 $\pm$ 0.074&--0.383 $\pm$ 0.074&2.354 $\pm$ 0.042&04 Mar 2012&2 $\times$ 300\\ %
1073209&[SWW2004] 52&05 38 09.95&--02 51 37.8&--0.23 $\pm$ 0.86&--0.08 $\pm$ 0.75&--0.73 $\pm$ 0.51&22 Dic 2012&4 $\times$ 150\\ %
757219&Haro 5-8&05 38 13.15&--02 45 51.1&1.199 $\pm$ 0.072&--1.175 $\pm$ 0.065&2.518 $\pm$ 0.044&15 Nov 2012&2 $\times$ 300\\ %
329261AB&[SWW2004] 207&05 38 23.08&--02 36 49.4&1.42 $\pm$ 0.25&--1.88 $\pm$ 0.21&1.92 $\pm$ 0.16&26 Dic 2012&2 $\times$ 250\\ %
609206&Haro 5-10&05 38 27.26&--02 45 09.7&1.129 $\pm$ 0.064&--0.675 $\pm$ 0.058&2.516 $\pm$ 0.039&18 Mar 2012&3 $\times$ 45\\ %
1207349&Haro 5-9&05 38 29.16&--02 16 15.8&0.345 $\pm$ 0.047&0.769 $\pm$ 0.042&2.412 $\pm$ 0.036&16 Nov 2012&3 $\times$ 120\\ %
180277$^{b}$&[W96] rJ053832-0235b&05 38 32.85&--02 35 39.3&0.546 $\pm$ 0.048&--0.32 $\pm$ 0.047&2.405 $\pm$ 0.028&04 Mar 2012&2 $\times$ 300\\ %
521199&TX Ori&05 38 33.69&--02 44 14.1&--1.29 $\pm$ 0.46&--2.70 $\pm$ 0.39&1.75 $\pm$ 0.26&16 Nov 2012&3 $\times$ 200\\ %
...&[HHM2007] 648&05 38 39.23&--02 53 08.4&2.218 $\pm$ 0.063&2.328 $\pm$ 0.056&0.364 $\pm$ 0.038&22 Dic 2012&2 $\times$ 150\\ %
156353&[W96] rJ053843-0233&05 38 43.56&--02 33 25.4&0.39 $\pm$ 0.13&--1.04 $\pm$ 0.12&2.149 $\pm$ 0.084&04 Mar 2012&2 $\times$ 300\\ %
1316178$^{a}$&S Ori J053847.2-025756&05 38 47.15&--02 57 55.7&1.92 $\pm$ 0.54&--1.87 $\pm$ 0.50&2.85 $\pm$ 0.35&06 Jan 2013&2 $\times$ 1200\\ %
528005AB$^{e}$&[W96] 4771-899&05 38 48.04&--02 27 14.2&11.410 $\pm$ 2.267&--5.078 $\pm$ 2.267&...&18 Mar 2012&3 $\times$ 45\\ %
...&[HHM2007] 829&05 38 59.46&--02 42 19.7&--7.23 $\pm$ 0.046&8.58 $\pm$ 0.043&1.144 $\pm$ 0.027&19 Dic 2012&2 $\times$ 150\\ %
1082013&Haro 5-16&05 39 01.37&--02 18 27.5&1.752 $\pm$ 0.055&--0.115 $\pm$ 0.053&2.379 $\pm$ 0.036&16 Nov 2012&3 $\times$ 200\\ %
687156&[WB2004] 26&05 39 03.58&--02 46 27.0&2.19 $\pm$ 0.19&--1.31 $\pm$ 0.16&2.698 $\pm$ 0.090&22 Dic 2012&2 $\times$ 200\\ %
...&Haro 5-17&05 39 13.01&-01 27 21.2&1.94 $\pm$ 0.042&--1.204 $\pm$ 0.042&2.791 $\pm$ 0.030&16 Nov 2012&3 $\times$ 200\\ %
872139&[BZR99] S Ori 28&05 39 23.19&--02 46 55.8&1.51 $\pm$ 0.71&--2.54 $\pm$ 0.67&2.55 $\pm$ 0.36&31 Mar 2012&1 $\times$ 900\\ %
1403026&Haro 5-22&05 39 26.40&--02 15 03.5&0.908 $\pm$ 0.040&0.475 $\pm$ 0.039&2.491 $\pm$ 0.026&16 Nov 2012&2 $\times$ 450\\ %
957055&[SWW2004] 163&05 39 37.30&--02 26 56.8&--1.226 $\pm$ 0.044&--1.463 $\pm$ 0.042&2.436 $\pm$ 0.028&04 Mar 2012&1 $\times$ 600\\ %
1400036&Haro 5-25&05 39 39.38&--02 17 04.5&2.285 $\pm$ 0.058&0.28 $\pm$ 0.056&2.548 $\pm$ 0.035&16 Nov 2012&2 $\times$ 300\\ %
931117&RW Ori&05 39 39.99&--02 43 09.8&2.19 $\pm$ 0.13&--0.72 $\pm$ 0.12&2.61 $\pm$ 0.13&18 Nov 2012&3 $\times$ 120\\ %
1233042&RV Ori&05 39 40.17&--02 20 48.0&2.405 $\pm$ 0.040&--0.167 $\pm$ 0.041&2.51 $\pm$ 0.025&16 Nov 2012&2 $\times$ 300\\ %
1626148AB${^f}$&Haro 5-33&05 39 42.77&--02 58 53.8&2.0 $\pm$ 5.6 &2.6 $\pm$ 5.6&...&18 Nov 2012&3 $\times$ 60\\ %
897077$^{a}$&S Ori J053943.2-023243&05 39 43.18&--02 32 43.4&1.82 $\pm$ 0.14&--0.34 $\pm$ 0.13&2.76 $\pm$ 0.10&23 Dic 2012&2 $\times$ 200\\ %
1120128&[BZR99] S Ori 32&05 39 43.58&--02 47 31.8&1.55 $\pm$ 0.95&--2.4 $\pm$ 1.0&2.40 $\pm$ 0.47&16 Jan 2013&3 $\times$ 600\\ %
1045067AB$^{a}$&$[$SWW2004$]$ 126/162&05 39 48.92&--02 29 10.5&--1.82 $\pm$ 0.16&1.23 $\pm$ 0.15&2.27 $\pm$ 0.10&23 Dic 2012&2 $\times$ 250\\ %
1279052&Haro 5-30&05 39 51.73&--02 22 47.3&2.24 $\pm$ 0.16&--0.23 $\pm$ 0.16&2.50 $\pm$ 0.10&18 Nov 2012&2 $\times$ 300\\ %
1042077&[HHM2007] 1250&05 39 52.47&--02 32 02.4&2.095 $\pm$ 0.046&--0.283 $\pm$ 0.042&2.477 $\pm$ 0.028&16 Jan 2013&2 $\times$ 150\\ %
1041082&Haro 5-32&05 39 53.62&--02 33 42.7&2.40 $\pm$ 0.12&--0.31 $\pm$ 0.10&2.580 $\pm$ 0.070&18 Nov 2012&2 $\times$ 150\\ %
1245057AB&Haro 5-31&05 39 54.27&--02 24 40.1&2.28 $\pm$ 0.10 &0.048 $\pm$ 0.094 &2.517 $\pm$ 0.097&18 Nov 2012&2 $\times$ 300\\ %
1223121&Haro 5-34&05 39 54.66&--02 46 34.1&2.123 $\pm$ 0.051&--0.725 $\pm$ 0.048&2.459 $\pm$ 0.030&18 Nov 2012&2 $\times$ 150\\ %
1446053&Haro 5-36&05 40 01.96&--02 21 32.6&2.42 $\pm$ 0.13&0.02 $\pm$ 0.13&2.683 $\pm$ 0.084&18 Nov 2012&2 $\times$ 200\\ %
1652046&Haro 5-35&05 40 03.66&--02 16 46.2&2.547 $\pm$ 0.059&0.044 $\pm$ 0.060&2.454 $\pm$ 0.044&18 Nov 2012&2 $\times$ 250\\ %
1196092$^{a}$&S Ori J054004.5-023642&05 40 04.53&--02 36 42.1&1.13 $\pm$ 0.79&--0.21 $\pm$ 0.88&1.58 $\pm$ 0.53&12 Mar 2012&1 $\times$ 900\\ %
1273081&[HHM2007] 1359&05 40 08.67&--02 32 43.3&2.09 $\pm$ 0.17&--0.24 $\pm$ 0.15&2.59 $\pm$ 0.10&03 Jan 2013&2 $\times$ 250\\ %
1564058&Haro 5-38&05 40 12.87&--02 22 02.2&--2.544 $\pm$ 0.051&--4.255 $\pm$ 0.052&2.404 $\pm$ 0.034&18 Nov 2012&2 $\times$ 200\\ %
1364078&V2754 Ori&05 40 13.96&--02 31 27.4&--1.10 $\pm$ 0.58&--4.84 $\pm$ 0.60&2.02 $\pm$ 0.39&14 Jan 2013&2 $\times$ 600\\ %
1748052&Haro 5-37&05 40 17.02&--02 18 11.3&2.629 $\pm$ 0.064&--0.196 $\pm$ 0.066&2.493 $\pm$ 0.045&18 Nov 2012&2 $\times$ 200\\ %
...&Haro 5-40&05 40 46.77&--02 10 50.1&3.40 $\pm$ 0.11&0.797 $\pm$ 0.095&2.554 $\pm$ 0.068&27 Dic 2012&2 $\times$ 150\\ %
...&Haro 5-44&05 40 51.48&--02 36 02.5&1.63 $\pm$ 0.16&--0.23 $\pm$ 0.16&2.48 $\pm$ 0.11&22 Dic 2012&2 $\times$ 150\\ %
...&Haro 5-46&05 41 20.69&--01 58 58.3&1.25 $\pm$ 0.10&0.358 $\pm$ 0.091&2.373 $\pm$ 0.066&22 Dic 2012&2 $\times$ 150\\ %
\noalign{\smallskip}
\end{longtable}
}
\begin{list}{}{}
\small{
\item[$^{a}$] The full alternative names of Mayrit 1316178, 897077, 1045067, and 1196092 are preceded by [BMZ2001].
\item[$^{b}$] Mayrit 180277 was also observed on 06 Mar 2012 at an airmass of 1.25 for 900\,s with the volume-phase holographic grating R2500V (R~$\sim$1400, $\Delta \lambda \sim$~4400--6000\,{\AA}).
\item[$^{c}$] 2E~1456 is a galaxy {previously} thought to be a star. We did not account for it in the analysis.
\item[\textbf{$^{d}$}] Mayrit 1298302 {was not tabulated by} \textit{Gaia} DR2 and its equatorial coordinates are from 2MASS.
\item[$^{e}$] Proper motions from \citet{2017A&A...600L...4A} (HSOY).
\item[$^{f}$] Proper motions from \citet{2005yCat.1297....0Z} (UCAC4).
%\item[$^{x}$] The spectrum of Mayrit 872139 was flagged as of poor quality (``PQC'') during the quality control due to the high air mass of the observation and the faintness of the target. 
}
\end{list}

%... ... ... ... ... ... ... ... ... ... ... ... ... ... ... ... ... ... ... ... ... ... ... ... ... ... ... ... ... ... ... ... ... ... ... ... ... ... ... ... ... ... ... ... ... ... ... ... ... ... ... ... 
\begin{longtable}{lcccccc}
\label{table.results.ids}\\ %
\caption[]{Equivalent widths of IDS/INT spectra of stars and brown dwarfs towards $\sigma$~Orionis.}\\ %
   \hline
   \hline
   \noalign{\smallskip}
Name	&EW(H$\beta$)		&EW(He~{\sc i} D$_{3}$)	&EW(H$\alpha$)	&EW(Li~{\sc i})		&SpT	&SpT\\
		&[\AA]			&[\AA]				&[\AA]			&[\AA]			&MILES	&PyHammer\\
\noalign{\smallskip}
    \hline
    \noalign{\smallskip}		
 \endfirsthead
\caption[]{Equivalent widths of IDS/INT spectra of stars and brown dwarfs towards $\sigma$~Orionis {(cont.)}.}\\ %
  \hline
  \hline
  \noalign{\smallskip}		
Name	&EW(H$\beta$)		&EW(He~{\sc i} D$_{3}$)	&EW(H$\alpha$)	&EW(Li~{\sc i})		&SpT	&SpT\\
		&[\AA]			&[\AA]				&[\AA]			&[\AA]			&MILES	&PyHammer\\
  \noalign{\smallskip}
  \hline
  \noalign{\smallskip}
  \endhead
  \noalign{\smallskip}
  \hline
  \endfoot

TYC~4770-1018-1&...&...&+1.4$_{-0.3}^{+0.2}$&...&K0:&K0 \\ %
\noalign{\smallskip}
Mayrit 1456284&...&...&+3.6$_{-0.3}^{+0.3}$&...&F7&G0 \\ %
\noalign{\smallskip}
Mayrit~1415279AB&...&...&--0.37$_{-0.07}^{+0.10}$&+0.24$_{-0.02}^{+0.04}$&K1&K5 \\ %
\noalign{\smallskip}
Mayrit~1374283&...&...&+2.1$_{-0.2}^{+0.4}$&+0.29$_{-0.02}^{+0.03}$&G7&K3 \\ %
\noalign{\smallskip}
HD~294270&...&...&+3.6$_{-0.4}^{+0.5}$&...&F6:&F6 \\ %
\noalign{\smallskip}
HD~294276&...&...&+3.0$_{-0.3}^{+0.5}$&...&G4&K0 \\ % 
\noalign{\smallskip}
2M J05372885-0255555&...&...&+1.8$_{-0.2}^{+0.2}$&...&K0&K3 \\ %
\noalign{\smallskip}
TYC~4770-1468--1&...&...&+1.4$_{-0.2}^{+0.1}$&...&K1&K5 \\ %
\noalign{\smallskip}
Mayrit~1227243&...&...&+11$_{-1}^{+1}$&...&A1&A3 \\ %
\noalign{\smallskip}
Mayrit~1116300&...&...&+10$_{-1}^{+2}$&...&A2&A3 \\ %
\noalign{\smallskip}
TYC 4771-621-1&...&...&+4.6$_{-0.6}^{+1.0}$&...&F3&F2 \\ %
\noalign{\smallskip}
Mayrit~968292&...&...&+5.4$_{-0.8}^{+0.7}$&...&F7&F4 \\ %
\noalign{\smallskip}
HD~294274&...&...&+2.6$_{-0.5}^{+0.1}$&...&G2&G7 \\ %
\noalign{\smallskip}
SO210868&...&...&+3.7$_{-0.6}^{+0.7}$&...&G4&F6 \\ %
\noalign{\smallskip}
Mayrit~797272$^{a}$&--4.2$_{-0.6}^{+0.5}$&...&--4.5$_{-0.2}^{+0.2}$&...&M2&M4 \\ %
\noalign{\smallskip}
Mayrit~789281&...&...&--0.66$_{-0.12}^{+0.23}$&+0.26$_{-0.04}^{+0.04}$&G4&K0 \\ %
\noalign{\smallskip}
Mayrit~783254&...&...&+0.66$_{-0.04}^{+0.06}$&+0.34$_{-0.04}^{+0.02}$&G7&K5 \\ %
\noalign{\smallskip}
[HHM2007] 244&...&...&+1.3$_{-0.1}^{+0.1}$&...&K1:&K5 \\ %
\noalign{\smallskip}
HD~294277&...&...&+1.4$_{-0.2}^{+0.1}$&...&K3&K5 \\ %
\noalign{\smallskip}
2M J05375789-0259536&...&...&+1.3$_{-0.1}^{+0.1}$&...&G7&K0 \\ %
\noalign{\smallskip}
TYC~4771-720-1&...&...&+2.7$_{-0.4}^{+0.4}$&...&G6&K2 \\ %
\noalign{\smallskip}
[W96]~4771-0950&...&...&+5.3$_{-0.8}^{+0.8}$&...&F2&F6 \\ %
\noalign{\smallskip}
Mayrit~662301&--14.2$_{-1.5}^{+0.7}$&-1.37$_{-0.63}^{+0.37}$&--32$_{-6}^{+1}$&...&M0&M7 \\ %
\noalign{\smallskip}
Mayrit~615296&...&...&--0.48$_{-0.09}^{+0.06}$&+0.69$_{-0.04}^{+0.06}$&K3&K4 \\ %
\noalign{\smallskip}
[HHM2007] 385&...&...&+1.2$_{-0.2}^{+0.2}$&...&M3&M2 \\ %
\noalign{\smallskip}
Mayrit~1285339&...&...&--0.40$_{-0.49}^{+0.71}$&+0.14$_{-0.02}^{+0.03}$&F8&K0 \\ %
\noalign{\smallskip}
TYC 4771-873-1&...&...&+8.4$_{-1.2}^{+1.5}$&...&F2&F4 \\ %
\noalign{\smallskip}
[SE2004] 10&...&...&+1.2$_{-0.3}^{+0.1}$&...&K1&K0 \\ %
\noalign{\smallskip}
IRAS~05358-0238&...&...&--0.61$_{-0.13}^{+0.09}$&...&M7&M7 \\ %
\noalign{\smallskip}
2M J05382265-0257421&...&...&+1.3$_{-0.1}^{+0.1}$&...&K0:&K5 \\ %
\noalign{\smallskip}
Mayrit~1449349$^{a}$&--13$_{-1}^{+1}$&--1.0$_{-0.6}^{+0.4}$&--30$_{-1}^{+1}$&+0.20$_{-0.07}^{+0.05}$&K5&M7 \\ %
\noalign{\smallskip}
Mayrit~609206$^{a}$&--11.1$_{-0.7}^{+0.6}$&--0.79$_{-0.16}^{+0.19}$&--57.5$_{-1.1}^{+0.8}$&...&K4:&... \\ %
\noalign{\smallskip}
Mayrit~521210&...&...&+10$_{-2}^{+2}$&...&A2&A9 \\ %
\noalign{\smallskip}
HD~294280&...&...&+1.2$_{-0.1}^{+0.1}$&...&K5&K5 \\ %
\noalign{\smallskip}
Mayrit~1207349&...&--2.2$_{-0.6}^{+0.5}$&--140$_{-11}^{+8}$&+0.54$_{-0.06}^{+0.06}$&K5&M7 \\ %
\noalign{\smallskip}
Mayrit~1275190&...&...&+5.1$_{-0.8}^{+0.8}$&...&F9&... \\ %
\noalign{\smallskip}
Mayrit~1160190&...&...&+7.3$_{-0.9}^{+0.9}$&...&F1&F4 \\ %
\noalign{\smallskip}
Mayrit~203283&...&...&--14$_{-1}^{+2}$&+0.66$_{-0.09}^{+0.15}$&M0&M8 \\ %
\noalign{\smallskip}
Mayrit~180277&...&--0.18$_{-0.03}^{+0.08}$&--2.3$_{-0.2}^{+0.6}$&+0.54$_{-0.08}^{+0.05}$&K5&K7 \\ %
\noalign{\smallskip}
Mayrit~521199&...&...&--13$_{-2}^{+2}$&+0.42$_{-0.02}^{+0.02}$&K1&... \\ %
\noalign{\smallskip}
Mayrit~165257$^{a}$&--16$_{-3.3}^{+2.8}$&...&--13.1$_{-0.5}^{+0.5}$&...&M4&M7 \\ %
\noalign{\smallskip}
Mayrit~189303&...&+0.18$_{-0.01}^{+0.02}$&+7.3$_{-0.6}^{+0.5}$&...&B7&A7 \\ %
\noalign{\smallskip}
Mayrit~168291AB&...&...&--1.2$_{-0.1}^{+0.4}$&+0.53$_{-0.04}^{+0.05}$&K5&K7 \\ %
\noalign{\smallskip}
StHa 50&...&...&--11.2$_{-0.8}^{+0.5}$&...&B6&... \\ %
\noalign{\smallskip}
[W96]~pJ053834-0239&...&...&+1.5$_{-0.1}^{+0.1}$&...&M0&K4 \\ %
\noalign{\smallskip}
Mayrit~182305&...&...&+8.1$_{-0.5}^{+0.7}$&...&A1&A3 \\ %
\noalign{\smallskip}
IDS 05335-0238 D&...&...&+1.7$_{-0.2}^{+0.1}$&...&G7&K0 \\ %
\noalign{\smallskip}
Mayrit~285331&...&--0.45$_{-0.07}^{+0.08}$&--3.6$_{-0.5}^{+0.6}$&+0.57$_{-0.05}^{+0.04}$&K4&K7 \\ %
\noalign{\smallskip}
Mayrit~344337AB&...&...&--3.1$_{-0.4}^{+0.2}$&+0.59$_{-0.07}^{+0.05}$&K4&K7 \\ %
\noalign{\smallskip}
Mayrit~489196&...&...&...&+0.47$_{-0.06}^{+0.05}$&K0:&... \\ %
\noalign{\smallskip}
Mayrit~208324&...&+0.44$_{-0.04}^{+0.08}$&+5.6$_{-0.7}^{+1.3}$&...&B5&... \\ %
\noalign{\smallskip}
Mayrit~105249&...&...&--1.9$_{-0.2}^{+0.5}$&+0.49$_{-0.06}^{+0.05}$&K5&K7 \\ %
\noalign{\smallskip}
Mayrit~114305AB&...&...&--0.94$_{-0.28}^{+0.24}$&+0.48$_{-0.05}^{+0.04}$&K2&K4 \\ %
\noalign{\smallskip}
HD~294278&...&...&+1.4$_{-0.2}^{+0.2}$&...&K0&K4 \\ %
\noalign{\smallskip}
[SE2004] 30&...&...&+1.1$_{-0.2}^{+0.1}$&...&K4&K5 \\ %
\noalign{\smallskip}
Mayrit~1248183AB&...&--0.27$_{-0.05}^{+0.08}$&--3.9$_{-0.4}^{+0.4}$&+0.57$_{-0.04}^{+0.07}$&K5&K7 \\ %
\noalign{\smallskip}
Mayrit~348349&...&--0.43$_{-0.04}^{+0.07}$&--11.1$_{-0.8}^{+1.0}$&+0.84$_{-0.16}^{+0.07}$&M0&K5 \\ %
\noalign{\smallskip}
Mayrit~97212&...&...&--2.8$_{-0.3}^{+0.3}$&+0.52$_{-0.05}^{+0.04}$&K5&K7 \\ %
\noalign{\smallskip}
Mayrit~83207$^{a}$&--3.9$_{-0.3}^{+0.3}$&...&--6.3$_{-0.3}^{+0.7}$&...&M0&M1 \\ %
\noalign{\smallskip}
Mayrit~156353&...&...&--4.5$_{-0.3}^{+0.4}$&+0.81$_{-0.11}^{+0.10}$&M0&M1 \\ %
\noalign{\smallskip}
Mayrit~11238&...&...&+10.2$_{-1.1}^{+0.9}$&...&A2&A3 \\ %
\noalign{\smallskip}
Mayrit~260182&...&...&--4.0$_{-0.5}^{+0.7}$&+0.62$_{-0.04}^{+0.03}$&K5&K7 \\ %
\noalign{\smallskip}
Mayrit~207358&...&...&--2.8$_{-0.6}^{+0.7}$&+0.57$_{-0.14}^{+0.14}$&K4&K7 \\ %
\noalign{\smallskip}
$\sigma$~Ori~AB&...&+0.86$_{-0.05}^{+0.08}$&+2.9$_{-0.2}^{+0.2}$&...&B1&... \\ %
\noalign{\smallskip}
[W96] pJ053844-0233&...&...&+1.4$_{-0.2}^{+0.1}$&...&K0&K5 \\ %
\noalign{\smallskip}
Mayrit~359179AB$^{a}$&--7.9$_{-0.5}^{+0.4}$&--0.45$_{-0.10}^{+0.15}$&--15.2$_{-0.5}^{+0.9}$&...&M1&M7 \\ %
\noalign{\smallskip}
Mayrit~13084&...&+0.85$_{-0.05}^{+0.06}$&+4.2$_{-0.3}^{+0.4}$&...&B2&O9 \\ %
\noalign{\smallskip}
Mayrit~42062&...&+0.94$_{-0.13}^{+0.08}$&+0.58$_{-0.27}^{+0.32}$&...&B2&O8 \\ %
\noalign{\smallskip}
Mayrit~53049&...&--1.3$_{-0.2}^{+0.1}$&--19$_{-1}^{+1}$&+0.55$_{-0.06}^{+0.04}$&M1&M7 \\ %
\noalign{\smallskip}
Mayrit~528005AB&...&...&--1.5$_{-0.6}^{+0.5}$&+0.53$_{-0.08}^{+0.11}$&K5&K7 \\ %
\noalign{\smallskip}
Mayrit~157155&...&...&--2.5$_{-0.3}^{+0.3}$&+0.72$_{-0.08}^{+0.05}$&K7&K7 \\ %
\noalign{\smallskip}
Mayrit~332168AB&...&...&--5.1$_{-0.6}^{+0.8}$&+0.71$_{-0.14}^{+0.10}$&M6&M7 \\ %
\noalign{\smallskip}
Mayrit~653170&...&--0.50$_{-0.10}^{+0.17}$&--17$_{-1}^{+1}$&+0.61$_{-0.16}^{+0.05}$&K7:&M7 \\ %
\noalign{\smallskip}
Mayrit~203039&...&...&--1.0$_{-0.1}^{+0.2}$&+0.56$_{-0.06}^{+0.05}$&K3&K4 \\ %
\noalign{\smallskip}
Mayrit~822170&...&...&--1.1$_{-0.3}^{+0.2}$&+0.43$_{-0.10}^{+0.06}$&K5:&K5 \\ %
\noalign{\smallskip}
Mayrit~707162AB&...&--0.23$_{-0.08}^{+0.18}$&--0.73$_{-0.07}^{+0.04}$&+0.50$_{-0.12}^{+0.06}$&K5&K5 \\ %
\noalign{\smallskip}
Mayrit~591158&...&...&+3.6$_{-0.6}^{+0.7}$&...&F3&G1 \\ %
\noalign{\smallskip}
Mayrit~1082013&...&--1.2$_{-0.2}^{+0.1}$&--23$_{-1}^{+1}$&+0.48$_{-0.11}^{+0.05}$&M0&M7 \\ %
\noalign{\smallskip}
HHM2007~846&...&...&+1.2$_{-0.2}^{+0.1}$&...&M3&M1 \\ %
\noalign{\smallskip}
Mayrit~306125AB&...&+0.33$_{-0.07}^{+0.05}$&+4.6$_{-1.0}^{+0.8}$&...&B5&... \\ %
\noalign{\smallskip}
Mayrit~374056&...&...&--0.98$_{-0.16}^{+0.12}$&+0.59$_{-0.08}^{+0.06}$&K5&K7 \\ %
\noalign{\smallskip}
Mayrit~397060&...&--0.61$_{-0.11}^{+0.17}$&--20.8$_{-2}^{+3}$&+0.53$_{-0.01}^{+0.04}$&K5&M7 \\ %
\noalign{\smallskip}
Mayrit~1011159$^{a}$&--2.3$_{-0.2}^{+0.2}$&...&--3.5$_{-0.2}^{+0.2}$&...&M0&M1 \\ %
\noalign{\smallskip}
Mayrit~1288163&...&...&+9.4$_{-1.3}^{+0.7}$&...&B8&A7 \\ %
\noalign{\smallskip}
Mayrit~497054$^{a}$&--7.0$_{-0.5}^{+1.2}$&--0.44$_{-0.21}^{+0.23}$&--9.6$_{-0.6}^{+0.4}$&...&M0&M7 \\ %
\noalign{\smallskip}
Mayrit~403090&...&...&--1.9$_{-0.2}^{+0.3}$&+0.62$_{-0.09}^{+0.05}$&K5&K7 \\ %
\noalign{\smallskip}
TYC~4771-1012-1&...&...&+0.96$_{-0.28}^{+0.12}$&...&K3&K4 \\ %
\noalign{\smallskip}
Mayrit~524060&...&...&+10$_{-2}^{+3}$&...&A3&A9 \\ %
\noalign{\smallskip}
[HHM2007]~961&...&...&+1.9$_{-0.5}^{+0.1}$&...&G6&K0 \\ %
\noalign{\smallskip}
[SE2004] 50&...&...&+1.6$_{-0.1}^{+0.2}$&...&K1:&K5 \\ %
\noalign{\smallskip}
Mayrit~634052&...&...&+0.40$_{-0.08}^{+0.06}$&+0.44$_{-0.08}^{+0.04}$&K0&K5 \\ %
\noalign{\smallskip}
Mayrit~596059&...&...&--32$_{-2}^{+2}$&+0.70$_{-0.12}^{+0.10}$&K7&M7 \\ %
\noalign{\smallskip}
TYC~4771-661-1&...&...&+2.4$_{-0.2}^{+0.4}$&...&G2&G0 \\ %
\noalign{\smallskip}
[HHM2007] 1009&...&...&+1.4$_{-0.2}^{+0.3}$&...&K1:&K4 \\ %
\noalign{\smallskip}
Mayrit~622103&...&--0.77$_{-0.09}^{+0.20}$&--36$_{-2}^{+1}$&+0.61$_{-0.05}^{+0.06}$&K7&M7 \\ %
\noalign{\smallskip}
Mayrit~1403026$^{a}$&--6.5$_{-0.6}^{+0.6}$&--0.61$_{-0.22}^{+0.37}$&--11.9$_{-0.4}^{+0.2}$&...&K5&M7 \\ %
\noalign{\smallskip}
[HHM2007] 1092&...&...&+1.4$_{-0.1}^{+0.1}$&...&K0&K3 \\ %
\noalign{\smallskip}
Mayrit~750107&...&...&--1.5$_{-0.2}^{+0.2}$&+0.75$_{-0.10}^{+0.09}$&K5&K7 \\ %
\noalign{\smallskip}
Mayrit~863116&...&...&--1.8$_{-0.9}^{+0.3}$&+0.13$_{-0.04}^{+0.03}$&G4&K0 \\ %
\noalign{\smallskip}
[HHM2007] 1129&...&...&+1.5$_{-0.2}^{+0.1}$&...&K1&K5 \\ %
\noalign{\smallskip}
Mayrit~957055&...&...&--2.3$_{-0.3}^{+0.5}$&+0.61$_{-0.03}^{+0.04}$&K7&K7 \\ %
\noalign{\smallskip}
Haro~5-28&...&...&+3.9$_{-0.3}^{+0.4}$&...&F6&F9 \\ %
\noalign{\smallskip}
Mayrit~1400036&...&--0.92$_{-0.1}^{+0.18}$&--12.3$_{-0.8}^{+1.3}$&+0.63$_{-0.06}^{+0.06}$&M0&M7 \\ %
\noalign{\smallskip}
Mayrit~871071$^{a}$&--38.7$_{-0.8}^{+0.6}$&--6.2$_{-1.7}^{+1.2}$&--123$_{-3}^{+2}$&...&K2&... \\ %
\noalign{\smallskip}
Mayrit~931117&...&...&--3.5$_{-0.6}^{+0.5}$&+0.43$_{-0.03}^{+0.04}$&K1&K4 \\ %
\noalign{\smallskip}
Mayrit~1233042&...&--1.2$_{-0.1}^{+0.1}$&--21.0$_{-0.9}^{+1.6}$&+0.41$_{-0.03}^{+0.05}$&K4&M7 \\ %
\noalign{\smallskip}
HD 294299&...&...&+9.5$_{-1.5}^{+1.5}$&...&A6&A7 \\ %
\noalign{\smallskip}
Mayrit~1626148AB&...&--1.4$_{-0.3}^{+0.2}$&--56$_{-6}^{+6}$&+0.39$_{-0.04}^{+0.04}$&M0&M7 \\ %
\noalign{\smallskip}
TYC~4771-934-1&...&...&+1.4$_{-0.3}^{+0.1}$&...&K0&K5 \\ %
\noalign{\smallskip}
[HHM2007] 1189&...&...&+1.2$_{-0.1}^{+0.1}$&...&K2&K5 \\ %
\noalign{\smallskip}
Mayrit~960106AB&...&...&+6.2$_{-0.8}^{+0.6}$&...&A3&A9 \\ %
\noalign{\smallskip}
Mayrit~1106058AB&...&...&--1.5$_{-0.2}^{+0.1}$&+0.44$_{-0.05}^{+0.07}$&K0&K0 \\ %
\noalign{\smallskip}
Mayrit~969077&...&...&--2.6$_{-0.5}^{+0.4}$&+0.63$_{-0.09}^{+0.13}$&K4&K7 \\ %
\noalign{\smallskip}
Mayrit~1082115&--1.4$_{-0.2}^{+0.3}$&--0.22$_{-0.12}^{+0.08}$&--2.0$_{-0.1}^{+0.2}$&...&M0	&M3 \\ %
\noalign{\smallskip}
[HHM2007] 1251&...&...&+1.3$_{-0.1}^{+0.1}$&...&K4&K4 \\ %
\noalign{\smallskip}
[HHM2007] 1256&...&...&+1.7$_{-0.1}^{+0.1}$&...&G:&... \\ %
\noalign{\smallskip}
Mayrit~1245057AB&...&...&--11.3$_{-1.9}^{+0.7}$&+0.53$_{-0.17}^{+0.04}$&M3&M6 \\ %
\noalign{\smallskip}
%Mayrit~1245057A&...&--0.97$_{-0.11}^{+0.06}$&--17.0$_{-1.1}^{+0.5}$&+0.70$_{-0.05}^{+0.10}$&M3&M7 \\ %
%\noalign{\smallskip}
%Mayrit~1245057B&...&--0.32$_{-0.04}^{+0.02}$&--13.1$_{-0.5}^{+0.5}$&+0.60$_{-0.04}^{+0.04}$&M6&M7 \\ %
%\noalign{\smallskip}
[HHM2007] 1269&...&...&+1.4$_{-0.2}^{+0.1}$&...&K0&K0 \\ %
\noalign{\smallskip}
Mayrit~1223121&...&--0.42$_{-0.11}^{+0.17}$&--16$_{-2}^{+3}$&+0.51$_{-0.15}^{+0.09}$&K5&M7 \\ %
\noalign{\smallskip}
Mayrit~1366055&...&...&+0.62$_{-0.17}^{+0.08}$&+0.24$_{-0.03}^{+0.02}$&G2&K0 \\ %
\noalign{\smallskip}
UCAC2 30800287&...&...&+1.2$_{-0.1}^{+0.1}$&...&M2&... \\ %
\noalign{\smallskip}
Mayrit~1250070&...&...&--2.8$_{-0.3}^{+0.5}$&+0.63$_{-0.05}^{+0.09}$&K7&K7 \\ %
\noalign{\smallskip}
Mayrit~1652046&...&--0.30$_{-0.07}^{+0.06}$&--10.5$_{-0.7}^{+0.5}$&+0.52$_{-0.09}^{+0.04}$&M0&M7 \\ %
\noalign{\smallskip}
Mayrit~1541051&...&...&--2.8$_{-0.3}^{+0.7}$&+0.65$_{-0.04}^{+0.05}$&M0&M1 \\ %
\noalign{\smallskip}
[HHM2007] 1347&...&...&+1.5$_{-0.2}^{+0.1}$&...&K0:&K0 \\ %
\noalign{\smallskip}
Mayrit~1311070&...&...&+4.4$_{-0.8}^{+0.7}$&...&F6&F7 \\ %
\noalign{\smallskip}
Mayrit~1273081&...&--0.80$_{-0.15}^{+0.15}$&--8.1$_{-0.6}^{+0.8}$&+0.45$_{-0.07}^{+0.04}$&M7&M7 \\ %
\noalign{\smallskip}
Mayrit~1269083&...&--0.88$_{-0.16}^{+0.36}$&--17.4$_{-1.4}^{+0.9}$&+0.50$_{-0.04}^{+0.07}$&K7&M7 \\ %
\noalign{\smallskip}
HD~294307&...&...&+4.0$_{-0.4}^{+0.5}$&...&F7&F6 \\ %
\noalign{\smallskip}
Mayrit~1396071&...&...&--8.5$_{-0.7}^{+0.4}$&+0.66$_{-0.24}^{+0.24}$&K7&M6 \\ %
\noalign{\smallskip}
Mayrit~1564058&...&...&--16$_{-1}^{+2}$&+0.61$_{-0.02}^{+0.05}$&K5&M7 \\ %
\noalign{\smallskip}
Mayrit~1359077&...&...&+10$_{-1}^{+1}$&...&B9&A3 \\ %
\noalign{\smallskip}
Mayrit~1500066$^{a}$&--2.4$_{-0.5}^{+0.5}$&...&--2.5$_{-0.1}^{+0.2}$&...&M3&M4 \\ %
\noalign{\smallskip}
Mayrit~1748052$^{a}$&--5.2$_{-0.3}^{+0.6}$&...&--12.7$_{-0.2}^{+0.3}$&...&M1&M7 \\ %
\noalign{\smallskip}
Mayrit~1548068&...&+0.55$_{-0.10}^{+0.10}$&+5.5$_{-0.4}^{+0.4}$&...&B2&B4 \\ %
\noalign{\smallskip}
Mayrit~1476077$^{a}$&--4.1$_{-0.5}^{+0.3}$&--0.38$_{-0.32}^{+0.25}$&--8.9$_{-1.7}^{+0.5}$&+0.17$_{-0.05}^{+0.03}$&K4&M1 \\ %
\noalign{\smallskip}
HD~294301&...&...&+6.6$_{-0.5}^{+0.9}$&...&F2&F3 \\ %
\noalign{\smallskip}
Mayrit~1471085&...&...&--1.5$_{-0.3}^{+0.4}$&+0.54$_{-0.07}^{+0.08}$&K5&K7 \\ %
\noalign{\smallskip}
HD~294297&...&...&+3.4$_{-1.1}^{+0.9}$&...&G2:&G3 \\ %
\noalign{\smallskip}
Mayrit~1679078&...&...&--2.0$_{-0.4}^{+0.5}$&+0.56$_{-0.14}^{+0.06}$&K5&K7 \\ %
\noalign{\smallskip}

\end{longtable}
\begin{list}{}{}
\small{
\item[$^{a}$] Stars observed {also} with the R150V grism.
}
\end{list}

%... ... ... ... ... ... ... ... ... ... ... ... ... ... ... ... ... ... ... ... ... ... ... ... ... ... ... ... ... ... ... ... ... ... ... ... ... ... ... ... ... ... ... ... ... ... ... ... ... ... ... ... 
\begin{longtable}{lccccccccc}
\label{table.results.gtc} \\ %
\caption[]{Equivalent widths of OSIRIS/GTC spectra of stars and brown dwarfs towards $\sigma$~Orionis.}\\ %
   \hline
   \hline
   \noalign{\smallskip}
Name	&EW(Ca K)	&EW(Ca H)	&EW(H$\gamma$)	&EW(H$\beta$)	&EW(He~{\sc i} D$_{3}$)	&EW(H$\alpha$)	&EW(Li~{\sc i})		&SpT 	    &SpT     \\
	   	&[\AA]		&[\AA]		&[\AA]			&[\AA]		    &[\AA]			    	&[\AA]			&[\AA]			    &Standards	&PyHammer\\
\noalign{\smallskip}
    \hline
    \noalign{\smallskip}		
 \endfirsthead
\caption[]{Equivalent widths of OSIRIS/GTC spectra of stars and brown dwarfs towards $\sigma$~Orionis.}\\ %
  \hline
  \hline
  \noalign{\smallskip}		
Name	&EW(Ca K)	&EW(Ca H)	&EW(H$\gamma$)	&EW(H$\beta$)	&EW(He~{\sc i} D$_{3}$)	&EW(H$\alpha$)	&EW(Li~{\sc i})		&SpT 	    &SpT\\
	   	&[\AA]		&[\AA]		&[\AA]			&[\AA]		    &[\AA]			    	&[\AA]			&[\AA]			    &Standards	&PyHammer\\
  \noalign{\smallskip}
  \hline
  \noalign{\smallskip}
  \endhead
  \noalign{\smallskip}
  \hline
  \endfoot

Mayrit~1329304	&--73$_{-12}^{+8}$	&--57$_{-9}^{+7}$	&--26$_{-5}^{+4}$	&--56$_{-6}^{+5}$	&--3.3$_{-0.4}^{+0.3}$	&--118$_{-15}^{+14}$	&+0.13$_{-0.08}^{+0.06}$		&M2.5 &M7 \\ % 
\noalign{\smallskip}
Mayrit~1298302$^{a}$ 	&$>$ --2.0		&$>$--2.0			&$>$ --1.0		&--2.6$\pm$0.2		&...	&--3.5$\pm$0.3		&$<$+0.2		&M3.0 & M3 \\ % Only flare in LC08
\noalign{\smallskip}
Mayrit~1129222	&...	&--25$_{-13}^{+7}$	&--14$_{-3}^{+3}$	&--12$_{-2}^{+2}$	&--1.4$_{-0.3}^{+0.3}$	&--22$_{-2}^{+1}$	&+0.31$_{-0.27}^{+0.18}$		&M2.5: & M6 \\ % 
\noalign{\smallskip}
Mayrit~783254$^{a}$	&+13$\pm$2		&+9$\pm$2		&+0.7$\pm$0.2		&+0.8$\pm$0.6		&...	&+0.8$\pm$0.2		&+0.34$\pm$0.09	&$\ll$K7 & K2 \\ %
\noalign{\smallskip}
Mayrit~873229AB	&--13$_{-5}^{+4}$	&--25$_{-4}^{+4}$	&--18$_{-3}^{+2}$	&--19$_{-2}^{+2}$	&--2.3$_{-0.5}^{+0.5}$	&--59$_{-6}^{+5}$	&...	&M4.5 & M7 \\ % 
\noalign{\smallskip}
Mayrit~662301$^{a}$	&--5.5$\pm$1.5		&--5.5$\pm$1.0        	&--9$\pm$2		&--7.0$\pm$1.0		&...	&--11.8$\pm$0.9	&+0.66$\pm$0.14	&M1.5 & M1 \\ % [O I], [N II], [S II]
\noalign{\smallskip}
Mayrit~1073209	&--18$_{-4}^{+7}$	&--15$_{-5}^{+4}$	&--9.2$_{-2.9}^{+2.2}$	&--9.2$_{-2.3}^{+1.2}$	&--1.2$_{-0.3}^{+0.3}$	&--24$_{-3}^{+2}$	&+0.57$_{-0.21}^{+0.11}$	&M3.0 & M4 \\ % 
\noalign{\smallskip}
Mayrit~757219	&--94$_{-21}^{+15}$		&--74$_{-11}^{+11}$	&--33$_{-5}^{+3}$	&--49$_{-5}^{+5}$	&--2.2$_{-0.3}^{+0.3}$	&--91$_{-9}^{+7}$	&+0.14$_{-0.06}^{+0.05}$	&M1.0 & M7 \\ % 
\noalign{\smallskip}
Mayrit~329261AB	&--18$_{-9}^{+5}$	&--29$_{-8}^{+5}$	&--17$_{-5}^{+3}$	&--52$_{-13}^{+10}$		&--5.7$_{-1.3}^{+0.8}$	&--148$_{-20}^{+16}$	&... 	&M4.5 & M8 \\ % 
\noalign{\smallskip}
Mayrit~609206$^{a}$	&--26$\pm$3		& --26$\pm$3		&--9$\pm$2	        	&--15.0$\pm$1.0	&...	&--46$\pm$2		&+0.40$\pm$0.07	&M0:	 & M4 \\ % [O I], [N II]?, [S II]
\noalign{\smallskip}
Mayrit~1207349	&--4.6$_{-2.0}^{+1.9}$	&--14$_{-3}^{+2}$	&--12$_{-4}^{+3}$	&--12$_{-1}^{+1}$	&--1.0$_{-0.2}^{+0.3}$	&--54$_{-3}^{+3}$	&+0.14$_{-0.18}^{+0.09}$		&M0.0 & M4 \\ % 
\noalign{\smallskip}
Mayrit~180277$^{a}$	&--7$\pm$2		&--8$\pm$2	        	&--2.5$\pm$1.5		&--3.0$\pm$1.0		&...	&--4.4$\pm$0.2		&+0.26$\pm$0.16	&M0.0 & M0 \\ %
\noalign{\smallskip}
Mayrit~521199 	&--5.3$_{-1.7}^{+1.6}$	&--1.1$_{-0.5}^{+0.3}$	&+1.3$_{-0.3}^{+0.4}$	&+0.81$_{-0.31}^{+0.25}$		&...	&--12.6$_{-0.6}^{+0.5}$	&+0.29$_{-0.05}^{+0.04}$		&M0: & M0 \\ % 
\noalign{\smallskip}
$[$HHM2007]~648	&+7.2$_{-1.0}^{+0.8}$	&+6.8$_{-0.2}^{+0.4}$	&+1.1$_{-0.3}^{+0.3}$	&+0.78$_{-0.12}^{+0.12}$		&--0.13$_{-0.08}^{+0.05}$		&+1.1$_{-0.1}^{+0.1}$	& ...	&$\ll$K7 & K5 \\ % 
\noalign{\smallskip}
Mayrit~156353$^{a}$ 	&--10$\pm$2		&--9.5$\pm$1.0        	&--2.0$\pm$0.5		&--2.5$\pm$0.2		&...	&--3.9$\pm$0.2		&+0.18$\pm$0.09	&M1.0 & M1 \\ %
\noalign{\smallskip}
Mayrit~1316178	&--13$_{-6}^{+4}$	&--54$_{-23}^{+14}$		&--70$_{-22}^{+13}$		&--74$_{-20}^{+13}$		&--14$_{-3}^{+3}$	&--80$_{-13}^{+10}$		& ...	&M6.5 & M8 \\ % 
\noalign{\smallskip}
Mayrit~528005AB$^{a}$	&+4.5$\pm$0.2		&+4.5$\pm$0.4		&$<$+0.4			&$<$ +0.2			&...	&--1.9$\pm$0.2		&+0.30$\pm$0.11		&M0.0 & K7 \\ %
\noalign{\smallskip}
$[$HHM2007]~829  	&+5.9$_{-0.3}^{+0.3}$	&+6.1$_{-0.4}^{+0.3}$	&+2.6$_{-0.2}^{+0.1}$	&+3.2$_{-0.3}^{+0.4}$	&--0.12$_{-0.08}^{+0.06}$		&+3.2$_{-0.3}^{+0.3}$	&... &$\ll$K7 & G5 \\ % 
\noalign{\smallskip}
Mayrit~1082013	&--15$_{-3}^{+2}$	&--32$_{-8}^{+7}$	&--18$_{-3}^{+2}$	&--16$_{-1}^{+1}$	&--1.0$_{-0.2}^{+0.2}$	&--23$_{-1}^{+1}$	&... 	&M1.0 & M4 \\ % 
\noalign{\smallskip}
Mayrit~687156	&--23$_{-6}^{+3}$	&--29$_{-7}^{+4}$	&--23$_{-4}^{+3}$	&--26$_{-4}^{+3}$	&--2.2$_{-0.5}^{+0.4}$	&--50$_{-6}^{+4}$	&+0.22$_{-0.08}^{+0.08}$ 	&M4.5 & M7 \\ % 
\noalign{\smallskip}
Haro~5-17	&--13.3$_{-0.7}^{+0.8}$	&--10.1$_{-0.6}^{+0.7}$	&--5.7$_{-1.0}^{+1.0}$	&--9.2$_{-1.0}^{+0.8}$	&--0.59$_{-0.32}^{+0.18}$		&--33$_{-2}^{+2}$	&+0.31$_{-0.06}^{+0.06}$	&$\ll$K7 & K5 \\ % 
\noalign{\smallskip}
Mayrit~872139$^{a}$	&...				&...				&...				&...				&...	&--13$\pm$3		&+1$\pm$1			&M5.0 & M7 \\ % Li I in absorption?
\noalign{\smallskip}
Mayrit~1403026	&--6.0$_{-0.8}^{+0.7}$	&--10.3$_{-0.8}^{+0.8}$	&--7.7$_{-1.0}^{+1.0}$	&--8.3$_{-0.8}^{+1.1}$	&--0.60$_{-0.21}^{+0.09}$		&--23$_{-1}^{+1}$	&+0.27$_{-0.08}^{+0.7}$	&M0.0 & K5 \\ % 
\noalign{\smallskip}
Mayrit~957055$^{a}$ 	&--13$\pm$2		&--12$\pm$2	        	&--3.5$\pm$1.5		&--3.5$\pm$0.5		&...	&--5.5$\pm$0.3		&+0.23$\pm$0.07	&M0.0 & M0 \\ %
\noalign{\smallskip}
Mayrit~1400036	&--9.7$_{-1.0}^{+0.8}$	&--12$_{-2}^{+2}$	&--2.0$_{-0.7}^{+0.7}$	&...	&...	&--1.2$_{-0.1}^{+0.1}$	&+0.24$_{-0.26}^{+0.16}$		&M0.5 & M1 \\ %
\noalign{\smallskip}
Mayrit~931117	&--0.73$_{-0.35}^{+0.29}$		&...	&+1.4$_{-0.3}^{+0.3}$	&+2.7$_{-0.4}^{+0.4}$	&--0.54$_{-0.38}^{+0.25}$		&+2.2$_{-0.2}^{+0.1}$	&+0.50$_{-0.09}^{+0.10}$		&$\ll$K7 & K5 \\ % 
\noalign{\smallskip}
Mayrit~1233042	&--8.9$_{-2.7}^{+3.0}$	&--6.5$_{-2.3}^{+1.5}$	&--3.6$_{-1.0}^{+1.2}$	&--3.1$_{-0.4}^{+0.3}$	&--0.93$_{-0.32}^{+0.27}$		&--14.8$_{-1.1}^{+0.9}$	&+0.36$_{-0.06}^{+0.05}$		&K7.0 & K5 \\ % 
\noalign{\smallskip}
Mayrit~1626148AB	&--28$_{-3}^{+2}$	&--34$_{-6}^{+5}$	&--18$_{-3}^{+2}$	&--15$_{-2}^{+2}$	&--1.4$_{-0.4}^{+0.3}$	&--14$_{-1}^{+1}$	&+0.18$_{-0.10}^{+0.12}$	 &M1.5 & M4 \\ % 
\noalign{\smallskip}
Mayrit~897077	&--8.8$_{-1.9}^{+2.0}$	&--11$_{-4}^{+3}$	&--4.5$_{-0.9}^{+0.6}$	&--6.1$_{-1.1}^{+0.8}$	&...	&--8.0$_{-1.1}^{+0.8}$	&+0.17$_{-0.12}^{+0.17}$	 &M4.5 & M5 \\ % 
\noalign{\smallskip}
Mayrit~1120128	&...	&...	&...	&...	&--3.2$_{-2.4}^{+1.0}$	&--7.9$_{-1.9}^{+1.4}$	&... 	&M6.0 & M8 \\ % 
\noalign{\smallskip}
Mayrit~1045067AB	&--20$_{-7}^{+4}$	&--15$_{-6}^{+3}$	&--11$_{-2}^{+2}$	&--11$_{-1}^{+1}$	&...	&--12.1$_{-1.0}^{+0.8}$	&--0.14$_{-0.12}^{+0.17}$	 	&M3.0 & M4 \\ % 
\noalign{\smallskip}
Mayrit~1279052	&--53$_{-16}^{+11}$		&--46$_{-13}^{+10}$		&--30$_{-5}^{+3}$	&--57$_{-13}^{+8}$	&--3.6$_{-0.8}^{+0.3}$	&--100$_{-25}^{+20}$	&... 	&M5.5 & M8 \\ % 
\noalign{\smallskip}
Mayrit~1042077	&--3.6$_{-1.0}^{+0.6}$	&--3.8$_{-0.9}^{+0.8}$	&...	&--0.61$_{-0.16}^{+0.03}$		&--0.44$_{-0.24}^{+0.11}$		&--2.0$_{-0.2}^{+0.1}$	&+0.41$_{-0.18}^{+0.15}$	 	&<K7 &K7 \\ % 
\noalign{\smallskip}
Mayrit~1041082	&--33$_{-6}^{+4}$	&--43$_{-6}^{+5}$	&--34$_{-6}^{+7}$	&--44$_{-8}^{+5}$	&--4.6$_{-0.4}^{+0.6}$	&--89$_{-19}^{+12}$		&+0.65$_{-0.48}^{+0.34}$	 	&M3.0 & M6 \\ % 
\noalign{\smallskip}
Mayrit~1245057AB	&--27$_{-3}^{+2}$	&--21$_{-3}^{+2}$	&--6.8$_{-1.0}^{+1.0}$	&--7.4$_{-0.5}^{+0.4}$	&...	&--14.3$_{-1.2}^{+0.9}$	&... 	&M2.5 & M4 \\ % 
\noalign{\smallskip}
Mayrit~1223121	&--5.9$_{-1.4}^{+1.4}$	&--14$_{-3}^{+3}$	&--8.0$_{-2.5}^{+1.6}$	&--6.1$_{-0.8}^{+0.7}$	&--0.65$_{-0.16}^{+0.19}$		&--12.7$_{-0.8}^{+0.7}$	&+0.31$_{-0.05}^{+0.05}$	 	&<K7 & K5 \\ % 
\noalign{\smallskip}
Mayrit~1446053	&--8.9$_{-2.9}^{+1.9}$	&--11$_{-4}^{+3}$	&--5.4$_{-1.7}^{+1.2}$	&--3.7$_{-0.9}^{+0.6}$	&--0.53$_{-0.36}^{+0.13}$		&--22$_{-2}^{+2}$	&...	 &M4.5 & M6 \\ % 
\noalign{\smallskip}
Mayrit~1652046	&--16$_{-2}^{+1}$	&--12$_{-2}^{+1}$	&--3.2$_{-0.9}^{+0.8}$	&--3.5$_{-0.3}^{+0.3}$	&...	&--18$_{-2}^{+1}$	&+0.18$_{-0.12}^{+0.08}$	 	&M0.5 & M1 \\ % 
\noalign{\smallskip}
Mayrit~1196092$^{a}$ 	&...				&...				&...				&...				&...	&--125$\pm$15		&+0.24$\pm$0.30		&M6.5 & M8 \\ % Li I in absorption?
\noalign{\smallskip}
Mayrit~1273081	&--14$_{-3}^{+3}$	&--7.2$_{-1.7}^{+1.1}$	&--6.7$_{-1.5}^{+1.0}$	&--5.7$_{-1.3}^{+0.8}$	&...	&--8.7$_{-0.9}^{+0.5}$	&...	 &M5.5 & M7 \\ % 
\noalign{\smallskip}
Mayrit~1564058	&--15$_{-2}^{+3}$	&--14$_{-2}^{+2}$	&--4.9$_{-2.1}^{+1.3}$	&--5.5$_{-0.9}^{+0.8}$	&--0.84$_{-0.20}^{+0.24}$		&--16$_{-1}^{+1}$	&+0.35$_{-0.04}^{+0.05}$	 &M0: & K5 \\ % 
\noalign{\smallskip}
Mayrit~1364078	&--16$_{-12}^{+6}$	&--13$_{-9}^{+5}$	&--51$_{-12}^{+24}$		&--49$_{-17}^{+13}$		&...	&--73$_{-19}^{+10}$		&... 	&M6.5 & M8 \\ % 
\noalign{\smallskip}
Mayrit~1748052	&--20$_{-4}^{+3}$	&--18$_{-4}^{+4}$	&--3.0$_{-0.8}^{+0.6}$	&--5.5$_{-0.9}^{+0.8}$	&--0.10$_{-0.06}^{+0.02}$		&--17$_{-1}^{+1}$	&+0.19$_{-0.09}^{+0.08}$	 	&M1.5 & M4 \\ % 
\noalign{\smallskip}
Haro~5-40	&--43$_{-15}^{+9}$	&--57$_{-17}^{+8}$	&--42$_{-7}^{+6}$	&--72$_{-8}^{+9}$	&--3.4$_{-0.3}^{+0.2}$	&--147$_{-21}^{+11}$	&...	 	&M4.0 & M8 \\ % 
\noalign{\smallskip}
Haro~5-44		&--25$_{-4}^{+3}$	&--27$_{-4}^{+3}$	&--15$_{-3}^{+3}$	&--31$_{-5}^{+3}$	&--5.8$_{-0.4}^{+0.3}$	&--160$_{-35}^{+28}$	&+0.07$_{-0.06}^{+0.05}$	 	&M3:	 & M6 \\ % 
\noalign{\smallskip}
Haro~5-46		&--5.9$_{-2.0}^{+1.5}$	&--18$_{-5}^{+3}$	&--12$_{-3}^{+3}$	&--14$_{-2}^{+1}$	&--1.0$_{-0.5}^{+0.2}$	&--30$_{-3}^{+2}$	&+0.28$_{-0.14}^{+0.08}$	 	&M2.5 & M4 \\ % 
\noalign{\smallskip}
%GJ205			&+0.83$_{-0.29}^{+0.33}$		&+5.4$_{-1.4}^{+0.9}$	&...	&...	&...	&...	&-0.46$_{-0.10}^{+0.15}$	 	&... \\ % Li not possible
%\noalign{\smallskip}
%GJ270			&...	&...	&+0.61$_{-0.16}^{+0.17}$		&...	&...	&+0.12$_{-0.03}^{+0.04}$	&...	 &... \\ % Ha almost nothing - remove
%\noalign{\smallskip}
%GJ328			&...	&+6.3$_{-1.0}^{+0.7}$	&+0.62$_{-0.13}^{+0.11}$		&...	&...	&+0.26$_{-0.04}^{+0.04}$		&--0.19$_{-0.05}^{+0.06}$	 	&... \\ % Li not possible
%\noalign{\smallskip}
%GJ380			&...	&...	&+1.2$_{-0.4}^{+0.2}$	&...	&...	&+0.32$_{-0.05}^{+0.05}$		&... 	&... \\ % Ha almost nothing - remove
%\noalign{\smallskip}
%GJ846			&...	&...	&+0.46$_{-0.13}^{+0.12}$		&...	&...	&...	&-0.46$_{-0.13}^{+0.10}$	 &... \\ % Li not possible
%\noalign{\smallskip}
\end{longtable}

\begin{list}{}{}
\small{
\item[$^{a}$] Sources observed by \citet{2012A&A...546A..59C}.
}
\end{list}

%... ... ... ... ... ... ... ... ... ... ... ... ... ... ... ... ... ... ... ... ... ... ... ... ... ... ... ... ... ... ... ... ... ... ... ... ... ... ... ... ... ... ... ... ... ... ... ... ... ... ... ... 
\normalsize{
\begin{longtable}{lccccccccccc}
\label{table.results.spt}\\ %
\caption[]{Membership and youth features of investigated stars$^{a}$.} \\ %
   \hline
   \hline
   \noalign{\smallskip}
Name	&SpT		&Ref.$^{b}$	&SpT 		&OB	&$\mu$ 	&$\varpi$	&Li~{\sc i}	&H$\alpha$	&mIR	&X rays	&Member	\\
		&Literature	&			&Adopted		&	&		&		&		&			&		&		&			\\
\noalign{\smallskip}
    \hline
    \noalign{\smallskip}		
 \endfirsthead
\caption[]{Membership and youth features of investigated stars$^{a}$ {(cont.)}.} \\ %
  \hline
  \hline
  \noalign{\smallskip}		
Name	&SpT		&Ref.$^{b}$	&SpT 		&OB	&$\mu$ 	&$\varpi$	&Li~{\sc i}	&H$\alpha$	&mIR	&X rays	&Member	\\
		&Literature	&			&Adopted		&	&		&		&		&			&		&		&			\\
  \noalign{\smallskip}
  \hline
  \noalign{\smallskip}
  \endhead
  \noalign{\smallskip}
  \hline
  \endfoot

TYC 4770-1018-1&...&...&K0:&&$\circ$&$\times$&$\times$&&&&$\times$ \\ % 
\noalign{\smallskip}
Mayrit 1456284&...&...&F7&&$\circ$&$\circ$&&&&$\circ$&$\star$ \\ % 
\noalign{\smallskip}
Mayrit 1415279AB&K2.0$\pm$1.0&Her14&K1&&$\bullet$&$\bullet$&$\circ$&&&$\circ$&$\star$ \\ % 
\noalign{\smallskip}
Mayrit 1374283&...&...&G7&&$\bullet$&$\bullet$&$\circ$&&&$\circ$&$\star$ \\ % 
\noalign{\smallskip}
HD 294270&G0&Nes95&F6:&&$\times$&$\bullet$&&&&&$\times$ \\ % 
\noalign{\smallskip}
HD 294276&G0&Nes95&G4&&$\times$&$\times$&&&&&$\times$ \\ % 
\noalign{\smallskip}
2M J05372885-0255555&...&...&K0&&$\bullet$&$\times$&$\times$&&&&$\times$ \\ % 
\noalign{\smallskip}
TYC 4770-1468-1&...&...&K1&&$\circ$&$\times$&$\times$&&&&$\times$ \\ % 
\noalign{\smallskip}
Mayrit 1329304&M2.0$\pm$0.5&Her14&M2.5&&$\bullet$&$\bullet$&$\bullet$&$\circ$&$\circ$&&$\star$ \\ % 
\noalign{\smallskip}
Mayrit 1298302&M3.0$\pm$0.5&Cab12&M3.0&&&&&&&$\circ$&$\star$ \\ % photometric candidate
\noalign{\smallskip}
Mayrit 1227243&A0&Nes95&A1&&$\circ$&$\circ$&&&&&$\star$ \\ % 
\noalign{\smallskip}
Mayrit 1116300&A3.5$\pm$2.5&Her14&A2&&$\circ$&$\circ$&&&&&$\star$ \\ % 
\noalign{\smallskip}
TYC 4771-621-1&A9.0$\pm$3.0&Her14&F3&&$\times$&$\bullet$&&&&&$\times$ \\ % 
\noalign{\smallskip}
Mayrit 968292&G8--K0 V&Cab08&F7&&$\circ$&$\bullet$&&&&&$\star$ \\ % 
\noalign{\smallskip}
HD 294274&G0&Nes95&G2&&$\times$&$\circ$&&&&&$\times$ \\ % 
\noalign{\smallskip}
SO210868&...&...&G4&&$\bullet$&$\times$&&&&&$\times$ \\ % 
\noalign{\smallskip}
Mayrit 797272&M1--3 V e&Cab08&M2&&$\bullet$&$\bullet$&&&&$\circ$&$\star$ \\ % 
\noalign{\smallskip}
Mayrit 789281&G2.5$\pm$2.0&Her14&G4&&$\bullet$&$\bullet$&$\circ$&$\circ$&&$\circ$&$\star$ \\ % 
\noalign{\smallskip}
Mayrit 1129222&...&...&M2.5:&&$\bullet$&$\bullet$&$\bullet$&$\circ$&&&$\star$ \\ % 
\noalign{\smallskip}
Mayrit 783254&K0:&Cab12&G7&&$\circ$&$\circ$&$\bullet$&&&$\circ$&$\star$ \\ % <<K7 from OSIRIS
\noalign{\smallskip}
[HHM2007] 244&G7.0$\pm$4.0&Her14&K1:&&$\bullet$&$\times$&$\times$&&&&$\times$ \\ % 
\noalign{\smallskip}
HD 294277&K2&Nes95&K3&&$\circ$&$\circ$&$\times$&&&&$\times$ \\ % 
\noalign{\smallskip}
2M J05375789-0259536&...&...&G7&&$\bullet$&$\times$&&&&&$\times$ \\ % 
\noalign{\smallskip}
TYC 4771-720-1&...&...&G6&&$\times$&$\bullet$&&&&&$\times$ \\ % 
\noalign{\smallskip}
Mayrit 873229AB&M4.5$\pm$0.5&Her14&M4.5&&$\bullet$&$\times^{c}$&&$\circ$&$\circ$&&$\star$ \\ % 
\noalign{\smallskip}
[W96] 4771-0950&F3.5$\pm$2.0&Her14&F2&&$\times$&$\circ$&&&&$\circ$&$\times$ \\ % 
\noalign{\smallskip}
Mayrit 662301&M1.0$\pm$1.0e&Cab12&M1.5&&$\bullet$&$\bullet$&$\bullet$&$\circ$&$\circ$&$\circ$&$\star$ \\ % M0 from IDS
\noalign{\smallskip}
Mayrit 615296&K3.0$\pm$1.0&Her14&K3&&$\bullet$&$\bullet$&$\bullet$&&&$\circ$&$\star$ \\ % 
\noalign{\smallskip}
Mayrit 1073209&M3.0$\pm$0.5&Her14&M3.0&&$\bullet$&$\times$&$\bullet$&$\bullet$&$\circ$&&$\star$ \\ % 
\noalign{\smallskip}
[HHM2007] 385&M1.5$\pm$1.0&Her14&M3&&$\times$&$\times$&$\times$&&&&$\times$ \\ % 
\noalign{\smallskip}
Mayrit 757219&M1.5$\pm$0.5&Her14&M1.0&&$\bullet$&$\bullet$&$\circ$&$\circ$&$\circ$&&$\star$ \\ % 
\noalign{\smallskip}
Mayrit 1285339$^{d}$&F7.5$\pm$2.5&Her14&F8&&$\circ$&$\circ$&$\circ$&$\circ$&$\circ$&$\circ$&$\star$ \\ % 
\noalign{\smallskip}
TYC 4771-873-1&...&...&F2&&$\circ$&$\times$&&&&&$\times$ \\ % 
\noalign{\smallskip}
[SE2004] 10&...&...&K1&&$\bullet$&$\times$&$\times$&&&&$\times$ \\ % 
\noalign{\smallskip}
IRAS 05358-0238&M6.0$\pm$0.5&Her14&M7&&$\bullet$&$\times$&$\times$&&$\circ$&&$\times$ \\ % low g 
\noalign{\smallskip}
2M J05382265-0257421&...&...&K0:&&$\bullet$&$\times$&$\times$&&&&$\times$ \\ % 
\noalign{\smallskip}
Mayrit 329261AB&...&...&M4.5&&$\bullet$&$\bullet$&&$\bullet$&$\circ$&&$\star$ \\ % low g 
\noalign{\smallskip}
Mayrit 1449349&K6.5$\pm$1.0&Her14&K5&&$\bullet$&$\bullet$&$\bullet$&$\bullet$&&&$\star$ \\ % 
\noalign{\smallskip}
Mayrit 609206&K7.0$\pm$1.0e&Cab12&M0:&&$\bullet$&$\bullet$&$\circ$&$\circ$&$\circ$&$\circ$&$\star$ \\ % K4: from IDS
\noalign{\smallskip}
Mayrit 521210&A8.0$\pm$2.5&Her14&A2&&$\circ$&$\bullet$&&&&&$\star$ \\ % 
\noalign{\smallskip}
HD 294280&K5&Nes95&K5&&$\circ$&$\times$&$\times$&&&&$\times$ \\ % 
\noalign{\smallskip}
Mayrit 1207349&K6.5$\pm$1.5&Her14&M0.0&&$\bullet$&$\bullet$&$\circ$&$\circ$&$\circ$&&$\star$ \\ % K5 from IDS
\noalign{\smallskip}
Mayrit 1275190$^{d}$&...&...&F9&&$\bullet$&$\bullet$&&&&&$\star$ \\ % 
\noalign{\smallskip}
Mayrit 1160190&A5&Nes95&F1&&$\circ$&$\circ$&&&&&$\star$ \\ % 
\noalign{\smallskip}
Mayrit 203283&M0.0&Sac08&M0&&&&$\circ$&$\circ$&$\circ$&$\circ$&$\star$ \\ % 
\noalign{\smallskip}
Mayrit 180277&M1.0$\pm$1.0&Cab12&M0.0&&$\bullet$&$\bullet$&$\bullet$&&&$\circ$&$\star$ \\ % K5 from IDS
\noalign{\smallskip}
Mayrit 521199&K4.5$\pm$1.5&Her14&K1&&$\bullet$&$\bullet$&$\bullet$&$\circ$&$\circ$&$\circ$&$\star$ \\ % M0: from OSIRIS
\noalign{\smallskip}
Mayrit 165257&M3.5&Sac08&M4&&$\bullet$&$\bullet$&$\circ$&$\circ$&$\circ$&$\circ$&$\star$ \\ % 
\noalign{\smallskip}
Mayrit 189303&B9.0$\pm$1.0&Her14&B7&$\circ$&$\circ$&$\circ$&&&&$\circ$&$\star$ \\ % 
\noalign{\smallskip}
Mayrit 168291AB&K1.0&Sac08&K5&&$\bullet$&$\bullet$&$\bullet$&&&$\circ$&$\star$ \\ % 
\noalign{\smallskip}
StHa 50&A2--6Ve&Cab08&B6&$\circ$&$\circ$&$\times$&&$\circ$&&&$\times$ \\ % 
\noalign{\smallskip}
[W96] pJ053834-0239&M1.0&Sac08&M0&&$\bullet$&$\times$&$\times$&&&&$\times$ \\ % 
\noalign{\smallskip}
Mayrit 182305&B9.5$\pm$1.0&Her14&A1&$\circ$&$\circ$&$\circ$&&&&$\circ$&$\star$ \\ % 
\noalign{\smallskip}
IDS 05335-0238 D&B9.0$\pm$1.5&Her14&G7&&$\times$&$\bullet$&&&&&$\times$ \\ % 
\noalign{\smallskip}
Mayrit 285331&K7.5&Sac08&K4&&$\bullet$&$\bullet$&$\bullet$&&&$\circ$&$\star$ \\ % 
\noalign{\smallskip}
Mayrit 344337AB&K8.0&Sac08&K4&&$\bullet$&$\bullet$&$\circ$&$\circ$&&$\circ$&$\star$ \\ % 
\noalign{\smallskip}
Mayrit 489196&K3.0$\pm$3.0&Her14&K0:&&$\bullet$&$\bullet$&$\bullet$&$\circ$&$\circ$&&$\star$ \\ % 
\noalign{\smallskip}
Mayrit 208324&B8&Nes95&B5&$\circ$&$\circ$&$\bullet$&&&&&$\star$ \\ % 
\noalign{\smallskip}
Mayrit 105249&K8.0&Sac08&K5&&$\bullet$&$\bullet$&$\circ$&$\circ$&&$\circ$&$\star$ \\ % 
\noalign{\smallskip}
Mayrit 114305AB&K2.0$\pm$1.0&Her14&K2&&$\bullet$&$\bullet$&$\circ$&$\circ$&$\circ$&$\circ$&$\star$ \\ % SB Bej01
\noalign{\smallskip}
HD 294278&K2&Nes95&K0&&$\times$&$\circ$&$\times$&&&&$\times$ \\ % 
\noalign{\smallskip}
[HHM2007] 648&G4.5$\pm$2.5&Her14&G5&&$\bullet$&$\times$&&&&&$\times$ \\ % 
\noalign{\smallskip}
[SE2004] 30&K5.5$\pm$1.5&Her14&K4&&$\bullet$&$\times$&$\times$&&&&$\times$ \\ % 
\noalign{\smallskip}
Mayrit 1248183AB&K7.5$\pm$0.5&Her14&K5&&$\bullet$&$\bullet$&$\circ$&&&&$\star$ \\ % 
\noalign{\smallskip}
Mayrit 348349&M0.0&Sac08&M0&&$\bullet$&$\bullet$&$\circ$&$\circ$&$\circ$&$\circ$&$\star$ \\ % 
\noalign{\smallskip}
Mayrit 97212&K9.5&Sac08&K5&&$\bullet$&$\bullet$&$\circ$&$\circ$&&$\circ$&$\star$ \\ % 
\noalign{\smallskip}
Mayrit 83207&M0.5&Sac08&M0&&$\bullet$&$\bullet$&&&$\circ$&$\circ$&$\star$ \\ %  
\noalign{\smallskip}
Mayrit 156353&M1.0$\pm$1.0&Cab12&M1.0&&$\bullet$&$\bullet$&$\bullet$&&&$\circ$&$\star$ \\ % M0 from IDS
\noalign{\smallskip}
Mayrit 11238&A2V&GK58&A2&&$\circ$&$\bullet$&&&&&$\star$ \\ % 
\noalign{\smallskip}
Mayrit 260182&K7.5&ZO02&K5&&$\bullet$&$\bullet$&$\circ$&$\circ$&$\circ$&$\circ$&$\star$ \\ % 
\noalign{\smallskip}
Mayrit 207358&K7.0$\pm$1.0&Her14&K4&&$\bullet$&$\bullet$&$\circ$&&&$\circ$&$\star$ \\ % 
\noalign{\smallskip}
$\sigma$ Ori AB&O9.5+&SD15&O9.5V&$\circ$&$\circ$&$\circ$&&&$\circ$&$\circ$&$\star$ \\ % 
\noalign{\smallskip}
[W96] pJ053844-0233&G5.0$\pm$2.5&Her14&K0&&$\bullet$&$\times$&$\times$&&&&$\times$ \\ % 
\noalign{\smallskip}
Mayrit 359179AB&M1.0&Sac08&M1&&$\bullet$&$\times^{c}$&&$\circ$&$\circ$&$\circ$&$\star$ \\ % 
\noalign{\smallskip}
Mayrit 13084&B2.0$\pm$1.5&Her14&B2&$\circ$&$\circ$&$\circ$&&&&$\circ$&$\star$ \\ % 
\noalign{\smallskip}
Mayrit 1316178&...&...&M6.5&&$\bullet$&$\bullet$&$\circ$&$\bullet$&$\circ$&&$\star$ \\ % low g 
\noalign{\smallskip}
Mayrit 42062&B2.0$\pm$1.5&Her14&B2&$\circ$&$\circ$&$\bullet$&&&&$\circ$&$\star$ \\ % 
\noalign{\smallskip}
Mayrit 53049&M1.0&Sac08&M1&&$\bullet$&$\bullet$&$\circ$&$\circ$&$\circ$&$\circ$&$\star$ \\ % 
\noalign{\smallskip}
Mayrit 528005AB$^{c}$&K7.0$\pm$1.0&Cab12&M0.0&&$\circ$&&$\circ$&$\circ$&$\circ$&$\circ$&$\star$ \\ % K5 from IDS
\noalign{\smallskip}
Mayrit 157155&M0.5&Sac08&K7&&$\bullet$&$\bullet$&$\circ$&$\circ$&&$\circ$&$\star$ \\ % 
\noalign{\smallskip}
Mayrit 332168AB&M3.5&Sac08&M6&&$\bullet$&$\bullet$&$\bullet$&&&$\circ$&$\star$ \\ % 
\noalign{\smallskip}
Mayrit 653170&K9.0&Sac08&K7:&&$\bullet$&$\bullet$&$\circ$&$\circ$&$\circ$&$\circ$&$\star$ \\ % 
\noalign{\smallskip}
Mayrit 203039&K5&Wol96&K3&&$\bullet$&$\bullet$&$\circ$&$\circ$&&$\circ$&$\star$ \\ % 
\noalign{\smallskip}
Mayrit 822170$^{d}$&K1.0$\pm$1.5&Her14&K5:&&$\bullet$&$\bullet$&$\circ$&$\circ$&&$\circ$&$\star$ \\ % 
\noalign{\smallskip}
Mayrit 707162AB&M2.0$\pm$1.0&Her14&K5&&$\bullet$&$\bullet$&$\circ$&$\circ$&$\circ$&$\circ$&$\star$ \\ % 
\noalign{\smallskip}
[HHM2007] 829&...&...&K0&&$\times$&$\times$&&&&&$\times$ \\ % 
\noalign{\smallskip}
Mayrit 591158&F4.5$\pm$2.0&Her14&F3&&$\bullet$&$\bullet$&$\circ$&&&$\circ$&$\star$ \\ % 
\noalign{\smallskip}
Mayrit 1082013&M0.5$\pm$0.5&Her14&M1.0&&$\bullet$&$\bullet$&$\circ$&$\circ$&$\circ$&&$\star$ \\ % M0 from IDS
\noalign{\smallskip}
[HHM2007] 846&M1.5$\pm$1.0&Her14&M3&&$\circ$&$\times$&$\times$&&&&$\times$ \\ % 
\noalign{\smallskip}
Mayrit 306125AB&B5.0$\pm$1.5&Her14&B5&$\circ$&$\circ$&$\circ$&&&&$\circ$&$\star$ \\ % 
\noalign{\smallskip}
Mayrit 687156&M3.5$\pm$1.0&Her14&M4.5&&$\bullet$&$\bullet$&$\bullet$&$\circ$&$\circ$&&$\star$ \\ % 
\noalign{\smallskip}
Mayrit 374056&K5&Man13&K5&&$\bullet$&$\bullet$&$\circ$&$\circ$&&$\circ$&$\star$ \\ % 
\noalign{\smallskip}
Mayrit 397060&K8.0&Sac08&K5&&$\bullet$&$\bullet$&$\bullet$&$\circ$&$\circ$&$\circ$&$\star$ \\ % 
\noalign{\smallskip}
Mayrit 1011159&M0.5$\pm$0.5&Her14&M0&&$\bullet$&$\bullet$&$\circ$&&$\circ$&&$\star$ \\ % 
\noalign{\smallskip}
Mayrit 1288163&B9.0$\pm$1.0&Her14&B8&$\circ$&$\circ$&$\bullet$&&&&&$\star$ \\ %
\noalign{\smallskip}
Mayrit 497054&M0.5&Sac08&M0&&$\bullet$&$\bullet$&$\circ$&$\circ$&$\circ$&$\circ$&$\star$ \\ % 
\noalign{\smallskip}
Mayrit 403090&K8.0&Sac08&K5&&$\bullet$&$\bullet$&$\circ$&$\circ$&&&$\star$ \\ % 
\noalign{\smallskip}
Haro 5-17&...&...&K0&&$\bullet$&$\bullet$&$\bullet$&$\bullet$&&&$\times$ \\ % 
\noalign{\smallskip}
TYC 4771-1012-1&...&...&K3&&$\times$&$\times$&$\times$&&&&$\times$ \\ % 
\noalign{\smallskip}
Mayrit 524060&A6.0$\pm$2.0&Her14&A3&&$\circ$&$\circ$&&&&$\circ$&$\star$ \\ % 
\noalign{\smallskip}
[HHM2007] 961&K1.5$\pm$3.5&Her14&G6&&$\times$&$\times$&&&&&$\times$ \\ % 
\noalign{\smallskip}
[SE2004] 50&G3.5$\pm$2.0&Her14&K1:&&$\bullet$&$\times$&$\times$&&&&$\times$ \\ % 
\noalign{\smallskip}
Mayrit 634052&G7.5$\pm$2.5&Her14&K0&&$\bullet$&$\bullet$&$\bullet$&&&$\circ$&$\star$ \\ % 
\noalign{\smallskip}
Mayrit 596059&K9.0&Sac08&K7&&$\bullet$&$\bullet$&$\bullet$&$\circ$&$\circ$&$\circ$&$\star$ \\ % 
\noalign{\smallskip}
TYC 4771-661-1&...&...&G2&&$\times$&$\times$&&&&&$\times$ \\ % 
\noalign{\smallskip}
[HHM2007] 1009&G2.0$\pm$3.5&Her14&K1:&&$\bullet$&$\times$&$\times$&&&&$\times$ \\ % 
\noalign{\smallskip}
Mayrit 872139&M5.5$\pm$1.0&Cab12&M5.0&&$\bullet$&$\bullet$&$\circ$&&&&$\star$ \\ % low g 
\noalign{\smallskip}
Mayrit 622103&M0.5&Sac08&K7&&$\bullet$&$\bullet$&$\circ$&$\circ$&$\circ$&$\circ$&$\star$ \\ % 
\noalign{\smallskip}
Mayrit 1403026&K4.5$\pm$2.0&Her14&M0.0&&$\bullet$&$\bullet$&$\circ$&$\circ$&&&$\star$ \\ % K5 from IDS
\noalign{\smallskip}
[HHM2007] 1092&G3.5$\pm$1.5&Her14&K0&&$\bullet$&$\times$&$\times$&&&&$\times$ \\ % 
\noalign{\smallskip}
Mayrit 750107&K7.0$\pm$1.0&Her14&K5&&$\bullet$&$\bullet$&$\circ$&&&$\circ$&$\star$ \\ % 
\noalign{\smallskip}
Mayrit 863116&G1.0$\pm$2.0&Her14&G4&&$\circ$&$\circ$&$\circ$&$\circ$&&$\circ$&$\star$ \\ % 
\noalign{\smallskip}
[HHM2007] 1129&G4.0$\pm$2.0&Her14&K1&&$\bullet$&$\times$&$\times$&&&&$\times$ \\ % 
\noalign{\smallskip}
Mayrit 957055&M0.0$\pm$1.0&Cab12&M0.0&&$\bullet$&$\bullet$&$\bullet$&&&$\circ$&$\star$ \\ % K7 from IDS
\noalign{\smallskip}
Haro 5-28&...&...&F6&&$\times$&$\bullet$&&&&&$\times$ \\ % 
\noalign{\smallskip}
Mayrit 1400036&M0.0$\pm$1.0&Her14&M0.5&&$\bullet$&$\bullet$&$\bullet$&$\circ$&$\circ$&&$\star$ \\ % M0 from IDS
\noalign{\smallskip}
Mayrit 871071&K5.5$\pm$1.0&Her14&K2&&$\bullet$&$\bullet$&$\circ$&$\circ$&$\circ$&&$\star$ \\ % 
\noalign{\smallskip}
Mayrit 931117&K1.0$\pm$2.5&Her14&K1&&$\bullet$&$\bullet$&$\bullet$&$\circ$&$\circ$&$\circ$&$\star$ \\ % <<K7 from OSIRIS
\noalign{\smallskip}
Mayrit 1233042&K5.0$\pm$1.0&Her14&K4&&$\bullet$&$\bullet$&$\circ$&$\circ$&$\circ$&&$\star$ \\ % K7.0 from OSIRIS
\noalign{\smallskip}
HD 294299&F2&Nes95&A6&&$\bullet$&$\times$&&&&&$\times$ \\ % 
\noalign{\smallskip}
Mayrit 1626148AB&...&...&M1.5&&&&$\circ$&$\bullet$&&&$\star$ \\ % M0 from IDS
\noalign{\smallskip}
TYC 4771-934-1&...&...&K0&&$\circ$&$\times$&$\times$&&&&$\times$ \\ % 
\noalign{\smallskip}
Mayrit 897077&...&...&M4.5&&$\bullet$&$\bullet$&$\bullet$&&$\circ$&&$\star$ \\ % 
\noalign{\smallskip}
Mayrit 1120128&...&...&M6.0&&$\bullet$&$\bullet$&$\circ$&&&&$\star$ \\ % low g 
\noalign{\smallskip}
[HHM2007] 1189&K1.5$\pm$2.5&Her14&K2&&$\bullet$&$\times$&$\times$&&&&$\times$ \\ % 
\noalign{\smallskip}
Mayrit 960106AB&B8.5$\pm$1.0&Her14&A3&&$\circ$&$\circ$&&&&$\circ$&$\star$ \\ % 
\noalign{\smallskip}
Mayrit 1106058AB&...&...&K0&&$\bullet$&$\bullet$&$\bullet$&&&$\circ$&$\star$ \\ % 
\noalign{\smallskip}
Mayrit 969077&M0.0$\pm$1.0&Her14&K4&&$\bullet$&$\bullet$&$\circ$&$\circ$&&$\circ$&$\star$ \\ % 
\noalign{\smallskip}
Mayrit 1045067AB&...&...&M3.0&&$\bullet$&$\bullet$&&&&&$\star$ \\ % low g 
\noalign{\smallskip}
Mayrit 1082115&M1.0$\pm$0.5&Her14&M0&&&&$\bullet$&&&&$\star$ \\ % 
\noalign{\smallskip}
Mayrit 1279052&M5.5&ZO02&M5.5&&$\bullet$&$\bullet$&$\circ$&$\circ$&$\circ$&&$\star$ \\ % 
\noalign{\smallskip}
Mayrit 1042077&K7.0$\pm$1.5&Her14&K7&&$\bullet$&$\bullet$&$\bullet$&&&&$\star$ \\ % 
\noalign{\smallskip}
[HHM2007] 1251&K5.0$\pm$2.5&Her14&K4&&$\bullet$&$\times$&$\times$&&&&$\times$ \\ % 
\noalign{\smallskip}
[HHM2007] 1256&G2.0$\pm$2.5&Her14&G:&&$\bullet$&$\times$&&&&&$\times$ \\ % 
\noalign{\smallskip}
Mayrit 1041082&M2.5$\pm$1.0&Her14&M3.0&&$\bullet$&$\bullet$&$\circ$&$\circ$&$\circ$&&$\star$ \\ % 
\noalign{\smallskip}
Mayrit 1245057AB&M1.0$\pm$0.5&Her14&M2.5&&$\bullet$&$\bullet$&$\bullet$&$\bullet$&$\circ$&&$\star$ \\ % M3 from IDS
\noalign{\smallskip}
%Mayrit 1245057A&M1.0$\pm$0.5&Her14&M6&&$\bullet$&$\bullet$&$\bullet$&&&&$\star$ \\ % 
%\noalign{\smallskip}
%Mayrit 1245057B&M1.0$\pm$0.5&Her14&M3&&$\bullet$&$\bullet$&$\bullet$&$\bullet$&&&$\star$ \\ % 
%\noalign{\smallskip}
[HHM2007] 1269&G5.0$\pm$3.0&Her14&K0&&$\bullet$&$\times$&$\times$&&&&$\times$ \\ % 
\noalign{\smallskip}
Mayrit 1223121&K6.0$\pm$1.0&Her14&K5&&$\bullet$&$\bullet$&$\circ$&$\circ$&$\circ$&&$\star$ \\ % <K7 from OSIRIS
\noalign{\smallskip}
Mayrit 1366055&G2.0$\pm$2.5&Her14&G2&&$\circ$&$\bullet$&$\circ$&&&$\circ$&$\star$ \\ % 
\noalign{\smallskip}
Mayrit 1446053&M4.0&ZO02&M4.5&&$\bullet$&$\bullet$&$\circ$&$\circ$&$\circ$&&$\star$ \\ % 
\noalign{\smallskip}
UCAC2 30800287&...&...&M2&&$\circ$&$\times$&$\times$&&&&$\times$ \\ % 
\noalign{\smallskip}
Mayrit 1250070&K7.0$\pm$1.0&Her14&K7&&$\bullet$&$\bullet$&$\circ$&&&&$\star$ \\ % 
\noalign{\smallskip}
Mayrit 1652046&...&...&M0.5&&$\bullet$&$\bullet$&$\circ$&$\circ$&&&$\star$ \\ % M0 from IDS
\noalign{\smallskip}
Mayrit 1196092&M6.0$\pm$1.0e&Cab12&M6.5&&$\bullet$&$\bullet$&$\bullet$&$\circ$&$\circ$&&$\star$ \\ % low g 
\noalign{\smallskip}
Mayrit 1541051&M0.5$\pm$0.5&Her14&M0&&$\bullet$&$\bullet$&$\circ$&$\circ$&&$\circ$&$\star$ \\ % 
\noalign{\smallskip}
[HHM2007] 1347&G2.0$\pm$2.0&Her14&K0:&&$\bullet$&$\times$&$\times$&&&&$\times$ \\ % 
\noalign{\smallskip}
Mayrit 1311070&F7.5$\pm$2.0&Her14&F6&&$\bullet$&$\bullet$&&&&&$\star$ \\ % 
\noalign{\smallskip}
Mayrit 1273081&...&...&M5.5&&$\bullet$&$\bullet$&$\bullet$&&&&$\star$ \\ % M7 from IDS
\noalign{\smallskip}
Mayrit 1269083&K7.5$\pm$1.0&Her14&K7&&$\bullet$&$\bullet$&$\circ$&$\circ$&$\circ$&&$\star$ \\ % 
\noalign{\smallskip}
HD 294307&F8&Nes95&F7&&$\times$&$\times$&&&&&$\times$ \\ % 
\noalign{\smallskip}
Mayrit 1396071&K6.5$\pm$1.5&Her14&K7&&$\bullet$&$\bullet$&$\bullet$&$\bullet$&&$\circ$&$\star$ \\ % 
\noalign{\smallskip}
Mayrit 1564058&K5.5$\pm$1.0&Her14&K5&&$\bullet$&$\bullet$&$\bullet$&$\circ$&$\circ$&&$\star$ \\ % M0: from OSIRIS
\noalign{\smallskip}
Mayrit 1359077&B9.5V& {HS99} &B9&$\circ$&$\circ$&$\bullet$&&&&&$\star$ \\ % 
\noalign{\smallskip}
Mayrit 1364078&...&...&M6.5&&$\bullet$&$\bullet$&$\circ$&$\circ$&&&$\star$ \\ % low g 
\noalign{\smallskip}
Mayrit 1500066&...&...&M3&&&&&&&&$\star$ \\ % photometric candidate
\noalign{\smallskip}
Mayrit 1748052&...&...&M1.5&&$\bullet$&$\bullet$&$\circ$&$\circ$&&&$\star$ \\ % M1 from IDS
\noalign{\smallskip}
Mayrit 1548068&B4.0$\pm$1.5&Her14&B2&$\circ$&$\circ$&$\circ$&&&$\circ$&$\circ$&$\star$ \\ % 
\noalign{\smallskip}
Mayrit 1476077&...&...&K4&&$\bullet$&$\bullet$&$\bullet$&$\bullet$&&&$\star$ \\ % 
\noalign{\smallskip}
HD 294301&...&...&F2&&$\times$&$\circ$&&&&&$\times$ \\ % 
\noalign{\smallskip}
Mayrit 1471085&...&...&K5&&$\bullet$&$\bullet$&$\circ$&$\circ$&&$\circ$&$\star$ \\ % 
\noalign{\smallskip}
HD 294297&G0&Nes95&G2:&&$\times$&$\times$&&&&&$\times$ \\ % 
\noalign{\smallskip}
Mayrit 1679078&...&...&K5&&$\bullet$&$\bullet$&$\bullet$&&&$\circ$&$\star$ \\ % 
\noalign{\smallskip}
Haro 5-40&...&...&M4.0&&$\bullet$&$\bullet$&&$\bullet$&&&$\times$ \\ % 
\noalign{\smallskip}
Haro 5-44&...&...&M3:&&$\bullet$&$\bullet$&&$\bullet$&&&$\times$ \\ % 
\noalign{\smallskip}
Haro 5-46&...&...&M2.5&&$\bullet$&$\bullet$&$\bullet$&$\bullet$&&&$\times$ \\ % 
\noalign{\smallskip}
\end{longtable}
}
\begin{list}{}{}
\small{
\item[$^{a}$] {\bf Symbols.} 
Filled circles: 
for $\mu$ and $\varpi$, stars inside the cluster member boundaries, and for Li~{\sc i} and H$\alpha$, stars for which we measured such lines for the first time;
open circles: stars with previous measurements from the literature; 
filled stars: $\sigma$~Orions members in this work; 
crosses: for $\mu$ and $\varpi$, stars outside the cluster member boundaries, for Li~{\sc i}, K and M stars for which we have not found Li~{\sc i} in absorption, and for the last column, fore- or background stars.
\item[$^{b}$] {\bf References.} 
{GK58}: \citet{1958ApJ...127..172G}; 
Gue81: \citet{1981AJ.....86.1057G}; 
Nes95: \citet{1995A&AS..110..367N}; 
Wol96: \citet{1996PhDT........63W}; 
{HS99}: \citet{1999mctd.book.....H}; 
ZO02: \citet{2002A&A...384..937Z}; 
Sac08: \citet{2008A&A...488..167S}; 
Cab08: \citet{2008A&A...491..515C}; 
Cab12: \citet{2012A&A...546A..59C}; 
Man13: \citet{2013A&A...551A.107M}; 
Her14: \citet{2014ApJ...794...36H}.
\item[$^{c}$] Parallax affected by close binarity.
\item[$^{d}$] Discordance in radial velocity.
}
\end{list}

\end{document}